\documentclass[11pt]{article} 
\usepackage{amsfonts,amsmath,amssymb,bbm,hyperref}
\usepackage [amsthm,thmmarks,amsmath]{ntheorem} 
\usepackage[arrow, matrix, curve, all, cmtip]{xy}
\usepackage{color,psfrag}
\usepackage[dvips]{graphicx}
\usepackage{slashed}
\usepackage{exscale}

\newtheoremstyle{breakdef}%
  {\item[\rlap{\vbox{\normalfont\bfseries\hbox{\llap{##2}\hskip\labelsep
          ##1:}\hbox{\\[0.1cm]}}}]}%
  {\item[\rlap{\vbox{\normalfont\bfseries\hbox{\llap{##2}\hskip\labelsep
          ##1 (##3):}\hbox{\\[0.1cm]}}}]}
\newtheoremstyle{breaksatz}%
  {\item[\rlap{\vbox{\normalfont\normalsize\bfseries\hbox{\llap{##2}\hskip\labelsep
          ##1:}\hbox{\\[0.1cm]}}}]}%
  {\item[\rlap{\vbox{\normalfont\normalsize\bfseries\hbox{\llap{##2}\hskip\labelsep
          ##1 (##3):}\hbox{\\[0.1cm]}}}]}
\newtheoremstyle{breaklem}%
  {\item[\rlap{\vbox{\normalfont\normalsize\bfseries\hbox{\llap{##2}\hskip\labelsep
          ##1:}\hbox{\\[0.1cm]}}}]}%
  {\item[\rlap{\vbox{\normalfont\normalsize\bfseries\hbox{\llap{##2}\hskip\labelsep
          ##1 (##3):}\hbox{\\[0.1cm]}}}]}
\newtheoremstyle{breakprop}%
  {\item[\rlap{\vbox{\normalfont\normalsize\bfseries\hbox{\llap{##2}\hskip\labelsep
          ##1:}\hbox{\\[0.1cm]}}}]}%
  {\item[\rlap{\vbox{\normalfont\normalsize\bfseries\hbox{\llap{##2}\hskip\labelsep
          ##1 (##3):}\hbox{\\[0.1cm]}}}]}
\newtheoremstyle{breakbem}%
  {\item[\rlap{\vbox{\hbox{\hskip\labelsep\normalfont\bfseries
          ##1 ##2:}\hbox{\\[0.1cm]}}}]}%
  {\item[\rlap{\vbox{\hbox{\hskip\labelsep\normalfont\bfseries
          ##1 ##2 (##3):}\hbox{\\[0.1cm]}}}]}
\newtheoremstyle{breakbsp}%
  {\item[\rlap{\vbox{\hbox{\hskip\labelsep\normalfont\bfseries
          ##1 ##2:}\hbox{\\[0.2cm]}}}]}%
  {\item[\rlap{\vbox{\hbox{\hskip\labelsep\normalfont\bfseries
          ##1 ##2 (##3):}\hbox{\\[0.2cm]}}}]}
\newtheoremstyle{breakkor}%
  {\item[\rlap{\vbox{\hbox{\hskip\labelsep\normalfont\bfseries
          ##1 ##2:}\hbox{\\[0.1cm]}}}]}%
  {\item[\rlap{\vbox{\hbox{\hskip\labelsep\normalfont\bfseries
          ##1 ##2 (##3):}\hbox{\\[0.1cm]}}}]}
\newtheoremstyle{proof}%
  {\item[\rlap{\vbox{\hbox{\hskip\labelsep\normalfont\bfseries
          \underline{##1:}}\hbox{\\[0.1cm]}}}]}%
  {\item[\rlap{\vbox{\hbox{\hskip\labelsep\normalfont\bfseries
          \underline{##1 (##3):}}\hbox{\\[0.1cm]}}}]}
\theoremstyle{breakkor} 			
\newtheorem{Definition}{Definition}[section] 	
\theoremstyle{breakkor}

\newtheorem{Theorem}[Definition]{Theorem}
\theoremstyle{breakkor}

\theoremstyle{breakkor}
\newtheorem{Proposition}[Definition]{Proposition}
\theoremstyle{breakkor}

\newtheorem{Corollary}[Definition]{Corollary}
\theoremstyle{breakkor}
\theorembodyfont{\normalfont}

\theoremstyle{proof}

\newtheorem{Proof}{Proof}

\newcommand{\id}{\textup{id}}

\newcommand{\m}{\mbox{}}
\newcommand{\be}{\begin{eqnarray}}
\newcommand{\ee}{\end{eqnarray}}

\newcommand{\ket}[1]{\left| #1 \right\rangle}

\title{{\sf On a partially reduced phase space quantisation of general relativity }\\
{\sf conformally coupled to a scalar field}} 
\author{
{\sf N. Bodendorfer}$^1$\thanks{{\sf 
norbert.bodendorfer@gravity.fau.de}},
{\sf A. Stottmeister}$^1$\thanks{{\sf 
alexander.stottmeister@gravity.fau.de}},
{\sf A. Thurn}$^1$\thanks{{\sf 
andreas.thurn@gravity.fau.de}}\\
\\
{\sf $^1$ Inst. for Theoretical Physics III, FAU Erlangen -- N\"urnberg,}\\
{\sf Staudtstr. 7, 91058 Erlangen, Germany}
}
\date{{\small\sf \today}}

\makeatletter
\@addtoreset{equation}{section}
\makeatother

\begin{document}

\maketitle

{\sf
\begin{abstract}
The purpose of this paper is twofold: On the one hand, after a thorough review of the matter free case, we supplement the derivations in our companion paper on ``loop quantum gravity without the Hamiltonian constraint'' with calculational details and extend the results to standard model matter, a cosmological constant, and non-compact spatial slices. On the other hand, we provide a discussion on the role of observables, focussed on the situation of a symmetry exchange, which is key to our derivation. Furthermore, we comment on the relation of our model to reduced phase space quantisations based on deparametrisation.  
\end{abstract}
}

\newpage

\tableofcontents

\newpage

\section{Introduction}
\label{sec:Introduction}

As was outlined in our companion paper \cite{BSTI}, general relativity conformally coupled to a scalar field can be quantised on a phase space which is reduced with respect to the Hamiltonian constraint $\mathcal{H}$ and the generator of local conformal\footnote{By a local conformal transformation, we mean a local rescaling of the phase space variables by a scalar function. With the conformal group in mind, one could also call this a dilatation, however, we will use the former expression in order to be consistent with the remaining literature on the subject.} transformations $\mathcal{D}$, acting on both the metric and the scalar field. Certain technical assumptions, such as compact spatial slices, a strictly positive scalar field, and the existence of a constant mean curvature (CMC) Cauchy slice were necessary in order to proof this claim.\\
Apart from detailing the calculations, we elaborate on certain conceptual points in this paper. The first question that one generally asks when confronted with this type of reformulation of general relativity is related to Dirac observables. When trading invariance under the flow generated by the Hamiltonian constraint for local conformal invariance, what is essentially what we will be doing, it is at first sight unclear how the same Dirac observables should be selected in both theories. The key to understand their relation is to gauge unfix the reduced theory with respect to $\mathcal{D}$, resulting in a first class formulation with local conformal invariance. We will show that, despite a rather complicated mapping, the two observable algebras are in fact isomorphic, which is a general, already known, result from gauge unfixing. The proof rests on the observation that the gauge unfixing projector used to construct gauge invariant extensions from arbitrary phase space functions is a Poisson homomorphism, i.e. it preserves the Poisson structure of the observable algebra\footnote{It is even a (weak) Poisson isomorphism w.r.t. the algebra of weak Dirac observables.}. A word of caution is in order at this point: Although the observable algebras are isomorphic, the mapping between them is very complicated and we observe that, although the constraint algebra is in this case significantly simplified by trading invariances, the complexity of general relativity is conserved for generic problems, e.g. time evolution, where this mapping is needed in order to recover a spacetime picture. Nevertheless, our framework still has its merits in special situations like black hole entropy state counting or in the definition of geometric operators whose classical counterparts serve as good initial data\footnote{The basic idea of the argument in \cite{BSTI} was that having initial data invariant under local conformal transformations generated by $\mathcal{D}$ is as good as having it satisfy $\mathcal{H}=0$.}.\\
It is important to stress the conceptual difference between this reduced phase space quantisation and others available in the literature \cite{GieselAQG4, DomagalaGravityQuantizedLoop, HusainTimeAndA}, which are based on using additional scalar fields or dust as clocks and rods in relation to which a physical Hamiltonian can be defined. While one could try to promote $\mathcal{D}$, the generator of local conformal transformations, to a time variable and compute a physical Hamiltonian which governs the evolution with respect to this clock, this is not our aim. Instead, we want to restrict ourselves to the spatial hypersurface on which $\mathcal{D}=0$ and use the Dirac bracket derived from this gauge fixing. When good initial data, i.e. satisfying $\mathcal{H}=0$, is given on such a hypersurface, the initial value problem for general relativity is well posed modulo technical assumptions (cf. e.g. \cite{WaldGeneralRelativity} for an exposition). Accordingly, the complete spacetime can be recovered from this initial data. When quantising this phase space of initial values, neglecting the spatial diffeomorphism constraint for a moment, there are no additional quantum Einstein equations to be satisfied as long as our representation is constructed according to the Dirac bracket. Thus, the spectral analysis of the operator representation of the reduced phase space functions yields the spectrum of the initial values allowed by the quantum theory. \\
At this point, one might object that nothing has been gained with respect to the deparametrised treatments, \cite{GieselAQG4, DomagalaGravityQuantizedLoop, HusainTimeAndA}, where explicit access to the physical Hilbert space is also available. However, our gauge choice is conceptually very different in that we are fixing the momenta instead of the fields themselves and that our gauge fixing (clock) is purely geometric. This allows the gauge to be accessible on the whole spatial slice in interesting applications like black hole entropy computations, where the access to the physical Hilbert space at a given ``time'' is necessary.\\
\\
This paper is organised as follows: \\
We start in section 2 with a general discussion on observables and the observable projectors used later in this paper. In section 3, we review results from shape dynamics, on which parts of the ideas of this work are based. Next, in section 4, we introduce the conformally coupled scalar field, perform a detailed canonical analysis and construct a new metric which is invariant under local conformal transformations. In section 5, we extend our framework to standard model matter and a cosmological constant. We conclude with a discussion on the implications and applicability of the proposed reduced phase space quantisation. In the appendices, we provide some further calculational details and helpful identities. 

\section{Partial and complete gauge (un)fixing}
\label{sec:PartialComplete}
In this section, we discuss some general facts involved in the mechanism of gauge fixing and unfixing (see \cite{MitraGaugeInvariantReformulationAnomalous, AnishettyGaugeInvarianceIn, VytheeswaranGaugeUnfixingIn, HenneauxQuantizationOfGauge, DittrichPartialAndComplete, RovelliWhatIsObservable, RovelliPartialObservables, ThiemannReducedPhaseSpace} for original literature) of (totally constrained, first class, irreducible)\footnote{An extension to more general situations is conceivable, but technically more involved.} Hamiltonian systems.\\
To fix the terminology, we recall that gauge fixing for such a systems refers to determining a surface that cuts the constraint surface transversally (w.r.t. the gauge flow). Gauge unfixing refers to the observation that the equations governing the (local) admissibility of a gauge fixing exactly reflect the structure of a second class constraint system, i.e. the transversal surface can be given in terms of another constraint systems which is second class with the first. Therefore, given a second class constraint system, it is possible (under favourable conditions) to select a subsystem of first class, and drop the remaining constraints, which are viewed as gauge fixing conditions of the former. An important point of this is that the choice of the first class subsystem is generally far from unique, which will be exploited in the following.\\
The aim of this section is to provide the general setup for the following chapters, in which we review, and extend, a certain gauge fixing for general relativity (in its ADM formulation), and show that it can further be supplemented by a gauge unfixing leading to an equivalent of general relativity with different constraint algebra, which, if coupled to a conformally coupled scalar field, admits a quantisation by Loop Quantum Gravity methods. Differences in interpretation arise in this approach when it comes to the question of observables, and how their dynamics are to be formulated. The methods involved should be viewed as complementary to the construction of a linking theory, as was done in \cite{GomesTheLinkBetween, GomesEinsteinGravityAs}, and on which the ideas for our approach are based. \\
After giving the general structure of the constraints, where we closely follow the notation in \cite{ThiemannModernCanonicalQuantum} (Chapter 2, p. 82 et seq.), necessary for the steps outlined above, we focus on the (classical) observables, and how their Poisson algebra is characterised  in the different pictures. Concerning technical details of symplectic geometry, especially on infinite dimensional manifolds, we refer the reader to the literature \cite{BinzGeometryOfClassical, ChoquetBruhatGeneralRelativityAnd}. In all that follows, we will assume that the geometrical structures can be modelled as suitable Banach manifolds.\\[0.25cm]
Let $\Gamma$ be the phase space equipped with a 
symplectic structure $\omega_{\Gamma}$, and $\mathcal{F}\left(\Gamma\right)$ a convenient space of functions on it. 
Let $\left\{\mathcal{H}_{j}\right\}_{j\in J}\subset\mathcal{F}\left(\Gamma\right)$ be a first class system
of constraints. 
Now we define the Hamiltonian vector fields $\{\chi_{\mathcal{H}_{j}}\}_{j\in J}$ associated with the constraint system as
\be
\chi_{\mathcal{H}_{j}}:=\left\{\mathcal{H}_{j},\ .\ \right\}_{\Gamma},\ j\in J \text{.}
\ee
Note that, due to the first class property of the constraints, these vector fields have well-defined restrictions to $S_{J(\mathcal{H})}$, thus giving rise to canonical transformations $\alpha_{\beta^{j}\chi_{\mathcal{H}_{j}}}, \left\{\beta_{j}\right\}\in\mathbb{R}^{J}$ via
\be
\dot{x}(t)=\beta^{j}\chi_{\mathcal{H}_{j}|x(t)},
\ee
which leave the constraint surface invariant.
A formal solution is given in terms of the (dual) action on $\mathcal{F}\left(\Gamma\right)$:
\be
\alpha_{t\beta^{j}\chi_{\mathcal{H}_{j}}}^{*}(f)=e^{t\beta^{j}\chi_{\mathcal{H}_{j}}}(f):=\sum_{n=0}^{\infty}\frac{t^{n}}{n!}\left\{\beta^{j}\mathcal{H}_{j},f\right\}_{\Gamma,(n)},\ \ \ f\in\mathcal{F}\left(\Gamma\right).
\ee
The orbits of these transformations, the gauge orbits, are given as
\be
\label{eq:gaugeorbit}
\left[x\right]_{J(\mathcal{H})}:=\left\{\alpha_{\beta^{j_{1}}\chi_{\mathcal{H}_{j_{1}}}}\circ\dots\circ\alpha_{\beta^{j_{N}}\chi_{\mathcal{H}_{j_{N}}}}(x),\ N<\infty,\ \left\{\beta^{j_{k}}\right\}\in\mathbb{R}^{J\times N}\right\}.
\ee
$N\neq1$ accounts for the possibility of structure functions, i.e. $\left\{\mathcal{H}_{j}\right\}_{j\in J}$ will in general only be a generating set of the Poisson algebra of constraints, which means that general gauge transformations, even those connected to the identity, cannot be written in the form $\alpha_{\beta^{j}\chi_{\mathcal{H}_{j}}},\ \left\{\beta^{j}\right\}_{j\in J}$.\\[0.25cm]
Ignoring the subtleties arising from the fact that we allow for infinite dimensional phase spaces, one obtains the intuitive picture that the reduced phase space $\Gamma_{R}$ of the system is given in terms of the gauge orbits, and the constraint surface is a principal bundle over the latter with gauge group given by the canonical transformations defined above (cf. \cite{HenneauxQuantizationOfGauge}).

\subsection{Geometry of the constraints}

Given a system as above, we explain in the following how, under certain conditions, systems or subsystems of constraints can be traded for one another.

\subsubsection{Complete gauge (un)fixing}

Additionally to the data given before, we consider in this subsection a second system of first class phase space functions $\left\{\mathcal{D}_{j}\right\}_{j\in J}\subset\mathcal{F}\left(\Gamma\right)$, which is assumed to give (at least locally) rise to a canonical gauge of the first system $\left\{\mathcal{H}_{j}\right\}_{j\in J}$\footnote{Note that the index sets of the constraints systems are intentionally chosen to coincide. This is indicates that one needs the ``same number'' of gauge conditions as there are gauge symmetry generators.}, i.e.
\be
\label{eq:gfloc}
\left\{\mathcal{D}_{i},\mathcal{H}_{j}\right\}_{\Gamma}=M_{ij},\ \ \ M=\left(M_{ij}\right)_{i,j\in J}\ \textup{invertible on}\ S_{J\left(\mathcal{D},\mathcal{H}\right)},
\ee
where $S_{J\left(\mathcal{D},\mathcal{H}\right)}$ denotes the submanifold defined by the common zeros of both constraint systems. The invertibility of $M$ ensures the (local) accessibility and uniqueness of the gauge\footnote{In the sense that the system of linear equations $M_{ij}\beta^{j}=0,\ i\in J$, admits a unique solution, and is not over-constrained.}. If one is dealing with a finite number of degrees of freedom $(|J|<\infty)$, $M$ will be given by a square matrix with non-vanishing determinant, which follows directly from a dimensionality argument. In case there is an infinite number of degrees of freedom $(|J|=\infty)$, the situation will be more subtle. Besides the uniqueness of the gauge (injectivity of $M$), one has to ensure that the gauge is (infinitesimally) accessible, which is tied to the range of $M$ (surjectivity being sufficient). The first class property for the family $\left\{\mathcal{D}_{j}\right\}_{j\in J}$ only becomes important when we come to the process of gauge unfixing. The gauge fixing itself does not require this assumption. \\
Now, an important observation is the (weak) \textit{Abelianisation Theorem} (s. \cite{DittrichPartialAndComplete} and \cite{ThiemannModernCanonicalQuantum}, p. 84, Theorem 2.2.1.) for the Hamiltonian vector fields:
\begin{Theorem}
\label{thm:AbelComp}
Given constraint systems $\left\{\mathcal{H}_{j}\right\}_{j\in J}$ and $\left\{\mathcal{D}_{j}\right\}_{j\in J}$ as stated, if the operator $M$ is invertible on some open subset $U_{S_{J\left(\mathcal{D},\mathcal{H}\right)}}$ intersecting the (second class) surface $S_{J\left(\mathcal{D},\mathcal{H}\right)}$, one obtains (weakly) commuting Hamiltonian vector fields $\{\chi_{\mathcal{H}'_{j}}\}_{j\in J}$, where $\mathcal{H}'_{j}:=\sum_{i\in J}\left(M^{-1}\right)_{ij}\mathcal{H}_{i}$:
\be
\left[\chi_{\mathcal{H}'_{i}},\chi_{\mathcal{H}'_{j}}\right] & \approx_{S_{J\left(\mathcal{H}\right)}\cap U_{S_{J\left(\mathcal{D},\mathcal{H}\right)}}} 0,\ \ \ i,j\in J,
\ee
i.e. the structure functions $\left\{f^{\mathcal{H}'\ \ k}_{ij}\right\}_{i,j,k\in J}$ vanish weakly, since 
\be
\left[\chi_{\mathcal{H}'_{i}},\chi_{\mathcal{H}'_{j}}\right]=\chi_{\left\{\mathcal{H}'_{i},\mathcal{H}'_{j}\right\}_{\Gamma}}\approx_{S_{J\left(\mathcal{H}\right)}\cap U_{S_{J\left(\mathcal{D},\mathcal{H}\right)}}} f^{\mathcal{H}'\ \ k}_{ij}\chi_{\mathcal{H}'_{k}},\ \ \ i,j,k\in J.
\ee
\end{Theorem}
Moreover, due to the first class property, the primed system of constraints is equivalent to the original one on $U_{S_{J\left(\mathcal{D},\mathcal{H}\right)}}$, and induces an equivalent Hamiltonian flow on (part of) the constraint surface $S_{J\left(\mathcal{H}\right)}$. We will drop the dependence on $U_{S_{J\left(\mathcal{D},\mathcal{H}\right)}}$ in the following, since we want keep the notation compact and the possible dependence is not essential to the stated formulae.\\
Taking advantage of the theorem, the condition for the constraint surface $S_{J\left(\mathcal{D}\right)}$ to define a (local) canonical gauge cut w.r.t. the Hamiltonian flow on $S_{J\left(\mathcal{H}\right)}$ can be stated as the existence and uniqueness of a solution to
\be
\label{eq:gfglob}
\mathcal{D}_{i}(\alpha_{\beta^{j}\chi_{\mathcal{H}'_{j}}}(x))=0,\ \forall i\in J,\ x\in S_{J\left(\mathcal{H}\right)},
\ee
thus defining a function
\be
\label{eq:betafunc}
\beta^{\mathcal{H}'}_{\mathcal{D}}:S_{J\left(\mathcal{H}\right)}
\rightarrow\mathbb{R}^{J}.
\ee
A (formal) differentiation of this condition w.r.t. the labels $\beta^{j},\ j\in J$, leads to an equivalent of (\ref{eq:gfloc}) in terms of $\mathcal{D}_{i},\ \mathcal{H}'_{j},\ i,j\in J$,
\be
\left\{\mathcal{D}_{i},\mathcal{H}'_{j}\right\}_{\Gamma}\approx_{S_{J\left(\mathcal{D},\mathcal{H}\right)}}
\delta_{ij}.
\ee
It is noteworthy that, although (\ref{eq:gfglob}) can be formulated using only unprimed constraints, the resulting family of equations will not be appropriate for the definition of a canonical gauge, unless one is in the case of structure constants $(N=1)$, due to the nature of the gauge orbits\footnote{(\ref{eq:gfglob}) can also be stated with the right hand side being any family of constants $\{t_{i}\}_{i\in J}$. (\ref{eq:gfloc}) then involves the constraint surface defined by $\mathcal{D}_{i}-t_{i}=0,\ i\in J$ }.\\[0.5cm]
The important point, regarding gauge unfixing, is that the role of the two constraint systems is completely interchangeable, and we may define a second set of (weakly) commuting, this time w.r.t. the intersection $S_{J\left(\mathcal{D}\right)}
$, Hamiltonian vector fields $\{\chi_{\mathcal{D}'_{j}}\}_{j\in J}$, $\mathcal{D}'_{j}:=\sum_{i\in J}\left(M^{-1}\right)_{ji}\mathcal{D}_{i}$\footnote{Note the different index structures in the definition of the primed constraints for the two families. This is due to the antisymmetry of the Poisson bracket.}. Furthermore, treating $\left\{\mathcal{H}_{j}\right\}_{j\in J}$ as a set of gauge conditions for the constraint system $\left\{\mathcal{D}_{j}\right\}_{j\in J}$, we may drop the former completely and work exclusively with the latter. However, a word of caution is needed at this point: Although equation (\ref{eq:gfloc}) and the invertibility of $M$ represent the appropriate conditions to either view $\left\{\mathcal{D}_{j}\right\}_{j\in J}$ as a canonical gauge fixing for $\left\{\mathcal{H}_{j}\right\}_{j\in J}$, or vice versa, there may be differences when it comes to finite gauge transformations, and global questions concerning the gauge (cf. \cite{HenneauxQuantizationOfGauge}), e.g. the appearance of Gribov copies (\cite{GribovQuantizationOfNon}), and the problem of ``large gauge transformations". 
The necessary equations to generalise to these cases are (\ref{eq:gfglob}) and:
\be
\label{eq:gfglobD}
\mathcal{H}_{j}(\alpha_{-\gamma^{i}\chi_{\mathcal{D}'_{i}}}(x))=0,\ \forall j\in J,\ x\in S_{J\left(\mathcal{H}\right)}
,
\ee
where we have to replace the $\alpha_{\beta^{j}\chi_{\mathcal{H}'_{j}}},\ \alpha_{-\gamma^{i}\chi_{\mathcal{D}'_{i}}}$ by arbitrary gauge transformations. 

\subsubsection{Partial gauge (un)fixing}

Motivated by the structure of the constraint algebra of general relativity and the idea of partially gauge fixing the Hamiltonian constraints, leaving the spatial diffeomorphism constraints untouched, we introduce a general structure reflecting the important properties of that case.\\
Consider therefore a third (first class) system $\left\{\mathcal{D}_{j_{2}}\right\}_{j_{2}\in J_{2}}\subset\mathcal{F}\left(\Gamma\right)$, which we assume to (locally) canonically gauge fix part of the first constraint system $\left\{\mathcal{H}_{j_{2}}\right\}_{j_{2\in J_{2}}},\ J_{2}\subset J$:
\be
\label{eq:gflocpart}
\left\{\mathcal{D}_{i_{2}},\mathcal{H}_{j_{2}}\right\}_{\Gamma}=M_{i_{2}j_{2}},\ \ \  M=(M_{i_{2}j_{2}})_{i_{2},j_{2}\in J_{2}}\ \textup{invertible on}\ S_{J_{2}\left(\mathcal{D},\mathcal{H}\right)}.
\ee
Additionally, we assume the split $J=J_{1}\cup J_{2},\ J_{1}:=J\setminus J_{2}$ to be such that the family $\left\{\mathcal{H}_{j_{1}}\right\}_{j_{1}\in J_{1}}$ is again first class among itself, and that
\be
\label{eq:consinv}
\left\{\mathcal{H}_{j_{1}},\mathcal{H}_{j_{2}}\right\}_{\Gamma} & = & f^{\mathcal{H}\ \ i_{2}}_{j_{1}j_{2}}\mathcal{H}_{i_{2}}, \\[0.25cm]
\label{eq:consinv2}
\left\{\mathcal{H}_{j_{1}},\mathcal{D}_{j_{2}}\right\}_{\Gamma} & = & f^{\mathcal{D}\ \ i_{2}}_{j_{1}j_{2}}\mathcal{D}_{i_{2}}
\ee
i.e. the Hamiltonian flow generated by the $J_{1}(\mathcal{H})$-subsystem leaves $S_{J_{2}\left(\mathcal{H}\right)}$ and $S_{J_{2}\left(\mathcal{D}\right)}$ invariant\footnote{In principle, relation (\ref{eq:gflocpart}) can be weakened to equality on $S_{J_{1}\left(\mathcal{H}\right),J_{2}\left(\mathcal{D},\mathcal{H}\right)}$. But since the stated (stronger) condition holds in the case we study in subsequent sections, we will not pursue this possibility here.}. Concerning the $J_{2}(\mathcal{H})$-subsystem, we distinguish between two important (non-exhaustive) cases:
\begin{itemize}
	\item[] \textbf{Case I}\\[0.1cm]
	The $J_{2}(\mathcal{H})$-subsystem is even, i.e. $\left\{\mathcal{H}_{i_{2}},\mathcal{H}_{j_{2}}\right\}_{\Gamma}=f^{\mathcal{H}\ \ k_{2}}_{i_{2}j_{2}}\mathcal{H}_{k_{2}}$ (first class).
	\item[] \textbf{Case II}\\[0.1cm]
	The $J_{2}(\mathcal{H})$-subsystem is odd, i.e. $\left\{\mathcal{H}_{i_{2}},\mathcal{H}_{j_{2}}\right\}_{\Gamma}=f^{\mathcal{H}\ \ k_{1}}_{i_{2}j_{2}}\mathcal{H}_{k_{1}}$.
\end{itemize}
Again, one observes that it is possible to choose between two sets of constraints, namely $\left\{\mathcal{H}_{j_{1}},\mathcal{H}_{j_{2}}\right\}_{j_{1}\in J_{1},j_{2}\in J_{2}}$ and $\left\{\mathcal{H}_{j_{1}},\mathcal{D}_{j_{2}}\right\}_{j_{1}\in J_{1},j_{2}\in J_{2}}$, which are both first class. Furthermore, this is independent of the fact whether we deal with case I or II, therefore opening the possibility to change between the two since the family $\left\{\mathcal{D}_{j_{2}}\right\}_{j_{2}\in J_{2}}$ is assumed to be first class (as in case I).\\[0.25cm]
An important difference between case I and II is that only in case I it is possible to generalise the \textit{Abelianisation theorem} to the $J_{2}(\mathcal{H})$-subsystem, since its proof requires the constraints to be first class among themselves. The proof in case I is completely analogous to the one given in \cite{DittrichPartialAndComplete, ThiemannModernCanonicalQuantum}, the only difference being that one has to make use of (\ref{eq:consinv}).
Although the theorem fails to hold in case II, which is the one relevant for general relativity and will be considered in the following sections of the paper, we want to stress that it is still possible to change between the two systems. This stems from the fact that both admit the same (local) partial gauge fixing surface. Furthermore, the theorem is still applicable to the dual constraint system $\left\{\mathcal{H}_{j_{1}},\mathcal{D}_{j_{2}}\right\}_{j_{1}\in J_{1},j_{2}\in J_{2}}$ which is always assumed to be case I (and will be so in the application to general relativity).
\begin{Theorem}
\label{th:AbelPart}
Given constraint system $\left\{\mathcal{H}_{j_{1}}\right\}_{j_{1}\in J_{1}}$, $\left\{\mathcal{H}_{j_{2}}\right\}_{j_{2}\in J_{2}}$ and $\left\{\mathcal{D}_{j_{2}}\right\}_{j_{2}\in J_{2}}$ as stated, and assuming that case I holds, if the operator $M$ is invertible on some open subset $U_{S_{J_{2}\left(\mathcal{D},\mathcal{H}\right)}}$ intersecting the (second class) surface $S_{J_{2}\left(\mathcal{D},\mathcal{H}\right)}$, one obtains (weakly) commuting Hamiltonian vector fields $\{\chi_{\mathcal{H}'_{j_{2}}}\}_{j_{2}\in J_{2}}$, where $\mathcal{H}'_{j_{2}}:=\sum_{i_{2}\in J_{2}}\left(M^{-1}\right)_{i_{2}k_{2}}\mathcal{H}_{i_{2}}$:
\be
\left[\chi_{\mathcal{H}'_{i_{2}}},\chi_{\mathcal{H}'_{j_{2}}}\right] & \approx_{S_{J_{2}\left(\mathcal{H}\right)}\cap U_{S_{J_{2}\left(\mathcal{D},\mathcal{H}\right)}}} 0,\ \ \ i_{2},j_{2}\in J,
\ee
i.e. the structure functions $\left\{f^{\mathcal{H}'\ \ k_{2}}_{i_{2}j_{2}}\right\}_{i_{2},j_{2},k_{2}\in J}$ vanish weakly, since 
\be
\left[\chi_{\mathcal{H}'_{i_{2}}},\chi_{\mathcal{H}'_{j_{2}}}\right]=\chi_{\left\{\mathcal{H}'_{i_{2}},\mathcal{H}'_{j_{2}}\right\}_{\Gamma}}\approx_{S_{J_{2}\left(\mathcal{H}\right)}\cap U_{S_{J_{2}\left(\mathcal{D},\mathcal{H}\right)}}} f^{\mathcal{H}'\ \ k_{2}}_{i_{2}j_{2}}\chi_{\mathcal{H}'_{k_{2}}},\ \ \ i_{2},j_{2},k_{2}\in J.
\ee
\end{Theorem}
Another important difference between case I and II is that of finite gauge transformations, and how to state the (canonical) gauge fixing condition w.r.t. those.\\
Case I is again identical to a situation involving a complete gauge fixing, since the $J_{2}(\mathcal{H})$-subsystem forms an ideal of the $J(\mathcal{H})$-system, and equations (\ref{eq:gfglob}), (\ref{eq:gfglobD}) are well-defined by replacing $(i,j,J)$ with $(i_{2},j_{2},J_{2})$.\\
Case II, relevant for general relativity, is more involved, as the $J_{2}(\mathcal{H})$-subsystem has no well-defined action on its constraint surface $S_{J_{2}\left(\mathcal{H}\right)}$. Similarly, a restriction to $S_{J_{1}\left(\mathcal{H}\right)}$ is impossible due to equation (\ref{eq:consinv}). Only the joint action on $S_{J\left(\mathcal{H}\right)}$ is sensible. The gauge flow transversal to the partial gauge fixing surface $S_{J_{2}\left(\mathcal{D}\right)}$ is generated in first order by $\left\{\mathcal{H}_{j_{2}}\right\}_{j_{2}\in J_{2}}$, since $S_{J_{2}\left(\mathcal{D}\right)}$ is preserved by $\left\{\mathcal{H}_{j_{1}}\right\}_{j_{1}\in J_{1}}$, but higher orders involve the latter as well. Therefore, a generalisation of equation (\ref{eq:gfglob}) might involve a different parametrisation of the gauge flow along $S_{J\left(\mathcal{H}\right)}$, or, if this is not possible, a different choice of $J_{2}(\mathcal{H})$-subsystem, which agrees with the one before in first order at the partial gauge fixing surface. Nevertheless, the equivalent of (\ref{eq:gfglobD}) remains valid, since the $J_{2}(\mathcal{D})$-system obeys case I.

\subsection{Observable projectors}

Let us now turn to the role of observables in the outlined setting. To this end, we stick to the general definition of observable in a classical theory, namely that of (smooth) functions $\mathcal{F}\left(\Gamma_{R}\right)$ on the reduced phase space $\Gamma_{R}$ (the space of orbits, cf. (\ref{eq:gaugeorbit})), which is equivalent to considering gauge invariant functions $\mathcal{O}^{w}_{J\left(\mathcal{H}\right)}$, w.r.t. $\left\{\mathcal{H}_{j}\right\}_{j\in J}$, on the constraint surface $S_{J\left(\mathcal{H}\right)}$ (weak Dirac observables).

\subsubsection{Complete observable projectors}

In case we are dealing with a complete gauge fixing, the observables $\mathcal{O}^{w}_{J\left(\mathcal{H}\right)}$ can be constructed (formally) with the help of a Taylor's series (cf. \cite{ThiemannModernCanonicalQuantum}, p.85), which makes explicit use of the \textit{Abelianisation Theorem}:
\be
\label{eq:comptaylor}
\nonumber & \mathbb{P}_{\mathcal{H}}^{\mathcal{D}}(f) & = \sum_{\left\{k_{j}\right\}_{j\in J}\in\mathbb{N}_{0}^{J}}\left(\prod_{j\in J}\frac{(\mathcal{D}_{j})^{k_{j}}}{k_{j}!}\right)\circ_{j\in J}\left(\chi_{\mathcal{H}'_{j}}\right)^{\circ k_{j}}(f) \\[0.5cm]
 & & \approx_{S_{J\left(\mathcal{H}\right)}}
\alpha_{\beta^{j}\chi_{\mathcal{H}'_{j}}}^{*}(f)_{|\beta=\beta^{\mathcal{H}'}_{\mathcal{D}}},\ \ \ f\in\mathcal{F}\left(\Gamma\right),
\ee
where $\beta^{\mathcal{H}'}_{\mathcal{D}}$ refers to (\ref{eq:betafunc}) and $\circ$ denotes the composition of vector fields..\\
A map $\mathbb{P}_{\mathcal{H}}^{\mathcal{D}}:\mathcal{F}\left(\Gamma\right)\rightarrow\mathcal{O}^{w}_{J\left(\mathcal{H}\right)}$ associated with two constraint systems $\left\{\mathcal{H}_{j}\right\}_{j\in J}$, $\left\{\mathcal{D}_{j}\right\}_{j\in J}$, satisfying the complete gauge fixing properties, will be called a \textit{complete observable projector}.

\newpage

\begin{figure}[h]
\be\nonumber
\xymatrix{
 & \left(\mathcal{F}\left(\Gamma\right),\left\{\ ,\ \right\}_{\textup{DB}\left(\mathcal{D},\mathcal{H}\right)}\right) \ar@/_1pc/[lddd]|{\mathbb{P}^{\mathcal{D}}_{\mathcal{H}}} \ar@/^1pc/[rddd]|{\mathbb{P}^{\mathcal{H}}_{\mathcal{D}}} & \\
 & \left(\mathcal{F}\left(\Gamma_{R}\right),\left\{\ ,\ \right\}_{\Gamma_{R}}\right) & \\
 & & \\
\left(\mathcal{O}^{w}_{J\left(\mathcal{H}\right)},\left\{\ ,\ \right\}_{\Gamma}\right) \ar@/^1pc/[rr]|{\mathbb{P}^{\mathcal{H}}_{\mathcal{D}}} \ar@/^/[uur]|{\pi_{J\left(\mathcal{H}\right)}} \ar@/_/[ddr]|{\iota^{*}_{S_{J\left(\mathcal{D},\mathcal{H}\right)}}} & & \left(\mathcal{O}^{w}_{J\left(\mathcal{D}\right)},\left\{\ ,\ \right\}_{\Gamma}\right) \ar@/^1pc/[ll]|{\mathbb{P}^{\mathcal{D}}_{\mathcal{H}}} \ar@/_/[uul]|{\pi_{J\left(\mathcal{D}\right)}} \ar@/^/[ddl]|{\iota^{*}_{S_{J\left(\mathcal{D},\mathcal{H}\right)}}} \\
 & & \\
 & \left(\mathcal{F}\left(S_{J\left(\mathcal{D},\mathcal{H}\right)}\right),\left\{\ ,\ \right\}_{|S_{J\left(\mathcal{D},\mathcal{H}\right)}}\right) & 
}
\ee
\caption{$\pi_{J\left(\mathcal{H}\right)},\ \pi_{J\left(\mathcal{D}\right)}$ and $\iota^{*}_{S_{J\left(\mathcal{D},\mathcal{H}\right)}}$ are the canonical projections and the restriction to the gauge fixing surface.}
\label{fig:diagcomp}
\end{figure}
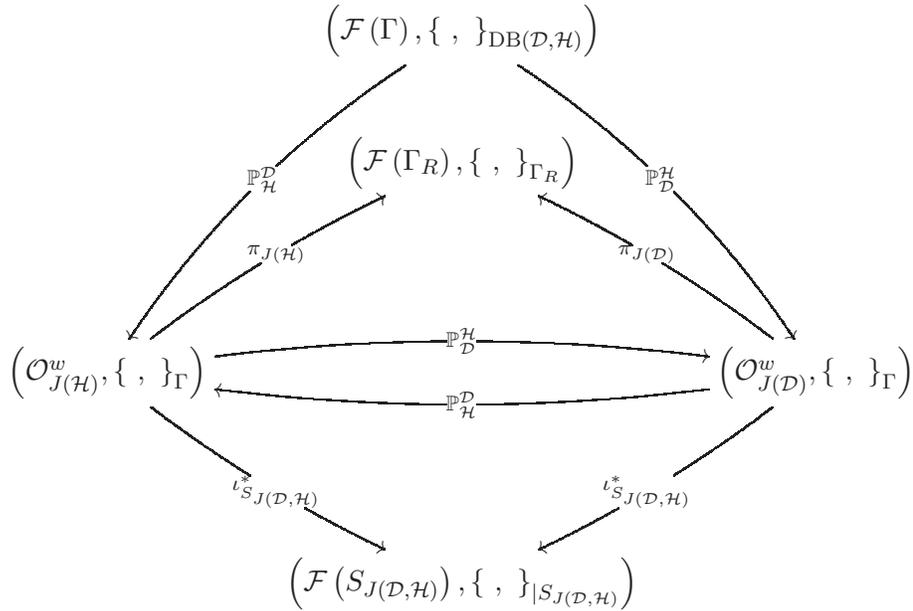

Such a map enjoys the following properties (cf. \cite{ThiemannModernCanonicalQuantum}, p.85, see also \cite{VytheeswaranGaugeUnfixingIn} for an earlier statement of part of the properties):
\begin{Proposition}
\label{prop:comptaylorprop}
A map $\mathbb{P}_{\mathcal{H}}^{\mathcal{D}}$ as above
\begin{enumerate}
	\item is a (weak, local) Poisson homomorphism, i.e.
		\be\label{eq:reducedbracket}
		\left\{\mathbb{P}_{\mathcal{H}}^{\mathcal{D}}(f),\mathbb{P}_{\mathcal{H}}^{\mathcal{D}}(g)\right\}_{\Gamma} & \approx_{S_{J\left(\mathcal{H}\right)}} & \mathbb{P}_{\mathcal{H}}^{\mathcal{D}}\left(\left\{f,g\right\}_{\textup{DB}(\mathcal{D},\mathcal{H})}\right),\\[0.5cm]\nonumber \mathbb{P}_{\mathcal{H}}^{\mathcal{D}}(f)\mathbb{P}_{\mathcal{H}}^{\mathcal{D}}(g) & \approx_{S_{J\left(\mathcal{H}\right)}} & \mathbb{P}_{\mathcal{H}}^{\mathcal{D}}(fg),\\[0.5cm]\nonumber
  \mathbb{P}_{\mathcal{H}}^{\mathcal{D}}(f)+\mathbb{P}_{\mathcal{H}}^{\mathcal{D}}(g) & =_{\ \ \ \ \ \ } & \mathbb{P}_{\mathcal{H}}^{\mathcal{D}}(f+g),\ \ \ f,g\in\mathcal{F}\left(\Gamma\right).
		\ee
		Here,  $\left\{\ .\ ,\ .\ \right\}_{\textup{DB}(\mathcal{D},\mathcal{H})}$ denotes the Dirac bracket associated with the constraint systems. It is important to recall that the Dirac bracket (weakly) coincides with $\left\{\ .\ ,\ .\ \right\}_{\Gamma}$ on (weak) Dirac observables.
	\item is (weakly, locally) onto, i.e. $\mathbb{P}_{\mathcal{H}}^{\mathcal{D}}(f)\approx_{S_{J\left(\mathcal{H}\right)}}
f,\ \ \ f\in\mathcal{O}^{w}_{J\left(\mathcal{H}\right)}$,
	\item defines a (weak, local) Poisson isomorphism $\mathbb{P}_{\mathcal{H}}^{\mathcal{D}}:\mathcal{O}^{w}_{J\left(\mathcal{D}\right)}\rightarrow\mathcal{O}^{w}_{J\left(\mathcal{H}\right)}$ with (weak) inverse $\mathbb{P}_{\mathcal{D}}^{\mathcal{H}}$, i.e.
		\be
		\mathbb{P}_{\mathcal{H}}^{\mathcal{D}}\circ\mathbb{P}_{\mathcal{D}}^{\mathcal{H}}\approx_{S_{J\left(\mathcal{H}\right)}}
\id_{\mathcal{O}^{w}_{J\left(\mathcal{H}\right)}},\ \ \  \mathbb{P}_{\mathcal{D}}^{\mathcal{H}}\circ\mathbb{P}_{\mathcal{H}}^{\mathcal{D}}\approx_{S_{J\left(\mathcal{D}\right)}} \id_{\mathcal{O}^{w}_{J\left(\mathcal{D}\right)}},
		\ee
	\item (weakly, locally) annihilates $\mathcal{N}_{J\left(\mathcal{D},\mathcal{H}\right)}:=\left\{f\in\mathcal{F}\left(\Gamma\right)\ |\ \forall x\in S_{J\left(\mathcal{D},\mathcal{H}\right)}:f(x)=0\right\}$\footnote{An element $f\in\mathcal{N}_{J\left(\mathcal{D},\mathcal{H}\right)}$ is (globally) of the form $f=\phi^{i}\mathcal{D}_{i}+\psi^{j}\mathcal{H}_{j}$ with arbitrary $\phi^{i},\psi^{j}\in\mathcal{F}\left(\Gamma\right),\ i,j\in J$ (cf. \cite{HenneauxQuantizationOfGauge}).}, i.e.
		\be
f\in\mathcal{N}_{J\left(\mathcal{D},\mathcal{H}\right)}\Rightarrow\mathbb{P}_{\mathcal{H}}^{\mathcal{D}}(f)\approx_{S_{J\left(\mathcal{H}\right)}}
0.
		\ee
\end{enumerate}
\textup{We only include a short proof of \textit{3.}, as this was, to our knowledge, not stated in the literature so far, at least not in this explicit form, although it is mentioned in \cite{BTTII} and implicit in the account on gauge invariant extensions of phase space functions in \cite{HenneauxQuantizationOfGauge}.\\
Additionally, we want to point out the importance of the first equation in (\ref{eq:reducedbracket})  which tells us that we should look for a set of phase space functions, such that $\mathbb{P}_{\mathcal{H}}^{\mathcal{D}}\left(\left\{f,g\right\}_{\textup{DB}(\mathcal{D},\mathcal{H})}\right)$ becomes simple, in view of a reduced phase space quantisation of the system.}
\begin{Proof}[of \textit{3.}]
Given $f\in\mathcal{O}^{w}_{J\left(\mathcal{H}\right)}$, this follows immediately from the first class property of the constraint systems:
\be
\nonumber & & \left(\mathbb{P}_{\mathcal{H}}^{\mathcal{D}}\circ\mathbb{P}_{\mathcal{D}}^{\mathcal{H}}\right)(f) \\[0.5cm]\nonumber
& = & \sum_{\left\{k_{j}\right\}_{j\in J}\in\mathbb{N}_{0}^{J}}\left(\prod_{j\in J}\frac{(\mathcal{D}_{j})^{k_{j}}}{k_{j}!}\right)\circ_{j\in J}\left(\chi_{\mathcal{H}'_{j}}\right)^{\circ k_{j}}\left(\mathbb{P}_{\mathcal{D}}^{\mathcal{H}}(f)\right) \\[0.5cm]\nonumber
& = & \sum_{\left\{k_{j}\right\}_{j\in J}\in\mathbb{N}_{0}^{J}}\left(\prod_{j\in J}\frac{(\mathcal{D}_{j})^{k_{j}}}{k_{j}!}\right)\circ_{j\in J}\left(\chi_{\mathcal{H}'_{j}}\right)^{\circ k_{j}} \left(\sum_{\left\{k_{i}\right\}_{i\in J}\in\mathbb{N}_{0}^{J}}\left(\prod_{i\in J}\frac{(-\mathcal{H}_{i})^{k_{i}}}{k_{i}!}\right)\circ_{i\in J}\left(\chi_{\mathcal{D}'_{i}}\right)^{\circ k_{i}}\left(f\right)\right) \\[0.5cm]\nonumber
& = & \sum_{\left\{k_{j}\right\}_{j\in J}\in\mathbb{N}_{0}^{J}}\left(\prod_{j\in J}\frac{(\mathcal{D}_{j})^{k_{j}}}{k_{j}!}\right)\sum_{\left\{k_{i}\right\}_{i\in J}\in\mathbb{N}_{0}^{J}}\circ_{j\in J}\left(\chi_{\mathcal{H}'_{j}}\right)^{\circ k_{j}} \left(\left(\prod_{i\in J}\frac{(-\mathcal{H}_{i})^{k_{i}}}{k_{i}!}\right)\circ_{i\in J}\left(\chi_{\mathcal{D}'_{i}}\right)^{\circ k_{i}}\left(f\right)\right) \\[0.5cm]\nonumber
& \approx & \sum_{\left\{k_{j}\right\}_{j\in J}\in\mathbb{N}_{0}^{J}}\left(\prod_{j\in J}\frac{(\mathcal{D}_{j})^{k_{j}}}{k_{j}!}\right)\sum_{\left\{k_{i}\right\}_{i\in J}\in\mathbb{N}_{0}^{J}}\left(\prod_{i\in J}\frac{(-\mathcal{H}_{i})^{k_{i}}}{k_{i}!}\right) \circ_{j\in J}\left(\chi_{\mathcal{H}'_{j}}\right)^{\circ k_{j}}\left(\circ_{i\in J}\left(\chi_{\mathcal{D}'_{i}}\right)^{\circ k_{i}}\left(f\right)\right) \\[0.5cm]\nonumber
& \approx & \sum_{\left\{k_{j}\right\}_{j\in J}\in\mathbb{N}_{0}^{J}}\left(\prod_{j\in J}\frac{(\mathcal{D}_{j})^{k_{j}}}{k_{j}!}\right)\circ_{j\in J}\left(\chi_{\mathcal{H}'_{j}}\right)^{\circ k_{j}}(f)\\[0.5cm]
& = & \mathbb{P}_{\mathcal{H}}^{\mathcal{D}}(f)\approx f,
\ee
where in line 4 and 5 $\approx$ denotes weak equality, i.e. equality on $S_{J\left(\mathcal{H}\right)}$, and we made use of the first class property of $\left\{\mathcal{H}_{j}\right\}_{j\in J}$. The different $(-1)^{k_{j}}$-factor for the two observable projectors is due to the antisymmetry of the Poisson bracket.\\
The corresponding calculation for $f\in\mathcal{O}^{w}_{J\left(\mathcal{D}\right)}$ is analogous, and will be omitted. The morphism properties directly follow by 1. of which a proof can be found in \cite{ThiemannModernCanonicalQuantum} (p.85 et seq.). $\Box$
\end{Proof}
\end{Proposition}
This allows us to conclude that if we change between the two constraint systems, there will be maps between the observable algebras, which we may construct (locally) in the above manner (at least formally). In general, the existence of a (local) Poisson isomorphism between the two systems follows from the fact that both constraint systems admit the same (local) gauge fixing surface $S_{J\left(\mathcal{D},\mathcal{H}\right)}$ and the observables are given by the functions on this surface together with the induced Poisson bracket, which is well-defined and non-degenerate due to the second class property of $\left\{\mathcal{H}_{j},\mathcal{D}_{j}\right\}_{j\in J}$.\\
Let us, at this point, shortly comment on the description of the reduced phase space in terms of gauge invariant functions on the constraint surface (cf. \cite{HenneauxQuantizationOfGauge}):\\[0.25cm]
The space of weak Dirac observables $\mathcal{O}^{w}_{J\left(\mathcal{H}\right)}$ contains the Poisson ideal $\mathcal{N}_{J\left(\mathcal{H}\right)}$ of functions that vanish on $S_{J\left(\mathcal{H}\right)}$, i.e. $\mathcal{O}^{w}_{J\left(\mathcal{H}\right)}\cdot\mathcal{N}_{J\left(\mathcal{H}\right)}\subset\mathcal{N}_{J\left(\mathcal{H}\right)}$ and $\left\{\mathcal{O}^{w}_{J\left(\mathcal{H}\right)},\mathcal{N}_{J\left(\mathcal{H}\right)}\right\}\subset\mathcal{N}_{J\left(\mathcal{H}\right)}$ (the same holds for the $J(\mathcal{D})$-system). Therefore, the quotients
\be
\left[\mathcal{O}^{w}_{J\left(\mathcal{H}\right)}\right]:=\mathcal{O}^{w}_{J\left(\mathcal{H}\right)}/\mathcal{N}_{J\left(\mathcal{H}\right)}, & & \left[\mathcal{O}^{w}_{J\left(\mathcal{D}\right)}\right]:=\mathcal{O}^{w}_{J\left(\mathcal{D}\right)}/\mathcal{N}_{J\left(\mathcal{D}\right)}
\ee
inherit well-defined Poisson structures, and proposition \ref{prop:comptaylorprop} tells us that the complete observable projectors descend to isomorphisms of these spaces.
\begin{Corollary}
\label{cor:obpro}
The maps
\be
\label{eq:classprojD}
\left[\mathbb{P}_{\mathcal{D}}^{\mathcal{H}}\right]:\left[\mathcal{O}^{w}_{J\left(\mathcal{H}\right)}\right]\longrightarrow\left[\mathcal{O}^{w}_{J\left(\mathcal{D}\right)}\right], & & \left[f\right]_{\mathcal{H}}\mapsto\left[\mathbb{P}_{\mathcal{D}}^{\mathcal{H}}(f)\right]_{\mathcal{D}}, \\[0.5cm]
\label{classprojH}
\left[\mathbb{P}_{\mathcal{H}}^{\mathcal{D}}\right]:\left[\mathcal{O}^{w}_{J\left(\mathcal{D}\right)}\right]\longrightarrow\left[\mathcal{O}^{w}_{J\left(\mathcal{H}\right)}\right], & & \left[f\right]_{\mathcal{D}}\mapsto\left[\mathbb{P}_{\mathcal{H}}^{\mathcal{D}}(f)\right]_{\mathcal{H}}
\ee
are well-defined and inverse to one another.
\end{Corollary}
The proof is straightforward and therefore omitted. The relations between the different structures are summarised in figure \ref{fig:diagcomp}.
\\[0.5cm]
Another important question concerning gauge fixing and gauge unfixing is that of dynamics of the system. For totally constrained system, this can be achieved by identifying appropriate parametrisations of the gauge flow along the constraint surface, which then allows for the construction of (weak) Dirac observables relative to it. The link between such parametrisations and  gauge fixing/unfixing will be outlined in the following.\\[0.25cm]
For any family of constants $\left\{t_{j}\right\}_{j\in J}$, we consider the constraint system $\left\{\mathcal{D}^{t}_{j}:=\mathcal{D}_{j}-t_{j}\right\}_{j\in J}$. If we assume that the analogue of equation (\ref{eq:gfloc}), i.e.
\be
\label{eq:gfloct}
\left\{\mathcal{D}^{t}_{i},\mathcal{H}_{j}\right\}_{\Gamma}=M^{t}_{ij},\ \ \ M^{t}=\left(M^{t}_{ij}\right)_{i,j\in J}\ \textup{invertible on}\ S_{J\left(\mathcal{D}^{t},\mathcal{H}\right)},
\ee
is valid, we may again apply the \textit{Abelianisation Theorem} (provided suitable invertibility conditions hold for $M^{t}$), which leads to a parametrisation of the gauge flow of the constraint system $\left\{\mathcal{H}_{j}\right\}_{J}$ in terms of gauge cuts $S_{J\left(\mathcal{D}^{t}\right)}$. Moreover, this parametrisation is weakly (on $S_{J\left(\mathcal{H}\right)}$) Abelian, and the dynamics of the system relative to it can (in principle) be extracted from
\be
\label{eq:reldyn}
\alpha^{*}_{\beta^{j}\chi_{\mathcal{H}'_{j}}}\left(f\right)_{|\left\{\beta^{j}\right\}_{j\in J}=\left\{\mathcal{D}_{j}\right\}_{j\in J}}\approx_{S_{J\left(\mathcal{H}\right)}}f,\ \ \ f\in\mathcal{O}^{w}_{J\left(\mathcal{H}\right)}.
\ee
This equation may as well be used as the definition of the elements $f\in\mathcal{O}^{w}_{J\left(\mathcal{H}\right)}$. If we are able to deparametrise the constraints
\be
\label{eq:depar}
\mathcal{H}_{j} & = & P_{\mathcal{D}_{j}}-H_{j},\ \ \ j\in J
\ee
w.r.t. to a canonically conjugate (to $\left\{\mathcal{D}_{j}\right\}_{j\in J}$) family $\left\{P_{\mathcal{D}_{j}}\right\}_{j\in J}$ of first class functions, such that the $H_{j},\ j\in J$, no longer depend on the $P_{\mathcal{D}_{j}},\mathcal{D}_{j},\ j\in J$, we may cast (\ref{eq:reldyn}) into the typical form of Hamilton's equations (with multi-fingered time generated by the (strong) Dirac observables $H_{j},\ j\in J$, cf. \cite{GieselManifestlyGaugeInvariantI, GieselManifestlyGaugeInvariantII, DomagalaGravityQuantizedLoop, ThiemannModernCanonicalQuantum}).\\
Additionally, we obtain a function $\beta^{\mathcal{H}'}_{\mathcal{D}^{t}}:S_{J\left(\mathcal{H}\right)}
\rightarrow\mathbb{R}^{J}$, and associated complete observable projectors $\mathbb{P}_{\mathcal{H}}^{\mathcal{D}^{t}},\ \mathbb{P}^{\mathcal{H}}_{\mathcal{D}^{t}}$ for all $\left\{t_{j}\right\}_{j\in J}$.\\
If we take the viewpoint that each of the gauge fixing surfaces $S_{J\left(\mathcal{D}^{t}\right)},\ t\in\mathbb{R}^{J}$ provides us with good, i.e. freely specifiable and independent, initial data for the ``dynamical system'' $\left(\left\{\mathcal{H}_{j}\right\}_{j\in J}, \Gamma, \left\{\ .\ ,\ .\ \right\}_{\Gamma}\right)$, we will call the system $\left\{\mathcal{D}_{j}\right\}_{j\in J}$ ``clocks''.\\[0.25cm]
Equivalent conclusions can be drawn if we treat $\left\{\mathcal{H}_{j}\right\}_{j\in J}$ as part of a family of gauge fixings $\left\{\mathcal{H}^{s}_{j}:=\mathcal{H}_{j}-s_{j}\right\}_{j\in J},\ \left\{s_{j}\right\}_{j\in J}\subset\mathbb{R}^{J}$, for $\left\{\mathcal{D}_{j}\right\}_{j\in J}$. But it should be pointed out that, although the equivalent of (\ref{eq:reldyn})
\be
\label{eq:reldynD}
\alpha^{*}_{-\gamma^{j}\chi_{\mathcal{D}'_{j}}}\left(f\right)_{|\left\{\gamma^{j}\right\}_{j\in J}=\left\{\mathcal{H}_{j}\right\}_{j\in J}}\approx_{S_{J\left(\mathcal{D}\right)}}f,\ \ \ f\in\mathcal{O}^{w}_{J\left(\mathcal{D}\right)}
\ee
defines the same (isomorphic by proposition \ref{prop:comptaylorprop}) algebra of (weak) Dirac observables, the relative dynamics are different since the two gauge transformations have (weakly) conjugate generators by the assumption that the constraint systems gauge fix one another.

\subsubsection{Partial observable projectors}
\label{sec:partialproj}

\begin{figure}[h]
\be\nonumber
\xymatrix{
 & \left(\mathcal{F}\left(\Gamma\right),\left\{\ ,\ \right\}_{\textup{DB}\left(\mathcal{D},\mathcal{H}\right)}\right) \ar@/_1pc/[lddd]|{_{2}\mathbb{P}^{\mathcal{D}}_{\mathcal{H}}} \ar@/^1pc/[rddd]|{_{2}\mathbb{P}^{\mathcal{H}}_{\mathcal{D}}} & \\
 & \left(\mathcal{F}\left(_{2}\Gamma_{R}\right),\left\{\ ,\ \right\}_{_{2}\Gamma_{R}}\right) & \\
 & & \\
\left(\mathcal{O}^{w}_{J_{2}\left(\mathcal{H}\right)},\left\{\ ,\ \right\}_{\Gamma}\right) \ar@/^1pc/[rr]|{_{2}\mathbb{P}^{\mathcal{H}}_{\mathcal{D}}} \ar@/^/[uur]|{\pi_{J_{2}\left(\mathcal{H}\right)}} \ar@/_/[ddr]|{\iota^{*}_{S_{J_{2}\left(\mathcal{D},\mathcal{H}\right)}}} & & \left(\mathcal{O}^{w}_{J_{2}\left(\mathcal{D}\right)},\left\{\ ,\ \right\}_{\Gamma}\right) \ar@/^1pc/[ll]|{_{2}\mathbb{P}^{\mathcal{D}}_{\mathcal{H}}} \ar@/_/[uul]|{\pi_{J_{2}\left(\mathcal{D}\right)}} \ar@/^/[ddl]|{\iota^{*}_{S_{J_{2}\left(\mathcal{D},\mathcal{H}\right)}}} \\
 & & \\
 & \left(\mathcal{F}\left(S_{J_{2}\left(\mathcal{D},\mathcal{H}\right)}\right),\left\{\ ,\ \right\}_{|S_{J_{2}\left(\mathcal{D},\mathcal{H}\right)}}\right) & 
}
\ee
\caption{$\pi_{J_{2}\left(\mathcal{H}\right)},\ \pi_{J_{2}\left(\mathcal{D}\right)}$ and $\iota^{*}_{S_{J_{2}\left(\mathcal{D},\mathcal{H}\right)}}$ are the canonical projections and the restriction to the gauge fixing surface.}
\label{fig:diagpart}
\end{figure}
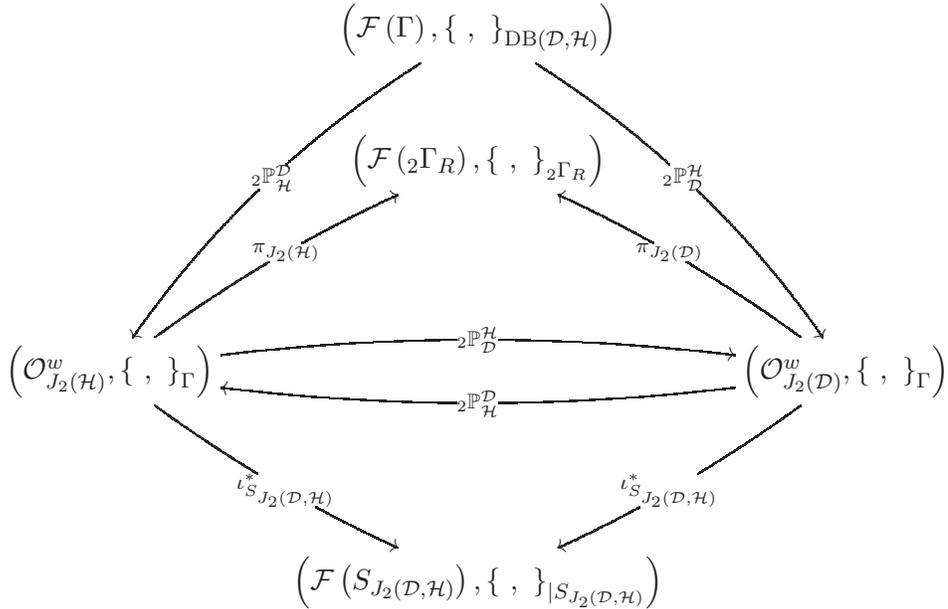

If we are dealing with a partial gauge fixing, the situation will possibly be more involved depending on the fact whether the $J_{2}(\mathcal{H})$-subsystem is even (case I) or odd (case II).\\[0.25cm]
Let us first analyse case I:\\[0.25cm]
Since both of the possible sets of constraints, $\left\{\mathcal{H}_{j_{1}},\mathcal{H}_{j_{2}}\right\}_{j_{1}\in J_{1},j_{2}\in J_{2}}$ and $\left\{\mathcal{H}_{j_{1}},\mathcal{D}_{j_{2}}\right\}_{j_{1}\in J_{1},j_{2}\in J_{2}}$, have the property that the respective $J_{2}$-subsystem generates an ideal within the constraint algebra, the construction of \textit{partial observable projectors} $_{2}\mathbb{P}_{\mathcal{H}}^{\mathcal{D}}:\mathcal{F}\left(\Gamma\right)\rightarrow\mathcal{O}^{w}_{J_{2}\left(\mathcal{H}\right)}$, $_{2}\mathbb{P}_{\mathcal{D}}^{\mathcal{H}}:\mathcal{F}\left(\Gamma\right)\rightarrow\mathcal{O}^{w}_{J_{2}\left(\mathcal{D}\right)}$ can be achieved with the help of theorem \ref{th:AbelPart}. Due to the ideal structure of the $J_{2}$-subsystems, the notion of partial observables and the partially reduced phase space $_{2}\Gamma_{R}$ (the space of orbits for the resp. $J_{2}$-subsystem, cf. (\ref{eq:gaugeorbit})) are well-defined. The formulas are identical to those for the complete observable projectors, but with $(i,j,J)$ replaced by $(i_{2},j_{2},J_{2})$:
\be
\label{eq:comptaylorpart}
\nonumber & _{2}\mathbb{P}_{\mathcal{H}}^{\mathcal{D}}(f) & = \sum_{\left\{k_{j_{2}}\right\}_{j_{2}\in J_{2}}\in\mathbb{N}_{0}^{J_{2}}}\left(\prod_{j_{2}\in J_{2}}\frac{(\mathcal{D}_{j_{2}})^{k_{j_{2}}}}{k_{j_{2}}!}\right)\circ_{j_{2}\in J_{2}}\left(\chi_{\mathcal{H}'_{j_{2}}}\right)^{\circ k_{j_{2}}}(f) \\[0.5cm]
 & & \approx_{S_{J_{2}\left(\mathcal{H}\right)}}
 \alpha_{\beta^{j_{2}}\chi_{\mathcal{H}'_{j_{2}}}}^{*}(f)_{|_{2}\beta=_{2}\beta^{\mathcal{H}'}_{\mathcal{D}}},\ \ \ f\in\mathcal{F}\left(\Gamma\right), \\[0.5cm]
 \label{eq:comptaylorpart2}
\nonumber & _{2}\mathbb{P}_{\mathcal{D}}^{\mathcal{H}}(f) & = \sum_{\left\{k_{j_{2}}\right\}_{j_{2}\in J_{2}}\in\mathbb{N}_{0}^{J_{2}}}\left(\prod_{j_{2}\in J_{2}}\frac{(-\mathcal{H}_{j_{2}})^{k_{j_{2}}}}{k_{j_{2}}!}\right)\circ_{j_{2}\in J_{2}}\left(\chi_{\mathcal{D}'_{j_{2}}}\right)^{\circ k_{j_{2}}}(f) \\[0.5cm]
 & & \approx_{S_{J_{2}\left(\mathcal{D}\right)}}
 \alpha_{-\gamma^{j_{2}}\chi_{\mathcal{D}'_{j_{2}}}}^{*}(f)_{|_{2}\gamma=_{2}\gamma^{\mathcal{D}'}_{\mathcal{H}}},\ \ \ f\in\mathcal{F}\left(\Gamma\right).
\ee
Furthermore, proposition \ref{prop:comptaylorprop} and the analysis concerning dynamics generalise as well (with the appropriate replacements), as one easily infers from the fact that the $J_{2}$-subsystems are ideals, e.g. the relative (partial) dynamics are contained in:
\be
\label{eq:reldynpart}
\alpha^{*}_{\beta^{j_{2}}\chi_{\mathcal{H}'_{j_{2}}}}\left(f\right)_{|\left\{\beta^{j_{2}}\right\}_{j_{2}\in J_{2}}=\left\{\mathcal{D}_{j_{2}}\right\}_{j_{2}\in J_{2}}}\approx_{S_{J_{2}\left(\mathcal{H}\right)}}f,\ \ \ f\in\mathcal{O}^{w}_{J_{2}\left(\mathcal{H}\right)}.
\ee
The only difference to the complete gauge fixing is the remaining gauge freedom due to the $J_{1}(\mathcal{H})$-subsystem. The description of the partially reduced phase space $_{2}\Gamma_{R}$ carries directly over from the complete gauge fixing case. We summarise the structural relations in figure \ref{fig:diagpart}. 
\\[0.5cm]
The situation in case II is quite different:\\[0.25cm]
Although the construction of the partial observable projector $_{2}\mathbb{P}_{\mathcal{D}}^{\mathcal{H}}$ for the constraint family $\left\{\mathcal{H}_{j_{1}},\mathcal{D}_{j_{2}}\right\}_{j_{1}\in J_{1},j_{2}\in J_{2}}$ can be achieved along the same lines, this is no longer possible for the (dual) family $\left\{\mathcal{H}_{j_{1}},\mathcal{H}_{j_{2}}\right\}_{j_{1}\in J_{1},j_{2}\in J_{2}}$. The main obstruction for the latter is the ill-defined gauge flow of the $J_{2}(\mathcal{H})$-subsystem on its constraint surface $S_{J_{2}\left(\mathcal{H}\right)}$, which is a necessary ingredient for the given formulae for $_{2}\mathbb{P}_{\mathcal{H}}^{\mathcal{D}}$ (cf. (\ref{eq:comptaylorpart})).\\
Therefore, an explicit correspondence between the two families on the level of observables in terms of partial observable projectors seems impossible, unless we choose a different set of generators $\left\{\mathcal{H}_{j}\right\}_{j\in J}$, or extend the family $\left\{\mathcal{D}_{j_{2}}\right\}_{j_{2}\in J_{2}}$ to a complete gauge fixing.\\
Nevertheless, an implicit map is achieved by gauge invariant extension, w.r.t to the resp. $(J_{1}, J_{2})$-system, of $J_{1}$-invariant functions on $S_{J_{1}\left(\mathcal{H}\right),J_{2}\left(\mathcal{D},\mathcal{H}\right)}$ (denoted by $\mathcal{O}^{w}_{J_{1}\left(\mathcal{H}\right)}\left(S_{J_{1}\left(\mathcal{H}\right),J_{2}\left(\mathcal{D},\mathcal{H}\right)}\right)$), which also enjoys the properties stated in proposition \ref{prop:comptaylorprop} as discussed in \cite{HenneauxQuantizationOfGauge}.\\
What is still true is that the remaining gauge freedom is given by the $J_{1}\left(\mathcal{H}\right)$-subsystem.
Again, a summary of the structures is given in figure \ref{fig:diagpart2}.
\begin{figure}[h]
\be\nonumber
\xymatrix{
 & \left(\mathcal{F}\left(\Gamma_{R}\right),\left\{\ ,\ \right\}_{\Gamma_{R}}\right) & \\
 & \left(\mathcal{F}\left(_{2}\Gamma_{R}\right),\left\{\ ,\ \right\}_{_{2}\Gamma_{R}}\right) & \\
 & & \\
\left(\mathcal{O}^{w}_{J_{1}\left(\mathcal{H}\right),J_{2}\left(\mathcal{H}\right)},\left\{\ ,\ \right\}_{\Gamma}\right) \ar@/^1pc/[uuur]|{\pi_{J_{1}\left(\mathcal{H}\right),J_{2}\left(\mathcal{H}\right)}} \ar@/_1pc/[ddr]|{\iota^{*}_{J_{2}\left(\mathcal{D},\mathcal{H}\right)}}  & & \left(\mathcal{O}^{w}_{J_{1}\left(\mathcal{H}\right),J_{2}\left(\mathcal{D}\right)},\left\{\ ,\ \right\}_{\Gamma}\right) \ar@/_1pc/[uuul]|{\pi_{J_{1}\left(\mathcal{H}\right),J_{2}\left(\mathcal{D}\right)}} \ar@/^1pc/[ddl]|{\iota^{*}_{J_{2}\left(\mathcal{D},\mathcal{H}\right)}} \ar@/_/[uul]|{\pi_{J_{2}\left(\mathcal{D}\right)}} \\
 & & \\
 & \left(\mathcal{O}^{w}_{J_{1}\left(\mathcal{H}\right)}\left(S_{J_{2}\left(\mathcal{D},\mathcal{H}\right)}\right),\left\{\ ,\ \right\}_{|S_{J_{2}\left(\mathcal{D},\mathcal{H}\right)}}\right) \ar@/_1pc/[uul]|{\textup{ext}^{J_{1}\left(\mathcal{H}\right),J_{2}\left(\mathcal{H}\right)}_{J_{2}\left(\mathcal{D},\mathcal{H}\right)}} \ar@/^1pc/[uur]|{\textup{ext}^{J_{1}\left(\mathcal{H}\right),J_{2}\left(\mathcal{D}\right)}_{J_{2}\left(\mathcal{D},\mathcal{H}\right)}} & 
}
\ee
\caption{$\pi_{J_{1}\left(\mathcal{H}\right),J_{2}\left(\mathcal{H}\right)},\ \pi_{J_{1}\left(\mathcal{H}\right),J_{2}\left(\mathcal{D}\right)}$, $\textup{ext}^{J_{1}\left(\mathcal{H}\right),J_{2}\left(\mathcal{H}\right)}_{J_{2}\left(\mathcal{D},\mathcal{H}\right)},\ \textup{ext}^{J_{1}\left(\mathcal{H}\right),J_{2}\left(\mathcal{D}\right)}_{J_{2}\left(\mathcal{D},\mathcal{H}\right)}$ and $\iota^{*}_{S_{J_{2}\left(\mathcal{D},\mathcal{H}\right)}}$ are the canonical projections, the gauge invariant extension maps and the restriction to the gauge fixing surface. The non-existence of $\pi_{J_{2}\left(\mathcal{H}\right)}$ is due to the fact that we are dealing with case II.}
\label{fig:diagpart2}
\end{figure}

\subsection{Quantisation}

After the discussion of some of the classical aspects of gauge (un)fixing, we briefly comment on the implications for quantisation. The two principal ways to achieve the latter in the presence of constraints are the reduced phase space quantisation and the Dirac quantisation, i.e. one either solves the constraints before or after associating a quantum-*-algebra of observables to a suitable Poisson subalgebra of $\mathcal{F}\left(\Gamma_{R}\right)$ resp. $\mathcal{F}\left(\Gamma\right)$.\\
Since quantisation is in general mathematically far from unique, it is necessary to understand the involved choices to restrict these by further physically and mathematically motivated assumptions. One of the possible choices, which we want to comment on from the perspective of gauge fixing/unfixing, is the one related to canonical transformations and Poisson isomorphisms, i.e. the choice of (generalised) position and momentum variables as starting point for quantisation, and how the latter depends on this.\\
A unique quantisation w.r.t. this freedom would ask for the following diagram to commute:
\be
\xymatrix{
\left(\Gamma, \left(p,q\right)\right) \ar[rr]^{\alpha} \ar[dd]_{\mathcal{A}} &  & \left(\Gamma, \left(P,Q\right)\right) \ar[dd]^{\mathcal{A}} \\
 &  &  \\
\mathcal{A}(\left(\Gamma\right),\left(\hat{p},\hat{q}\right)) \ar[rr]^{\mathcal{A}\left(\alpha\right)} &  & \mathcal{A}(\left(\Gamma\right),(\hat{P},\hat{Q})),
}
\ee
i.e. there would be, associated with every canonical transformation $\alpha$, an isomorphism $\mathcal{A}(\alpha)$ of the quantum-*-algebras. If this fails to be true, one typically speaks of a type of ``quantisation ambiguity" which can only be removed by further structural restrictions, or finally be decided upon  by experiments. A prominent example of such an ambiguity in the context of loop quantum gravity is given by the Immirzi parameter (cf. \cite{BarberoRealAshtekarVariables, ImmirziQuantumGravityAnd, RovelliTheImmirziParameter}). \\[0.25cm]
This aspect of the quantisation of (totally constrained) Hamiltonian system manifests itself in the mechanism of (complete\footnote{In case we are dealing with a partial gauge fixing, the appropriate index replacements are understood.}) gauge fixing/unfixing in the choice of constraint system, either $\left\{\mathcal{H}_{j}\right\}_{j\in J}$ or $\left\{\mathcal{D}_{j}\right\}_{j\in J}$, and therefore in the concrete realisation of (weak) Dirac observables $\mathcal{O}^{w}_{J\left(\mathcal{H}\right)}$ resp. $\mathcal{O}^{w}_{J\left(\mathcal{D}\right)}$.
\be
\label{eq:HDpic}
\xymatrix{
\left(\Gamma,\ \left\{\ ,\ \right\}_{\Gamma},\ \left\{\mathcal{H}_{j}\right\}_{j\in J},\ \mathcal{O}^{w}_{J\left(\mathcal{H}\right)}\right) \ar@/^1pc/[rr]^{\mathbb{P}^{\mathcal{H}}_{\mathcal{D}}} \ar[dd]_{\mathcal{A}} &  & \left(\Gamma,\ \left\{\ ,\ \right\}_{\Gamma},\ \left\{\mathcal{D}_{j}\right\}_{j\in J},\ \mathcal{O}^{w}_{J\left(\mathcal{D}\right)}\right) \ar@/^1pc/[ll]^{\mathbb{P}^{\mathcal{D}}_{\mathcal{H}}} \ar[dd]^{\mathcal{A}} \\
 &  &  \\
\left(\mathcal{A}_{\mathcal{H}},\ \frac{i}{\hbar}\left[\ ,\ \right],\ \left\{\hat{\mathcal{H}}_{j}\right\}_{j\in J},\hat{\mathcal{O}}^{w}_{J\left(\mathcal{H}\right)}\right) \ar@/^1pc/[rr]^{``\hat{\mathbb{P}}^{\mathcal{H}}_{\mathcal{D}}"} &  &  \left(\mathcal{A}_{\mathcal{D}},\ \frac{i}{\hbar}\left[\ ,\ \right],\ \left\{\hat{\mathcal{D}}_{j}\right\}_{j\in J},\hat{\mathcal{O}}^{w}_{J\left(\mathcal{D}\right)}\right) \ar@/^1pc/[ll]^{``\hat{\mathbb{P}}^{\mathcal{D}}_{\mathcal{H}}"}
}
\ee
We put the quantum observable projectors $\hat{\mathbb{P}}^{\mathcal{H}}_{\mathcal{D}}$, $\hat{\mathbb{P}}^{\mathcal{D}}_{\mathcal{H}}$ in quotation marks since their existence in a Dirac quantisation is not ensured, neither at the level of algebra automorphisms nor at the level of unitary transformations (w.r.t. a certain representation). Although we have, at least locally, explicit formulae for the observable projectors in terms of the constraint families, a quantisation cannot be directly conceived from those as this would involve the quantisation of the inverse of the Dirac matrix $\left\{\mathcal{D}_{i},\mathcal{H}_{j}\right\}_{\Gamma}=M_{ij},\ i,j\in J$. If we are dealing with a field theory at the classical level, we expect even further complications of this issue due to Haag's theorem (cf. \cite{HaagOnQuantumField, HaagLocalQuantumPhysics}). Nevertheless, we may take any quantisation of a representation of the classical theory as a viable quantum theory, letting convenience w.r.t. calculations and finally experiments discriminate between the possibilities. Moreover, it is possible that a quantisation of the classical system can only be achieved in one of its representations, which renders the question whether the quantum observable projectors exist meaningless. The latter compares to the use of Ashtekar-Babero variables in loop quantum gravity as opposed to ADM-variables, since up to now only the first allows for a rigorous definition of a quantum theory, at least at the kinematical level.\\
If we utilise the observable projectors only in an intermediate step to achieve a reduced phase space quantisation, we will be in no need to implement them by explicit formulae since we would quantise the algebra of Dirac observables (cf. proposition \ref{prop:comptaylorprop}) directly. In this case we will aim for an algebra of phase space functions with a simple expression on the right hand side of 
\be
\left\{\mathbb{P}_{\mathcal{H}}^{\mathcal{D}}(f),\mathbb{P}_{\mathcal{H}}^{\mathcal{D}}(g)\right\}_{\Gamma} & \approx_{S_{J\left(\mathcal{H}\right)}}
 & \mathbb{P}_{\mathcal{H}}^{\mathcal{D}}\left(\left\{f,g\right\}_{\textup{DB}(\mathcal{D},\mathcal{H})}\right),\ \ \ f,g\in\mathcal{F}\left(\Gamma\right),
\ee
or its dual in terms of $\mathbb{P}_{\mathcal{D}}^{\mathcal{H}}$, to make it possible to find representations of the quantum algebra.\\
Concerning the implementation of dynamics at the quantum level, one faces the usual problems of extracting it from the (weak) Dirac observables (cf. \cite{HenneauxQuantizationOfGauge, ThiemannModernCanonicalQuantum}) when dealing with totally constrained systems. One method to accomplish this is deparametrisation, which was already mentioned above (cf. equations (\ref{eq:reldyn}), (\ref{eq:depar})). If we try to combine deparametrisation with gauge fixing/unfixing in the sense of first changing the constraint system and then employing a suitable parametrisation of the gauge flow, we face complications regarding the interpretation as the gauge flows of the two constraint systems are inequivalent. For example, a ``time evolution"-picture might be tied to the geometrical action of the first, especially if we think of the hypersurface deformation algebra, arising e.g. in general relativity (see also section \ref{sec:RemarkOnQuantisationAndObservables}).

\subsection{Example}

We conclude this section by giving an instructive example of the mechanisms described above. To keep things simple, we consider a quantum mechanical model with $n$ degrees of freedom defined on $(\mathbb{R}^{2n},\ \left\{(p_{i},q_{i})\right\}_{i=1,...,n})$ together with its standard symplectic structure and a, possibly time-dependent, Hamiltonian $H=H(t,\{(p_{i},q_{i})\}_{i=1,...,n})$.\\
This system can be cast into the form of a totally constrained, first class system by considering the extended phase space $(\mathbb{R}^{2(n+1)},\ (p_{t},t),\left\{(p_{i},q_{i})\right\}_{i=1,...,n})$, again with its standard symplectic structure, and a constraint of the form $\mathcal{H}=p_{t}-H(t,\{(p_{i},q_{i})\}_{i=1,...,n})$. The constraint surface $S_{J\left(H\right)}=\left\{\left((p_{t},t),\left\{(p_{i},q_{i})\right\}_{i=1,...,n}\right)\in\mathbb{R}^{2(n+1)}\ |\ p_{t}=H(t,\{(p_{i},q_{i})\}_{i=1,...,n})\right\}$ admits a gauge flow defined by the elementary Poisson brackets of $\mathcal{H}$ with the coordinate functions $(p_{t},t),\ \left\{(p_{i},q_{i})\right\}_{i=1,...,n}$:
\be
\label{eq:qmgaugeflowcoord}\nonumber
\left\{\mathcal{H},p_{i}\right\}=-\frac{\partial H}{\partial q_{i}}, &  & \left\{\mathcal{H},q_{i}\right\}=\frac{\partial H}{\partial p_{i}}, \\
\left\{\mathcal{H},p_{t}\right\}=-\frac{\partial H}{\partial t}, &  & \left\{\mathcal{H},t\right\}=-1.
\ee
For arbitrary phase space functions $f\in\mathcal{F}\left(\mathbb{R}^{2(n+1)}\right)$ we have to integrate the equation
\be
\label{eq:qmgaugeflow}
\frac{\partial f}{\partial s}=\left\{\mathcal{H},f\right\},
\ee
leading, as before, to the (formal) solution
\be
\label{eq:qmgaugeflowsol}
\alpha^{*}_{s\chi_{\mathcal{H}}}(f)=\sum_{n=0}^{\infty}\frac{s^{n}}{n!}\left\{\mathcal{H},f\right\}_{n}.
\ee
It is important to point out that this gives the evolution of a general, non-observable phase space function w.r.t. the gauge flow, which should not be confused with the evolution of an observable w.r.t. the ``time flow'' of the original unconstrained system. Formulating the original system in terms of gauge invariant functions on the constraint surface (reduced phase space formulation) amounts to a description in terms of first integrals of Hamilton's equations (cf. \cite{HenneauxQuantizationOfGauge}). The latter are defined by
\be
\label{eq:qmdeparflow}
\frac{\partial F}{\partial t}=\left\{H,F\right\},\ \ \ F\in\mathcal{F}\left(\mathbb{R}^{2n+1}\right).
\ee
Such an observable is commonly termed a ``constant of motion''. We recall at this point that a system with $n$ degrees of freedom is integrable by quadratures if we know $n$ independent ``constants of motions'' that are in involution (cf. \cite{ArnoldMathematicalMethodsOf}, p.271 et seq.). In general, a (time-dependent) function on phase space evolves according to $\frac{d F}{dt}=\left\{F,H\right\}+\frac{\partial F}{\partial t}$. We also observe that $S_{J\left(\mathcal{H}\right)}$ is foliated by ``surfaces of constant energy'' ($H=E=p_{t}$). These will be preserved if $H$ is time-independent. Furthermore, H will be a Dirac observable in such a case.\\
The passage between (\ref{eq:qmgaugeflow}) and (\ref{eq:qmdeparflow}) can be achieved by deparametrisation of the equation which defines (weak) Dirac observables, and the identification $F=f_{|S_{J\left(\mathcal{H}\right)}},\ f\in\mathcal{F}(\mathbb{R}^{2(n+1)})$:
\be
\label{eq:qmdepargaugeflow}\nonumber
f\in\mathcal{O}^{w}_{S_{J\left(\mathcal{H}\right)}} & \Leftrightarrow & \left\{\mathcal{H},f\right\}\approx_{S_{J\left(\mathcal{H}\right)}}0 \\[0.25cm] \nonumber
 & \Leftrightarrow & \left\{p_{t},f\right\}\approx_{S_{J\left(\mathcal{H}\right)}}\left\{H,f\right\} \\[0.25cm] \nonumber
 & \Leftrightarrow & \frac{\partial f}{\partial t}\approx_{S_{J\left(\mathcal{H}\right)}}\left\{H,f\right\}\\[0.25cm] \nonumber
 & \Leftrightarrow & \frac{\partial f_{|S_{J\left(\mathcal{H}\right)}}}{\partial t}=\left\{H,f_{|S_{J\left(\mathcal{H}\right)}}\right\}_{\left\{(p_{i},q_{i})\right\}_{i=1,...,n}}\\[0.25cm]
 & \Leftrightarrow & \frac{\partial F}{\partial t}=\left\{H,F\right\}.
\ee
We want stress here that a deparametrisation of the constraint $\mathcal{H}$ w.r.t. the auxiliary momentum $p_{t}$ is in no way forced upon us. It is only natural because the system was originally in a form suggesting this. If $\mathcal{H}$ is solvable for any other of the phase space coordinates in a similar way, e.g. $\mathcal{H}=p_{1}-H_{1}(q_{1},(p_{t},t),\left\{(p_{i},q_{i})\right\}_{i=2,...,n})$, this will give an equally valid description of the behaviour of the system.\\
If the deparametrisation is only possible on subsets of the phase space or involves multi-valued inversions of non-linear functions, e.g. taking square roots as in $\mathcal{H}=p_{t}-\frac{1}{2}\sum_{i=1}^{n}(p_{i}^{2}+q_{i}^{2})$, which could be solved for $p_{1}$, one will encounter situations where the ``time'' variable associated with the deparametrisation breaks down (``branch points'').\\
If we choose the ``time'' variable such that $H$ is time-independent, the latter will be a (strong) Dirac observable, that can be used to implement dynamics on $\mathcal{O}^{w}_{J\left(\mathcal{H}\right)}$ (cf. \cite{GieselManifestlyGaugeInvariantI, GieselManifestlyGaugeInvariantII}).\\[0.25cm]
Coming to the gauge fixing/unfixing procedure, we observe that, due to (\ref{eq:qmgaugeflowcoord}), a family of canonical gauge fixings is given by the constraints $\mathcal{D}^{\tau}:=t-\tau,\ \tau\in\mathbb{R}$. The corresponding constraint surfaces $S_{J\left(\mathcal{D}^{\tau}\right)}$ define initial data representations at ``time'' $t=\tau$ of the system.  The complete observable projectors associated with these gauge fixings are:
\be
\label{eq:qmcompprojH}\nonumber
\mathbb{P}^{\mathcal{D}^{\tau}}_{\mathcal{H}}(f) & = & \sum_{n=0}^{\infty}\frac{\left(\mathcal{D}^{\tau}\right)^{n}}{n!}\left\{\mathcal{H},f\right\}_{n} \\[0.25cm]
 & = & \alpha^{*}_{s\chi_{\mathcal{H}}}(f)_{|s=s^{\mathcal{H}}_{\mathcal{D}^{\tau}}},\ \ \ f\in\mathcal{F}\left(\Gamma\right), \\[0.25cm]
\label{eq:qmcompprojD}\nonumber
\mathbb{P}^{\mathcal{H}}_{\mathcal{D}^{\tau}}(f) & = & \sum_{n=0}^{\infty}\frac{\left(-\mathcal{H}\right)^{n}}{n!}\left\{\mathcal{D}^{\tau},f\right\}_{n}\\[0.25cm]
 & = & \alpha^{*}_{-r\chi_{\mathcal{D}^{\tau}}}(f)_{|r=r^{\mathcal{D}^{\tau}}_{\mathcal{H}}},\ \ \ f\in\mathcal{F}\left(\Gamma\right).
\ee
where $s^{\mathcal{H}}_{\mathcal{D}^{\tau}}=D^{\tau}$ and $r^{\mathcal{D}^{\tau}}_{\mathcal{H}}=\mathcal{H}$. Since the gauge flows of the  family $\left\{\mathcal{D}^{\tau}\right\}_{\tau\in\mathbb{R}}$ generate translations in $p_{t}$, we can write the $\mathbb{P}^{\mathcal{H}}_{\mathcal{D}^{\tau}},\ \tau\in\mathbb{R}$, in closed form:
\be
\label{eq:qmcompprojDclosed}\nonumber
\mathbb{P}^{\mathcal{H}}_{\mathcal{D}^{\tau}}(f)(p_{t},t,\left\{p_{i},q_{i}\right\}_{i=1,...,n}) & = & f(p_{t}-\mathcal{H}(p_{t},t,\left\{p_{i},q_{i}\right\}_{i=1,...,n}),t,\left\{p_{i},q_{i}\right\}_{i=1,...,n}) \\[0.25cm]
 & = & f(H(t,\left\{p_{i},q_{i}\right\}_{i=1,...,n}),t,\left\{p_{i},q_{i}\right\}_{i=1,...,n}).
\ee
Thus, we see that $\mathbb{P}^{\mathcal{H}}_{\mathcal{D}^{\tau}},\ \tau\in\mathbb{R}$ indeed computes Dirac observables w.r.t. $\mathcal{D}^{\tau}$.\\
Moreover, the restriction of $\mathbb{P}^{\mathcal{H}}_{\mathcal{D}^{\tau}}(f)$ to the gauge fixing surface $S_{J\left(\mathcal{D}^{\tau},\mathcal{H}\right)}$ can be used as initial values for equation (\ref{eq:qmdepargaugeflow}, line 4) to compute a Dirac observable for $\mathcal{H}$. This is what is accomplished by $\mathbb{P}^{\mathcal{D}^{\tau}}_{\mathcal{H}}$.\\
The reduced phase space for $\mathcal{D}^{\tau}$ is also easily computed, since this constraint just tells us to drop the variables $(p_{t},t)$ completely, and work solely with the $\left\{p_{i},q_{i}\right\}_{i=1,...,n}$ and functions thereof.\\
At this point, one might wonder where the actual information about the dynamics of the system, namely type of interactions and their coupling constants, has gone, since we are only left with functions of the $\left\{p_{i},q_{i}\right\}_{i=1,...,n}$. The resolution of this apparent problem is that the observable projectors, which are used to identify the different structures, keep track of it. Indeed, the gauge fixing/unfixing by $\mathcal{D}^{\tau}$ is not tied to any special form of the Hamiltonian $H$, thus defining a general structure to build the quantum theory upon. The implementation of the dynamics is then brought in as a second step corresponding to the problem of defining the observable projector $\mathbb{P}^{\mathcal{D}^{\tau}}_{\mathcal{H}}$ at the quantum level. If we try to implement the observable projector of an interacting Hamiltonian $H_{\textup{int}}$ relative to that of the free Hamiltonian $H_{\textup{free}}=\sum_{i=1}^{n}\frac{p_{i}^{2}}{2}$, we are led to a sort of ``interaction'' picture.\\
If we want to return to the reduced phase space of $\mathcal{H}$ in the first place, we have to treat functions $f=f(\left\{p_{i},q_{i}\right\}_{i=1,...,n})$ on the reduced phase space, via their correspondence with the functions on the gauge fixing surface $S_{J\left(\mathcal{D},\mathcal{H}\right)}$, as initial data for equation (\ref{eq:qmdepargaugeflow}, line 5).

\section{Matterfree case}
\label{sec:MatterfreeCase}
After having worked out in detail what gauge fixing and unfixing imply for general Hamiltonian systems, we will now turn to specific applications. In this section, we will discuss vacuum general relativity in the constant mean curvature (CMC) gauge, which upon gauge unfixing turns into a theory with local conformal symmetry, which has been coined shape dynamics \cite{GomesEinsteinGravityAs}. We want to stress that many of the results in this section are not new (cf. the literature cited), but are included for a more coherent and comprehensive exposition. Moreover, our derivation using gauge unfixing will be different from and, in a sense, dual to the usual derivation of shape dynamics using a linking theory, see also \cite{WangConformalGeometrodynamicsTrue, WangTowardsConformalLoop} for an earlier account of introducing conformal symmetry into canonical quantum gravity. In contrast to the standard treatment of shape dynamics in the spatially compact case, we will not restrict to volume preserving conformal transformations and therefore not retain a global Hamiltonian\footnote{This possibility has also been pointed out by Tim Koslowski.}, which is more convenient for quantisation. We furthermore give a brief discussion of the spatially compact case with boundaries and the asymptotically anti-de Sitter case, which has not been spelled out in the literature so far, and the asymptotically flat case, which was discussed in \cite{GomesTheLinkBetween} only incompletely. We will conclude this section with remarks on a possible connection formulation of shape dynamics and a quantisation thereof. A hurdle here will be the complicated transformation behaviour of the connection under the new conformal symmetry\footnote{We thank Sean Gryb for pointing out this problem, which led to this paper.}. We will turn to a specific, matter coupled system in section \ref{sec:ConformallyCoupledScalarField}, which bypasses this hurdle, allowing for a connection formulation with a conformally invariant connection.

\subsection{Reminder: ADM formulation}
We start with the Einstein Hilbert action in $D+1$ spacetime dimensions
\be
	S_{EH} := \frac{1}{\kappa} \int_{M} d^{D+1}X ~ \sqrt{|g|} R^{(D+1)} \text{.}
\ee
The notation is as follows: $M$ denotes a $D+1$ dimensional, globally hyperbolic (connected) spacetime manifold\footnote{That is, $(M,g_{\mu\nu)}$ is isometric to $(\mathbb{R}\times\sigma,-\beta dt^{2}+g^{t})$, where $(\sigma,g^{t})$ is a (smooth) family of Riemannian manifolds and $\beta$ a (smooth) function on $M$ (cf. \cite{BernalSmoothnessOfTime, GerochDomainOfDependence, BaerWaveEquationsOn}, especially theorem 1.1. of \cite{BernalSmoothnessOfTime}).}, $g_{\mu \nu}$ denotes the metric tensor field and $g$ its determinant ($\mu, \nu, ... = 0,1,...,D$). Our signature convention is mostly plus  and $\kappa = 16 \pi G$ ($c=1$). The Riemann tensor is defined by $[\nabla_{\mu}, \nabla_{\nu}]\lambda_{\rho} =: R^{(D+1)}_{\mu \nu \rho}\m^{\sigma} \lambda_{\sigma}$, where $\nabla$ is torsion free and metric compatible. We furthermore define $R^{(D+1)}_{\mu \nu} := R^{(D+1)}_{\mu \rho \nu}\m^{\rho}$, $R^{(D+1)}:= R^{(D+1)}_{\mu}\m^{\mu}$.\\
\\
After a $D+1$ split and application of the Dirac algorithm \cite{DiracLecturesOnQuantum}, we arrive at the well-known ADM formulation \cite{ArnowittTheDynamicsOf}. The phase space is coordinatised by the canonical pair
\be
	\{q_{ab}(x), P^{cd}(y)\} := \delta_{(a}^c \delta_{b)}^{d} \delta^{(D)}(x,y) \text{,}
\ee
where $q_{ab}$ denotes the $D$ metric field on the spatial manifold $\sigma$ ($a,b,... = 1,...,D$) and $P^{ab}$ its conjugate momentum. The constraints of the system, the Hamiltonian and spatial diffeomorphism constraint, are given by
\be
	\kappa \mathcal{H}[N] &:=& \int_{\sigma}d^Dx~ N \left[ \frac{\kappa^2}{\sqrt{q}} P^{\text{tf}}_{ab} P_{\text{tf}}^{ab} - \sqrt{q} R^{(D)}  - \frac{\kappa^2}{(\Delta^g)^2 D(D-1)\sqrt{q}} \mathcal{D}^2 \right] \text{,} \\
	\mathcal{H}_a[N^a] &:=& \int_{\sigma}d^Dx~ P^{ab} (\mathcal{L}_N q)_{ab} \text{,}
\ee
where we, in contrast to the usual presentation of the Hamiltonian constraint, completely separated the trace ($P := P^{ab} q_{ab}$) and trace free parts $(P^{ab}_{\text{tf}} := P^{ab} - \frac{1}{D} q^{ab} P)$ of $P^{ab}$ and defined
\be
	\mathcal{D} := \Delta^g P \text{.}	
\ee
with $\Delta^{g}$ denoting the conformal weight of the metric (see also below).
The constraints satisfy the hypersurface deformation algebra
\be
\{\mathcal{H}_a[N^a], \mathcal{H}_b[M^b] \} &=& \mathcal{H}_a[(\mathcal{L}_N M)^a]  \text{,} \nonumber \\
\{\mathcal{H}_a[N^a], \mathcal{H}[M] \} &=& \mathcal{H}[\mathcal{L}_N M] \text{,} \nonumber \\
\{\mathcal{H}[N], \mathcal{H}[M] \} &=& \mathcal{H}_a[q^{ab} (N D_bM-M D_bN)] \text{.}  \label{eq:HDA}
\ee

\subsection{Gauge fixing}
\label{sec:GF}
The role of $\mathcal{D}$ is twofold. On the one hand, it is easy to verify that $\mathcal{D}$ generates conformal transformations (dilatations) on the spatial metric $q_{ab}$ and its momentum $P^{ab}$, with conformal weights $\Delta^g$ and $-\Delta^g$ respectively. Defining $\mathcal{D}[\rho] := \int_{\sigma} d^Dx ~ \rho(x) \mathcal{D}(x)$ for some smearing function $\rho$ of compact support, we find
\be
	\{q_{ab}(x), \mathcal{D}[\rho]\} = \Delta^g \rho(x) q_{ab}(x) \text{,} ~~~ \{P^{ab}(x), \mathcal{D}[\rho]\} = - \Delta^g \rho(x) P^{ab}(x) \text{.}
\ee
On the other hand, up to constant factors, $\mathcal{D} \propto P \propto \sqrt{q} q_{ab} K^{ab} := \sqrt{q} K$, where $K_{ab}$ denotes the extrinsic curvature. We now want to introduce the gauge fixing condition
\be
	\mathcal{D}_{\delta} := \mathcal{D} - \sqrt{q} \delta
\ee
for some constant $\delta \in \mathbb{R}$, corresponding to the well-known constant mean curvature (CMC) gauge $K = - \frac{\kappa}{(D-1)\Delta^g} \delta = \text{const.}$, which was already considered by Dirac \cite{DiracFixationOfCoordinates}. While this constraint, being a scalar density, obviously is covariant under spatial diffeomorphisms, it fixes the freedom to choose a foliation, i.e. the (on-shell) transformations generated by the Hamiltonian constraint, and therefore constitutes a partial gauge fixing. In view of the algebra of constraints (\ref{eq:HDA}), we are in case II according to the terminology of section \ref{sec:PartialComplete}. The question if $\mathcal{D}_{\delta}$ really constitutes a good gauge fixing for $\mathcal{H}$ will be investigated in the following.

\subsubsection{Invertibility of the Dirac matrix}
\label{sec:Invertibility}
As was discussed in section \ref{sec:PartialComplete}, locally, we need to check if the Dirac matrix corresponding to $\mathcal{H}$ and $\mathcal{D}_\delta$ is invertible on $S_{J(\mathcal{D}_{\delta}, \mathcal{H}, \vec{\mathcal{H}})}$. Using
\be
\int_{\sigma} d^Dx~ N \sqrt{q} \{R^{(D)}, \mathcal{D}[\rho]\} &=& \int_{\sigma} d^Dx ~ \rho \sqrt{q} \Delta^g \left[ - (D-1) D_a D^a - R^{(D)} \right] N \text{,} \\
\int_{\sigma} d^Dx~ N \{\sqrt{q}, \mathcal{D}[\rho]\} &=& \int_{\sigma} d^Dx ~ \frac{D}{2} \Delta^g N\rho \sqrt{q} \text{,} \\
\{\kappa \mathcal{H}[N], \int_{\sigma} d^Dx ~ \rho \sqrt{q} \delta \} &=& \frac{\delta \kappa^2}{\Delta^g (D-1)}  \mathcal{D}[N \rho] \text{,}
\ee
we find
\be
	\{\kappa \mathcal{H}[N], \mathcal{D}_\delta[\rho]\} &=& \int_{\sigma} d^Dx ~ \rho \left[ -\frac{D \kappa^2}{2\sqrt{q}} \Delta^g P^{ab}_{\text{tf}} P_{ab}^{\text{tf}} - \frac{D-2}{2} \Delta^g \sqrt{q} R^{(D)} \right.\nonumber \\ & & \left. \hspace{1.8cm} + \Delta^g \frac{\kappa^2}{2 (\Delta^g)^2 (D-1)\sqrt{q}} \mathcal{D}^2 - \frac{\kappa^2 \delta}{\Delta^g (D-1)} \mathcal{D}\right]N \nonumber \\
	&=& \kappa \mathcal{H}\left[-\frac{1}{2}\Delta^g N \rho\right] + \mathcal{D}_{\delta}\left[\frac{\kappa^2 N \rho}{\Delta^g}\left(\frac{1}{2 D\sqrt{q}} \mathcal{D}_{\delta} - \frac{\delta}{D(D-1)}\right)\right] \nonumber \\ & & + \int_{\sigma} d^Dx~ (D-1)\sqrt{q}\Delta^g \rho \left[ D_a D^a - \mathcal{P}(x)\right] N \text{,} \label{eq:DiracMatrix}
\ee
where
\be
	\mathcal{P}(x) := \frac{\kappa^2}{2q}P^{ab}_{\text{tf}} P_{ab}^{\text{tf}} + \frac{1}{2} R^{(D)} + \delta^2 \frac{\kappa^2(D+1)}{2(\Delta^g)^2D(D-1)^2} \text{.} \label{eq:Pvac}
\ee
In the matter free case, we can furthermore simplify $\mathcal{P}(x)$ by replacing $R^{(D)}$ using $\mathcal{H} = \mathcal{D}_{\delta} = 0$ to obtain\footnote{This further simplification will not be possible in the matter coupled case, cf. section \ref{sec:ConformallyCoupledScalarField}. The definition of $\mathcal{P}(x)$ in (\ref{eq:Pvac}) is chosen such that it can be easily compared with the matter coupled case.}
\be
	\mathcal{P}(x) \approx \frac{\kappa^2}{q} P^{ab}_{\text{tf}} P_{ab}^{\text{tf}} + \delta^2 \frac{\kappa^2}{(\Delta^g)^2 D(D-1)^2} \left\{\begin{array}{cl} > 0, & \mbox{if } \delta \neq 0 \\ \geq 0, & \mbox{if } \delta = 0  \text{.} \end{array}\right.
\ee
By (\ref{eq:DiracMatrix}) and the (formal) self-adjointness $D^a D_a$, we see that local invertibility of the Dirac matrix on the constraint surface $S_{J(\mathcal{D}_{\delta}, \mathcal{H})}$ is equivalent to the requirement that the elliptic PDE 
\be 
	[D^a D_a - \mathcal{P}(x)]N(x) = 0 \label{eq:EPDE}
\ee
has the unique solution $N(x) = 0$ (cf. \cite{ChoquetBruhatGeneralRelativityAnd}, Appendix II, Theorem 2.3, p.549). We will study this equation both, in the spatially compact, as well as in the asymptotically flat case, and comment on it in the asymptotically anti-de Sitter case.

\paragraph{Spatially compact case:}
We have to separately discuss the cases $\delta \neq 0$ and $\delta=0$.

\subparagraph{Case $\delta \neq 0$:}
If the spatial manifold $\sigma$ is compact without boundary, a sufficient condition for (\ref{eq:EPDE}) to have a unique solution is that $\mathcal{P}(x) > 0$. If this holds, it is easy to show that the obvious solution $N(x) = 0$ is unique, because suppose there would be a solution $\tilde{N}(x)$ which does not vanish everywhere. Then,
\be
	0 < \int_{\sigma} d^Dx ~ \sqrt{q} \tilde{N}^2 \mathcal{P} = \int_{\sigma} d^Dx ~ \sqrt{q} \tilde{N} D_a D^a \tilde{N} =  - \int_{\sigma} d^Dx ~ \sqrt{q} (D_a \tilde{N}) (D^a \tilde{N}) \leq 0
	\label{eq:Estimation}
\ee 
is a contradiction. Note that if the spatial manifold has a boundary $\partial \sigma \neq \emptyset$ (e.g. an inner horizon), we need to impose additional boundary conditions in order to deal with the surface terms appearing when partially integrating in (\ref{eq:Estimation}). For isolated horizons, it is known that Dirichlet boundary conditions $N|_{\partial \sigma} = 0$ have to be imposed \cite{AshtekarIsolatedHorizonsThe}, which automatically leads to a vanishing of the surface terms. Therefore, the above discussion extends to the case with isolated horizons as boundaries.

\subparagraph{Case $\delta=0$:} In this case, for a non-trivial solution $\tilde{N}(x)$, we have
\be
	0 \leq \int_{\sigma} d^Dx ~ \sqrt{q} \tilde{N}^2 \mathcal{P} = \int_{\sigma} d^Dx ~ \sqrt{q} \tilde{N} D_a D^a \tilde{N} =  - \int_{\sigma} d^Dx ~ \sqrt{q} (D_a \tilde{N}) (D^a \tilde{N}) \leq 0 \text{.}
	\label{eq:Estimation2}
\ee 
If $\partial \sigma = \emptyset$, we will make a further restriction by demanding that a certain conformal invariant, the Yamabe constant\footnote{The Yamabe constant \cite{LeeTheYamabeProblem} of a compact smooth manifold $m$ of dimension $n$ is defined by
\be
	Y(g) = \inf_{\rho} \mathcal{E}(e^{\Delta^g \rho} g) \text{,} \nonumber
\ee
where $g$ is a Riemannian metric on $m$,
\be
	\mathcal{E}(g) := \frac{\int_m d^nx~ \sqrt{g} R^{(n)}_g}{(\int_m d^nx~ \sqrt{g})^{\frac{n-2}{n}}} \text{,} \nonumber
\ee
and the infimum is taken over smooth functions $\rho$ on $m$.}, of the spatial Riemannian manifold $(\sigma, q)$ is positive. Note that to this end, $\sigma$ has to be of positive Yamabe type, which implies a restriction on the topology of $\sigma$ \cite{GromovTheClassificationOf}. This restriction excludes constant vanishing scalar curvature $R^{(D)}(x) = 0$ on $\sigma$.\\
\\
The only way that (\ref{eq:Estimation2}) is not a contradiction is if each expression vanishes identically. By the right hand side of (\ref{eq:Estimation2}), we see that $\tilde{N}$ needs to be constant, say, $\tilde{N}(x) = \tilde{N}_0 \neq 0$. Then, the leftmost integral in (\ref{eq:Estimation2}) is only vanishing if $\mathcal{P}$ vanishes identically. Using $\mathcal{H} = \mathcal{D}_{\delta = 0} = \mathcal{D} = 0$, we conclude that $\frac{\kappa^2}{q}P^{ab}_{\text{tf}} P_{ab}^{\text{tf}} - R^{(D)} = 0$ and therefore $R^{(D)} \geq 0$. By (\ref{eq:Pvac}), this implies everywhere vanishing scalar curvature $R^{(D)}$, which we excluded by restricting to positive Yamabe constant. Note that if the spatial manifold has a boundary and we impose Dirichlet boundary conditions as before, we do not need the restriction to positive Yamabe constant, since a constant value of $\tilde{N}$ is incompatible with the boundary conditions unless $\tilde{N}$ is the trivial solution.

\paragraph{Asymptotically flat case:}
\label{sec:AsympFlat}
Although the focus of this paper is on the spatially compact case, we add some remarks on asymptotically flat spacetimes\footnote{In the context of $D+1$ splits, one often speaks of asymptotically Euclidean initial data (cf. \cite{ChoquetBruhatGeneralRelativityAnd}).} (for simplicity, we consider the 3+1 case only). Since the latter are generically non-compact, everything we worked out so far only applies to smearing functions of compact support. This is mainly because we completely ignored the appearance of (possible) boundary terms as well as questions of convergence of all our integrals. If we want to allow for more general Lagrange multipliers in the definition of the constraints, we will have to specify their behaviour and that of our phase space variables at infinity in such a way that the expressions for the constraints and their Hamiltonian vector fields remain well-defined\footnote{A mathematically accurate treatment of these issues makes it necessary to delve into functional analysis on suitable Sobolev spaces (cf. \cite{ChoquetBruhatGeneralRelativityAnd}).}. Moreover, this will lead to the problem of determining whether some of the allowed multipliers still represent gauge transformations or correspond to actual symmetries of the theory (cf. \cite{HenneauxQuantizationOfGauge, ThiemannModernCanonicalQuantum}). For asymptotically flat spacetimes this rather generic issue in the presence of boundary conditions leads to the existence of asymptotic symmetries and their associated conserved charges, prominently the \textit{ADM (3+1)-momentum}, as well as the so-called \textit{supertranslation ambiguity} (the latter only appear in the 2+1 and 3+1 case, see e.g. \cite{AshtekarAUnifiedTreatment, BeigEinsteinsEquationsNear}). More precisely, one may define functionals extending the Hamiltonian and the spatial diffeomorphism constraint (\cite{ThiemannModernCanonicalQuantum}, p.62 et seq.) to multipliers incorporating supertranslations $(N_{S},N^{a}_{S})$ (additional gauge symmetries) and asymptotic symmetries $(N_{P},N^{a}_{P})$ (Poincar\'e transformations at infinity):
\be
\label{eq:impgen}\nonumber
\mathcal{J}\left[N\right] & := & \mathcal{H}\left[N\right]+\kappa\mathcal{E}\left[N\right]\\[0.25cm]
\mathcal{J}_{a}\left[N^{a}\right] & := & \mathcal{V}_{a}\left[N^{a}\right]+\kappa\mathcal{P}_{a}\left[N^{a}\right],
\ee
where $\mathcal{E}\left[N\right]$, $\mathcal{P}_{a}\left[N^{a}\right]$ are boundary terms leading to the \textit{ADM energy} and the \textit{ADM 3-momentum} when $(N,N^{a})$ represent an asymptotic translation
\be
\label{eq:ADMmom}\nonumber
\kappa\mathcal{E}\left[N\right] & := &\ \ \ \ \int_{\partial\sigma}d^{3}x~ \sqrt{q}q^{cd}q^{ef}\left((D_{c}N)(dS_{d}(q_{ef}-\delta_{ef}))-(D_{e}N)(dS_{c}(q_{df}-\delta_{df}))\right) \\[0.25cm] \nonumber
&  &+\ \int_{\partial\sigma}d^{3}x~ \sqrt{q}q^{cd}N\left(-dS_{c}\Gamma^{e}_{ed}+dS_{e}\Gamma^{e}_{cd}\right) \text{,} \\[0.25cm]
\mathcal{P}_{a}\left[N^{a}\right] & := &\ \ \ \frac{2}{\kappa}\int_{\partial\sigma}dS_{b}N^{a}P^{b}_{a},
\ee
and $\mathcal{V}_{a}\left[N^{a}\right]$ denotes the spatial diffeomorphism constraint in the form\footnote{The diffeomorphism constraint $\mathcal{H}_{a}\left[N^{a}\right]$ defined above agrees with the extended functional $\mathcal{J}_{a}\left[N^{a}\right]$.}
\be
\label{eq:diffconalt}
\mathcal{V}_{a}\left[N^{a}\right] & := &\ \ \ -2\int_{\sigma}N^{a}q_{ac}D_{b}P^{bc}.
\ee
The boundary terms (\ref{eq:ADMmom}) have the effect that the functionals $\mathcal{J}\left[N\right]$, $\mathcal{J}_{a}\left[N^{a}\right]$ and their variations $\delta\mathcal{J}\left[N\right]$, $\delta\mathcal{J}_{a}\left[N^{a}\right]$ are well defined for the extended set of multipliers. Furthermore, since the $\mathcal{J}\left[N\right]$, $\mathcal{J}_{a}\left[N^{a}\right]$ extend $\mathcal{H}\left[N\right]$, $\mathcal{V}_{a}\left[N^{a}\right]$, one finds that
\be
\label{eq:constraintext}
\mathcal{J}\left[N\right]=\mathcal{H}\left[N\right], & \mathcal{J}_{a}\left[N^{a}\right]=\mathcal{V}_{a}\left[N^{a}\right]
\ee
for compactly supported $(N,N^{a})$. Additionally, we have
\be
\label{eq:constraintextsupertrans}
\mathcal{J}\left[N_{S}\right]=\mathcal{H}\left[N_{S}\right], & \mathcal{J}_{a}\left[N^{a}_{S}\right]=\mathcal{V}_{a}\left[N^{a}_{S}\right]
\ee
for supertranslations $(N_{S},N^{a}_{S})$ which justifies treating them as constraints rather than symmetry generators. The algebra of the extended functionals $\mathcal{J}\left[N,\vec{N}\right]:=\mathcal{J}\left[N\right]+\mathcal{J}_{a}\left[N^{a}\right]$ has been worked out e.g. in \cite{ThiemannModernCanonicalQuantum}:
\be
\label{eq:extendedalg}
\left\{\mathcal{J}\left[N_{1},\vec{N}_{1}\right],\mathcal{J}\left[N_{2},\vec{N}_{2}\right]\right\}=\mathcal{J}\left[\mathcal{L}_{\vec{N}_{2}}N_{1}-\mathcal{L}_{\vec{N}_{1}}N_{2},\mathcal{L}_{\vec{N}_{2}}\vec{N}_{1}-\vec{N}_{12}(q)\right],
\ee
where $\vec{N}^{a}_{12}(q):=q^{ab}(N_{1}\partial_{b}N_{2}-N_{2}\partial_{b}N_{1})$. This algebra has the property that the compactly supported together with the supertranslations form an ideal w.r.t. all allowed multipliers, which renders the split into constraints and asymptotic symmetry generators consistent.\\[0.25cm]
Coming to the idea of partially gauge fixing the constraints with the functionals $\mathcal{D}_{\delta}[\rho]$, we face similar problems. Namely, we have to prescribe appropriate boundary conditions for the multipliers $\rho$ to make the functionals and their Hamiltonian vector fields well-defined. Moreover, as we require the phase space variables to describe asymptotically flat spacetimes, the only sensible gauge fixing functionals appear to be $\mathcal{D}_{\delta=0}[\rho]$ (cf. \cite{MurchadhaExistenceAndUniqueness}). For compactly supported smearing functions, everything works as in the spatially compact case, i.e. the operator $D^{a}D_{a}-\mathcal{P}$ is (formally) self-adjoint and injective\footnote{If we can complete the compactly supported (smooth) multipliers in a suitable (weigthed) Sobolev $p$-norm, we will even get a surjective operator $D^{a}D_{a}-\mathcal{P}:W^{p}_{s,\delta}\rightarrow L^{p}$ (cf. \cite{ChoquetBruhatGeneralRelativityAnd}, especially Appendix II, Theorem 3.7).}. For supertranslations the arguments we presented above do not work anymore, which leaves us with the possibility of a remaining gauge freedom. Still, we can conclude from equation (\ref{eq:extendedalg}) that it is consistent to gauge fix only the generators with compactly supported multipliers, since they generate an ideal within the whole constraint algebra\footnote{If we complete the compactly supported (smooth) multipliers to a Sobolev space, we will have to ensure that the ideal property remains valid. Otherwise a partial gauge fixing without the supertranslations might be inconsistent.}. The Hamiltonian constraints corresponding to multipliers in the kernel of $D^{a}D_{a}-\mathcal{P}$ have to be kept.

\paragraph{Asymptotically anti-de Sitter case:}
\label{sec:AsympAdS}

The Hamiltonian treatment of asymptotically anti-de Sitter spacetimes has been performed in \cite{HenneuaxHamiltonianTreatmentOf, HenneuaxAsymptoticallyAntiDe, AshtekarAsymptoticallyAntiDe}, see also \cite{AshtekarAsymptoticallyAntiDeConserved}. As opposed to the asymptotically flat case, no supertranslations appear and the asymptotic symmetry transformations generated by the constraints corresponds to the isometry group O$(3,2)$ of anti-de Sitter space. In order to generate this asymptotic symmetry, the lapse and shift functions have to approach the asymptotic anti-de Sitter killing vectors sufficiently fast. A derivation of the necessary fall-off behaviour of possible deviations has been given in \cite{HenneuaxAsymptoticallyAntiDe} for $3+1$ dimensions. Since we do not want to fix any asymptotic transformations using $\mathcal{D}_\delta=0$, we can restrict the fall-off behaviour of the lapse functions in all calculations accordingly. It follows that all boundary terms which were dropped before, e.g. in (\ref{eq:Estimation}), are still vanishing, and our results are thus still valid. We leave the case of higher dimensions to the interested reader, as the $3+1$ dimensional case suffices for the application to the $3+1$ dimensional, locally asymptotically anti-de Sitter black hole proposed in \cite{BSTI}. 
As in the asymptotically flat case, it might be that not all values of $\delta$ are in agreement with the asymptotic behaviour and have to be restricted accordingly.

\subsubsection{Global aspects}
So far, we only have been discussing the quality of the gauge fixing $\mathcal{D}_{\delta} = 0$ locally. In this section, we want to comment on global aspects of the chosen gauge. To assure a globally good gauge fixing, we ideally would like to calculate the finite Hamiltonian flow $\alpha_{N \mathcal{H}}^{*}(\mathcal{D}_{\delta})$ of $\mathcal{H}$ on $\mathcal{D}_{\delta}$ on the surface $S_{J(\mathcal{H})}$ and solve $\alpha_{N \mathcal{H}}^{*}(\mathcal{D}_{\delta} ) = 0$ uniquely for the gauge parameter $N$. This would guarantee both, uniqueness (up to large gauge transformations) of the gauge cut and accessibility of the chosen gauge fixing. However, due to the form of the algebra (\ref{eq:HDA}), in particular $\{\mathcal{H}, \mathcal{H}\} \propto \mathcal{H}_a$, $\alpha_{N \mathcal{H}}$ does not leave $S_{J(\mathcal{H})}$ invariant, and even if it would, the complicated form of $\mathcal{H}$ would probably preclude the calculation of finite $\mathcal{H}$ transformations. What, instead, can be calculated are the finite $\mathcal{D}_{\delta}$ transformations, since we know how conformal transformations act on the individual fields in the phase space. But even if $\alpha_{\rho \mathcal{D}_{\delta}}^{*}(\kappa \mathcal{H}) = 0$ could be uniquely solved for $\rho$ on $S_{J(\mathcal{D}_{\delta})}$ (we will study this question in section \ref{sec:GlobalAspects2}), this would not necessarily assure accessibility (there may still be $\mathcal{H}$ gauge orbits which do not intersect $S_{J(\mathcal{D}_{\delta})}$ at all) nor uniqueness of the gauge cut (even if each $\mathcal{D}_{\delta}$ orbit intersects the surface $S_{J(\mathcal{H})}$ only once, we have no argument which excludes the possibility that one $\mathcal{H}$ orbit on $S_{J(\mathcal{H})}$ intersects $S_{J(\mathcal{D}_{\delta})}$ more than once). Therefore, we leave this question for further research.\\
\\
In conclusion, we found that in the spatially compact case (with or without boundaries), $\mathcal{D}_{\delta} $ locally is a good gauge fixing condition for the Hamiltonian constraint.\\
In the asymptotically flat case, we are left with the possibility of a remaining gauge freedom coming from Hamiltonian constraints with non-trivial multipliers that are solutions to equation (\ref{eq:EPDE}).\\
To assure accessibility, we have to restrict to solutions of Einstein's equations which allow for a CMC Cauchy slice. Although this requirement is weak, there are solutions known admitting no CMC Cauchy surfaces at all, already pointed out in \cite{BartnikRemarksOnCosmological}. Global uniqueness of the gauge cut is unsettled.

\subsection{Gauge unfixing: shape dynamics}
\label{sec:GU:SD}
The algebra of constraints now reads, apart from the relations given in (\ref{eq:HDA}, \ref{eq:DiracMatrix}), as follows:
\be
	\{\mathcal{H}_a[N^a], \mathcal{H}_b[M^b] \} &=& \mathcal{H}_a[(\mathcal{L}_N M)^a]  \text{,} \nonumber \\
	\{\mathcal{H}_a[N^a], \mathcal{D}_{\delta}[\rho]\} &=& \mathcal{D}_{\delta}[\mathcal{L}_N\rho] \text{,} \nonumber \\
	\{\mathcal{D}_{\delta}[\rho], \mathcal{D}_{\delta}[\rho']\} &=& 0  \text{,} \label{eq:SDA}
\ee
where we repeated the first equation for convenience. Of course, we can gauge unfix this system in the trivial way, by dropping the gauge fixing condition we introduced. However, because in this case $\{\mathcal{H}_a, \mathcal{D}_{\delta}\}$ form a subalgebra, we can as well trivially gauge unfix the system by dropping $\mathcal{H}$, which now is interpreted as unnecessary gauge fixing condition for $\mathcal{D}_{\delta}$. The resulting algebra of constraints is given in (\ref{eq:SDA}). Compared to (\ref{eq:HDA}), it is far more convenient for quantisation, since it actually is a Lie algebra, whereas in (\ref{eq:HDA}) structure functions appear. This property is shared by deparametrised models \cite{GieselAQG4, DomagalaGravityQuantizedLoop, HusainTimeAndA}. As already mentioned, the above theory of gravity has been derived in \cite{GomesEinsteinGravityAs}\footnote{Actually, shape dynamics in the spatially compact case is recovered for the choice $\delta = \langle P \rangle$, where $\langle . \rangle$ denotes the spatial mean. Since $\langle P \rangle$ is not constant on the phase space, the resulting gauge fixing is different from the ones considered here and our calculations do not apply to that case. In particular, $\mathcal{D}_{\langle P \rangle}$ does not fix the Hamiltonian constraint completely and one retains a global Hamiltonian.} and is called shape dynamics.\\
\\
However, in order that the gauge unfixed theory and the original one (when restricted to solutions admitting at least one CMC slice) are identical, we have to check if the original constraint $\mathcal{H}$ is a good gauge fixing condition for the gauge unfixed theory. To the best of the authors' knowledge, it has not been shown that this requirement is related to $\mathcal{D}_{\delta}$ being a good gauge fixing condition for $\mathcal{H}$ and has to be discussed separately. E.g. even if $\mathcal{D}_{\delta}$ would globally be a good gauge fixing for $\mathcal{H}$, there might be finite $\mathcal{D}_{\delta}$ transformations connecting different points on the gauge cut $S_{J(\mathcal{D}_{\delta},\mathcal{H})}$. In this case, these configurations, which were distinct in the original $\mathcal{H}$ theory, would be erroneously interpreted as physically equivalent in the $\mathcal{D}_{\delta}$ theory.

\subsubsection{Local considerations}
Locally, the question if $\mathcal{H}$ is a good gauge fixing condition for $\mathcal{D}_{\delta}$ leads to (\ref{eq:DiracMatrix}) as before. The differential operator appearing is (formally) self-adjoint on $C^{\infty}(\sigma)$ (or some suitable Sobolev space $W^{p}_{s}(\sigma)$) and since, at least in the spatially compact case, the function spaces for $N$ and $\rho$ can be taken the same, we find that $\mathcal{H}$ locally is a good gauge fixing condition (cf. \cite{ChoquetBruhatGeneralRelativityAnd}, Appendix II, Theorem 2.3, p.549).\\
In the asymptotically flat case the situation is less clear due to the unresolved issue concerning the supertranslations (see section \ref{sec:Invertibility} above), which might even lead to a modification of the algebra (\ref{eq:SDA}), i.e. one would have to add the gauge unfixed versions of the Hamiltonian constraint with multipliers that solve equation (\ref{eq:EPDE}). The latter can be achieved, at least formally, by applying the observable projector $_{2}\mathbb{P}^{\mathcal{H}}_{D}$ (see section \ref{sec:partialproj}) for compactly supported smearing functions to the remaining constraints, although this seems not explicitly tractable, as it involves the inverse to the operator $D^{a}D_{a}-\mathcal{P}$. 

\subsubsection{Global aspects}
\label{sec:GlobalAspects2}
In order to study the global properties of the gauge fixing $\mathcal{H} = 0$, we have to investigate existence and uniqueness of the solution to the equation $\alpha_{\rho \mathcal{D}_{\delta}}^{*}(\mathcal{H}) = 0$ on $S_{J(\mathcal{D}_{\delta})}$. Using the formulas in appendix \ref{app:A}, it is straightforward to show that
\be
	\alpha_{\rho \mathcal{D}_{\delta}}^{*}(\kappa \mathcal{H}) &=\hspace{9mm}& \Omega^{-\frac{2D}{D-2}} \frac{\kappa^2}{\sqrt{q}} P^{ab}_{\text{tf}} P_{ab}^{\text{tf}}  - \Omega^2 \sqrt{q} \left[R^{(D)} - \frac{4(D-1)}{(D-2)} \Omega^{-1} D_a D^a \Omega \right] \nonumber \\ & & - \frac{\kappa^2}{(\Delta^g)^2 D(D-1)} \left[\Omega^{- \frac{2D}{D-2}} \frac{1}{\sqrt{q}}\mathcal{D}_{\delta} ^2 + 2\mathcal{D}_{\delta}  \delta + \Omega^{\frac{2D}{D-2}} \sqrt{q}\delta^2 \right] \nonumber \\
	&\approx_{S_{J(\mathcal{D}_{\delta})}}&  \Omega^{-\frac{2D}{D-2}} \frac{\kappa^2}{\sqrt{q}} P^{ab}_{\text{tf}} P_{ab}^{\text{tf}}  - \Omega^2 \sqrt{q} \left[R^{(D)} - \frac{4(D-1)}{(D-2)} \Omega^{-1} D_a D^a \Omega \right] \text{,} \label{eq:Global}
\ee
where $q_{ab} \rightarrow e^{\Delta^g \rho} q_{ab}$ and $\Omega := e^{\frac{(D-2)}{4} \Delta^g \rho}$. For the sake of a simpler exposition, restrict to $D=3$, multiply the above equation by $(\sqrt{q} \Omega)^{-1}$ 
and use $P^{ab} = \frac{1}{\kappa} \sqrt{q} (K^{ab} - q^{ab} K)$. We obtain
\be
	8 D_a D^a \Omega - R^{(3)} \; \Omega + K^{ab}_{\text{tf}} K_{ab}^{\text{tf}} \; \Omega^{-7} - \frac{2}{3}  K^2 \; \Omega^5 = 0\text{,} \label{eq:York}
\ee
which is the well-known Lichnerowicz-York equation. The existence and uniqueness of solutions have been studied extensively, both in the spatially compact case with and without boundary in any dimension $D \geq 3$ and including various matter coupling: A solution almost always exists (up to a measure zero set in the function space of all initial data sets) and then is generically unique. Under simple assumptions, the results continue to hold in the asymptotically flat case. For details, we refer the interested reader to the recent book \cite{ChoquetBruhatGeneralRelativityAnd} and the original literature cited therein.\\
\\
As will be shown in the next subsection, a quantisation of $\mathcal{D}_{\delta}$ for pure gravity is possible, however, the kernel of the constraint operator is very difficult to analyse due to the volume operator which would appear in a proper regularisation. In the next section, we will show how to circumvent this problem by introducing an additional matter degree of freedom and constructing an invariant connection at the classical level. In the light of this invariant connection, it does not really matter if we are using the gauge unfixed version of this theory or the if we are calculating the Dirac bracket, since on the one hand, the Dirac bracket will coincide with the Poisson bracket for the invariant connection, and on the other hand, the invariant connection is a Dirac observable with respect to $\mathcal{D}_{\delta}$ for the gauge unfixed system. In order to understand better the relation to shape dynamics, it is however useful to employ the gauge unfixing point of view and we have included it for completeness. Also, the relation between the observable algebras becomes more clear this way.

\subsection{Remarks on a connection formulation and quantisation}
Finally, we want to give some comments on a connection formulation of shape dynamics and its quantisation. Since the phase space is coordinatised, like in the ADM case, by a spatial metric $q_{ab}$ and its conjugate momentum $P^{ab}$, we can derive a connection formulation in complete analogy to the usual Ashtekar-Barbero treatment \cite{AshtekarNewVariablesFor, BarberoRealAshtekarVariables} or, in dimensions $D \geq 2$, following the construction in \cite{BTTI}, and quantise it using the standard LQG methods \cite{RovelliQuantumGravity,ThiemannModernCanonicalQuantum}. However, since the connection variable does not transform nicely under conformal transformations, solving the constraint $\mathcal{D}_{\delta}$ in the quantum theory is a problem of the same complexity as solving $\mathcal{H}$ in standard LQG.\\
\\
Exemplarily, we will discuss the case $D=3$ using Ashtekar-Barbero variables. We introduce an extended ADM phase space coordinatised by a densitised triad $E^{ai}$ and conjugate momentum $K_{ai}$ subject to an su(2) Gau{\ss} constraint $\mathcal{G}^{ij} := K_{a}^{[i} E^{a|j]}$, followed by a canonical transformation to Ashtekar-Barbero connection variables
\be
	\{K_{ai}, E^{bj}\} \longrightarrow \{\m^{(\gamma)} A_{ai} := - \frac{1}{2} \epsilon_{ijk} \Gamma^{jk}(e) + \gamma K_{ai}, \m^{(\gamma)}E^{bj}:= \frac{1}{\gamma} E^{bj}\} \text{,}
\ee
where $\Gamma_{aij}(e)$ denotes the spin connection annihilating the tetrad. It is easy to see that $E^{ai}$ and $K_{ai}$ have conformal weight $\Delta^g$ and $-\Delta^g$, respectively. While $\Gamma_{aij}(e)$, being a homogeneous rational function of degree zero of $e$ and its derivatives, is invariant under constant rescalings, under local ones it transforms as
\be
	\alpha_{\rho}^{\mathcal{D}}(\Gamma_{aij}(e)) = \Gamma_{aij}(e) + \Delta^g e^b\m_{[i} e_{a|j]} D_b \rho \text{.}
\ee
Therefore, also the connection $\m^{(\gamma)}A_{ai}$ has a rather complicated transformation behaviour under transformations generated by $\mathcal{D}_{\delta} = \Delta^g K - \sqrt{E} \delta$ (here, $K := K_{ai} E^{ai}$ and $E$ denotes the determinant of $E^{ai}$),
\be
	\alpha_{\rho}^{\mathcal{D}_{\delta}}(\m^{(\gamma)}A_{ai}) &=&-\frac{1}{2} \epsilon_{ijk} ~ \alpha_{\rho}^{\mathcal{D}}(\Gamma^{jk}_a(e)) + \gamma ~ \alpha_{\rho}^{\mathcal{D}_{\delta}}(K_{ai}^{\text{tf}} + \frac{1}{D \sqrt{E}} e_{ai} K) \nonumber \\
	&=&-\frac{1}{2} \epsilon_{ijk} ~ \alpha_{\rho}^{\mathcal{D}}(\Gamma^{jk}_a(e)) + \gamma ~ \alpha_{\rho}^{\mathcal{D}_{\delta}}(K_{ai}^{\text{tf}} + \frac{1}{\Delta^g D\sqrt{E}} e_{ai} (\mathcal{D}_{\delta} + \sqrt{E}\delta)) \nonumber \\
	&=&-\frac{1}{2} \epsilon_{ijk} ~ (\Gamma^{jk}_a(e) + \Delta^g e^b\m_{[i} e_{a|j]} D_b \rho) \nonumber \\
	& & + \gamma \left[ e^{-\Delta^g \rho} K_{ai} + \frac{\delta}{\Delta^g D} e^{-\frac{1}{2}\Delta^g \rho}(1 - e^{-\frac{1}{2}\Delta^g\rho})e_{ai} \right] \text{,}
\ee
which even in the simplest case, $\delta=0$, makes it questionable if the corresponding constraint can be solved after loop quantisation. However, using Thiemann's methods \cite{ThiemannModernCanonicalQuantum}, one can rewrite $K_{ai}$ according to
\be
	K_{ai}(x) \propto \{\m^{(\gamma)}A_{ai}(x), \{\mathcal{H}_E(1), V(\sigma)\}\} \text{,}
\ee
where $\mathcal{H}_E$ denotes the Euclidean Hamiltonian constraint and $V$ the volume, and represent the conformal constraint $\mathcal{D}_{\delta}$ as an operator on the kinematical Hilbert space, which unfortunately will be of the same complexity as the Hamiltonian constraint in standard LQG.

\section{Conformally coupled scalar field}
\label{sec:ConformallyCoupledScalarField}
As we have seen at the end of the previous section \ref{sec:MatterfreeCase}, while a canonical transformation to a connection formulation exists, the resulting canonical variables will not transform nicely under conformal transformations and it is unclear how to solve the constraint both classically and quantum mechanically. Of course, one might try to find a different connection formulation classically or modify the existing ones such that the connection in the end has a nice transformation behaviour, but there is at least no obvious way to do so in the matter free case. In this section, we will show how this problem can be solved for the case of a matter coupled system, namely general relativity conformally coupled to a scalar field. While it has been argued that matter fields in shape dynamics are most naturally coupled with conformal weight zero \cite{GomesCouplingShapeDynamics}, in the case at hand the scalar field with non-trivial conformal weight will be central for the construction of the connection formulation with nice conformal transformation properties. The model will be extended to other standard model matter fields in section \ref{sec:ExtensionToStandardModelMatterFields}. \\
\\
The action for the conformally coupled scalar field is given by
\be
 S_{\Phi} := \frac{1}{2\lambda} \int_{M} d^{D+1}X \sqrt{|g|} \left( g^{\mu \nu} (\nabla_{\mu} \Phi) (\nabla_{\nu} \Phi) - \frac{1}{D} \frac{\Delta^{\Phi}}{\Delta^g} R^{(D+1)} \Phi^2 \right) \text{,} \label{eq:ScalarFieldAction}
\ee
Using the formulas in appendix \ref{app:A}, it can be easily checked that (\ref{eq:ScalarFieldAction}) is invariant under the conformal transformation
\be
	g_{\mu \nu} \rightarrow \Omega^{\Delta^g} g_{\mu \nu}, ~~~ \Phi \rightarrow \Omega^{\Delta^{\Phi}} \Phi \text{,}
\ee
for a smooth, strictly positive function $\Omega$, provided that the conformal weights $\Delta^{\Phi}$ and $\Delta^g$ for the scalar field and the space time metric satisfy
\be
	\Delta^{\Phi} = \frac{1-D}{4} \Delta^g \text{.}
\ee
In the following, we will restrict to that case. \\
\\
The conformal invariance of the above action will be of central importance in what follows. It implies that the matter contribution of the Hamiltonian constraint $\mathcal{H}_{\Phi}$ transforms with weight $- \frac{1}{2} \Delta^g$, which is the negative conformal weight of the lapse function. This is what one would expect, since the invariance of the action suggests that $N \mathcal{H}_{\Phi}$ is conformally invariant. This intuitive result will be confirmed in this section by direct calculation and will allow for a direct generalisation to coupling of standard model matter fields in $D=3$ in section \ref{sec:ExtensionToStandardModelMatterFields}.
\subsection{Canonical analysis}
The action of the scalar field conformally coupled to general relativity is given by
\be
\label{eq:CombinedAction}
	S = S_{GR} + S_{\Phi} = \frac{1}{\kappa} \int_{M} d^{D+1}X \sqrt{|g|} R^{(D+1)} a(\Phi) +  \frac{1}{2\lambda} \int_{M} d^{D+1}X \sqrt{|g|} g^{\mu \nu} (\nabla_{\mu} \Phi) (\nabla_{\nu} \Phi) \text{,}
\ee
where we defined
\be
	a(\Phi) &:=& 1 - \alpha \Phi^2 \text{,}\\
	\alpha &:=& \frac{\kappa}{2 \lambda} \frac{1}{D} \frac{\Delta^{\Phi}}{\Delta^{g}} = - \frac{\kappa (D-1)}{8 \lambda D} \text{,}
\ee
for convenience. The corresponding equations of motion are given by
\be
	0 &=& \nabla_\mu \nabla^{\mu} \Phi + \frac{1}{D} \frac{\Delta^{\Phi}}{\Delta^g} R^{(D+1)} \Phi \text{,}	\\
	T_{\mu \nu} &=& \frac{\kappa}{16 \pi \lambda} \left[\frac{1}{D} \frac{\Delta^{\Phi}}{\Delta^g} \left(G_{\mu \nu} - \nabla_{\mu} \nabla_{\nu} + g_{\mu \nu} \nabla^{\rho} \nabla_{\rho} \right) \Phi^2 + (\nabla_{\mu} \Phi)(\nabla_{\nu} \Phi) - \frac{1}{2} g_{\mu \nu} (\nabla^{\rho}\Phi) (\nabla_{\rho} \Phi)\right]  \nonumber \text{,}\\
	G_{\mu \nu} &=& 8 \pi T_{\mu \nu} \text{.}
\ee
In particular, the trace of the energy momentum tensor is vanishing, a necessary consequence of the conformal invariance of the action (cf. e.g. \cite{WaldGeneralRelativity}). The corresponding Hamiltonian formulation has been derived in \cite{KieferNonminimallyCoupledScalar} for $3+1$ dimensions. We include the analysis in arbitrary dimension for completeness. The split of this action is rather different from the ADM case. The surface terms, which are usually dropped when performing the $D+1$ decomposition, here lead to extra terms because of the presence of $a(\Phi)$. After a tedious calculation, one finds
\be
	S &=& \frac{1}{\kappa} \int_{\mathbb{R}} dt \int_{\sigma} d^Dx ~ \left\{ N \left[a(\Phi) \sqrt{q} R^{(D)}  + \frac{\kappa}{2\lambda} \sqrt{q} \left( \frac{1}{D} q^{\mu \nu} \Phi_{,\mu} \Phi_{,\nu} - \frac{(D-1)}{D} \Phi \Delta \Phi\right)\right] \right. \nonumber \\ 
	& &\left. \hspace{2.5cm}+~ \frac{1}{2} M^{\mu \nu, \rho \sigma} (\mathcal{L}_n q)_{\mu \nu}(\mathcal{L}_n q)_{\rho \sigma} + M^{\mu \nu} (\mathcal{L}_n q)_{\mu \nu} (\mathcal{L}_n \Phi)+\frac{1}{2} M (\mathcal{L}_n \Phi)^2 \right. \nonumber \\ 
	& & \left. \hspace{2.5cm} - ~ 4 \partial_{\mu} \left[ a(\Phi) \sqrt{|g|} n^{[\nu} \nabla_{\nu} n^{\mu]}+ \alpha \sqrt{|g|} \Phi q^{\mu \nu}  \Phi_{,\nu} \right] \right\} \text{,}
\ee
where the terms in the last line constitute surface terms which will be dropped in the following, $n^{\mu}$ denotes the future pointing timelike unit normal to the spatial slices of the foliation and
\be
	M^{\mu \nu, \rho \sigma}  &:=& \frac{1}{2} a(\Phi) N \sqrt{q} \left[q^{(\mu| \rho} q^{\nu) \sigma} - q^{\mu \nu} q^{\rho \sigma} \right] \text{,}\\
	M^{\mu \nu} &:=& 2 \alpha N \sqrt{q} \Phi q^{\mu \nu}\text{,}\\
	M &:=& -\frac{\kappa}{\lambda} N \sqrt{q} \text{.}
\ee 
Using $\frac{\delta}{\delta \dot{X}} = \frac{1}{N} \frac{\delta}{\delta (\mathcal{L}_n X)}$, we find for the canonical momenta
\be
	\pi_{\Phi} &:=& \frac{\delta L}{\delta \dot{\Phi}} = \frac{1}{N\kappa} \left(M (\mathcal{L}_n \Phi) + M^{ab} (\mathcal{L}_n q)_{ab}\right) = -\frac{1}{\lambda} \sqrt{q} (\mathcal{L}_n\Phi) + \frac{4 \alpha}{\kappa} \sqrt{q} \Phi K \text{,} \label{eq:ScalarFieldMomentum} \\ 
	P^{ab} &:=& \frac{\delta L}{\delta \dot{q}_{ab}} = \frac{1}{N \kappa} \left( M^{ab,cd} (\mathcal{L}_n q)_{cd} + M^{ab} (\mathcal{L}_n \Phi) \right) \nonumber \\ &=& \frac{1}{\kappa} a(\Phi) \sqrt{q} \left(K^{ab} - q^{ab} K\right) +  \frac{2\alpha}{\kappa} \sqrt{q} q^{ab} \Phi (\mathcal{L}_n \Phi)\text{.} \label{eq:MetricMomenta}
\ee
Solving these equations for the velocities is simplified if we introduce the short hand notation for trace part  $X := X^{ab} q_{ab}$ and trace free part $X^{ab}_{\text{tf}} := X^{ab} - \frac{1}{D} q^{ab} X$ of tensor fields $X^{ab}$ and decompose the gravitational momenta accordingly. We obtain
\be
	(\mathcal{L}_n \Phi) &=& -\frac{\lambda}{\sqrt{q}} \left[a(\Phi) \pi_{\Phi} + \frac{4\alpha}{D-1} \Phi P\right] \text{,} \\
	(\mathcal{L}_n q)_{ab}  &=& \frac{2 \kappa}{\sqrt{q}~ a(\Phi)} P_{ab}^{\text{tf}} + \frac{2\kappa}{\sqrt{q}D(1-D)} q_{cd} \left[P + \frac{1-D}{4} \pi_{\Phi} \Phi \right] \text{,} \label{eq:SolvedForMetric} \\
	\dot{\Phi} &=& N (\mathcal{L}_n \Phi) - (\mathcal{L}_N \Phi) \text{,} \\
	\dot{q}_{ab} &=&N (\mathcal{L}_n q)_{ab} - (\mathcal{L}_N q)_{ab} \text{.}
\ee
To complete the split of the action, it is useful to note that
\be
	P^{ab} (\mathcal{L}_n q)_{ab} + \pi_{\Phi} (\mathcal{L}_n \Phi) = \frac{2 \kappa}{\sqrt{q} a(\Phi)} P^{\text{tf}}_{ab} P^{ab}_{\text{tf}} - \frac{\kappa}{2 \sqrt{q}D(D-1)} \left[\Delta^g P + \Delta^{\Phi} \pi_{\Phi} \Phi\right]^2 - \frac{\lambda}{\sqrt{q}} \pi_{\Phi}^2 \text{.}
\ee
Substituting back into the action, we obtain
\be
	S &=& \frac{1}{\kappa} \int_{\mathbb{R}} dt \int_{\sigma} d^Dx ~ \left\{ N \left[a(\Phi) \sqrt{q} R^{(D)}  + \frac{\kappa}{2\lambda} \sqrt{q} \left( \frac{1}{D} q^{ab} \Phi_{,a} \Phi_{,b} - \frac{(D-1)}{D} \Phi \Delta \Phi\right)\right] \right. \nonumber \\ 
	& &\left. \hspace{2.5cm}+~ \frac{1}{2} \left[M^{ab,cd} (\mathcal{L}_n q)_{cd} + M^{ab} (\mathcal{L}_n) \Phi \right] (\mathcal{L}_n q)_{ab} \right. \nonumber \\
	& &\left. \hspace{2.5cm}+~ \frac{1}{2} \left[ M (\mathcal{L}_n \Phi) + M^{ab} (\mathcal{L}_n q)_{ab}\right](\mathcal{L}_n \Phi) \right\} \nonumber \\ 
	&=&\int_{\mathbb{R}} dt \int_{\sigma} d^Dx ~ \left\{ \frac{N}{\kappa} \left[a(\Phi) \sqrt{q} R^{(D)}  + \frac{\kappa}{2\lambda} \sqrt{q} \left( \frac{1}{D} q^{ab} \Phi_{,a} \Phi_{,b} - \frac{(D-1)}{D} \Phi \Delta \Phi\right)\right] \right. \nonumber \\ 
	& &\left. \hspace{2.5cm}+~ \frac{N}{2} P^{ab} (\mathcal{L}_n q)_{ab} + \frac{N}{2} \pi_{\Phi} (\mathcal{L}_n \Phi) \right\} \nonumber \\ 
	&=&\int_{\mathbb{R}} dt \int_{\sigma} d^Dx ~ \left\{ N \left[\frac{1}{\kappa} a(\Phi) \sqrt{q} R^{(D)} + \frac{1}{2\lambda} \sqrt{q} \left( \frac{1}{D} q^{ab} \Phi_{,a} \Phi_{,b} - \frac{(D-1)}{D} \Phi \Delta \Phi\right)\right. \right. \nonumber \\ 
	&&\left. \left. \hspace{3.5cm}-~ \frac{1}{2} P^{ab} (\mathcal{L}_n q)_{ab}- \frac{1}{2} \pi_{\Phi} (\mathcal{L}_n \Phi) \right] \right. \nonumber \\ 
	& &\left. \hspace{2.5cm}+~ P^{ab} \dot{q}_{ab} + \pi_{\Phi} \dot{\Phi} - P^{ab} (\mathcal{L}_N q)_{ab} - \pi_{\Phi} (\mathcal{L}_N \Phi) \right\} \nonumber \\ 
	&=& \int_{\mathbb{R}} dt \int_{\sigma} d^Dx \left[P^{ab} \dot{q}_{ab} + \pi_{\Phi} \dot{\Phi} - N^a \mathcal{H}_a - N \mathcal{H} \right] \text{,}
\ee
where 
\be
	\mathcal{H}_a[N^a] &:=& \int_{\sigma} d^Dx N^a \mathcal{H}_a = \int_{\sigma} d^Dx \left[ P^{ab} (\mathcal{L}_N q)_{ab} + \pi_{\Phi} (\mathcal{L}_N \Phi) \right]\text{,} \label{eq:Ha} \\
	\mathcal{H}[N] &:=& \int_{\sigma} d^Dx N \mathcal{H} =  \int_{\sigma} d^Dx N \left[ \mathcal{H}_{\text{Grav}} + \mathcal{H}_{\Phi} - \frac{\kappa}{(\Delta^g)^2 D(D-1)\sqrt{q}} \mathcal{D}^2 \right]\text{,}  \label{eq:Hparts}\\
	\kappa \mathcal{H}_{\text{Grav}} &:=&  \frac{\kappa^2}{\sqrt{q} a(\Phi)} P^{\text{tf}}_{ab} P_{\text{tf}}^{ab} - \sqrt{q} R^{(D)} \text{,}\\
	\kappa \mathcal{H}_{\Phi} &:=& \frac{\kappa}{2\lambda} \sqrt{q} \left[ - \frac{\lambda^2}{q} \pi_{\Phi}^2 - \frac{1}{D} q^{ab} (D_a \Phi) (D_b \Phi) + \frac{D-1}{D} \Phi D_a D^a \Phi + \frac{1}{D} \frac{\Delta^{\Phi}}{\Delta^g} R^{(D)} \Phi^2\right]\text{,}~~~~~~~\\
	\mathcal{D} &:=& \Delta^g P + \Delta^{\Phi} \pi_{\Phi} \Phi\text{.}
\ee
Note that the matter part $\mathcal{H}_{\Phi}$ of the Hamiltonian constraint is, when partially integrating the second to last term and dropping the last term, the same as the matter part of a minimally coupled scalar field up to a term $\propto \partial N$. The gravitational contribution $\mathcal{H}_{\text{Grav}}$ is very similar to the standard ADM terms, up to $a(\Phi)$ in the denominator of the first term and up to the terms containing the trace part $P:=P^{ab}q_{ab}$ of the ADM momentum. The trace terms now are defined to contribute to $\mathcal{D}$, which obviously is the generator of conformal transformations with correct conformal weights and will again play a central role when gauge fixing $\mathcal{H}[N]$ in the next section. The constraints satisfy the standard hypersurface deformation algebra (\ref{eq:HDA}), which can be verified by direct calculation (cf. appendix \ref{app:3}). Interestingly, a short calculation shows that the generator of conformal transformations $\mathcal{D}$ again corresponds to the trace of the extrinsic curvature,
\begin{eqnarray}
	\mathcal{D} &=& \Delta^g P + \Delta^{\Phi} \pi_{\Phi} \Phi \nonumber \\
			   &=& \Delta^g \left(- \frac{D-1}{\kappa} a(\phi) \sqrt{q} K + \frac{2 D \alpha}{\kappa} \sqrt{q} \Phi (\mathcal{L}_n \Phi) \right) + \Delta^\Phi \left( - \frac{1}{\lambda} \sqrt{q} (\mathcal{L}_n \Phi) \Phi+ \frac{4 \alpha}{\kappa} \sqrt{q} \Phi^2 K \right) \nonumber \\
			   &=& -\frac{(D-1)\Delta^g}{\kappa} \sqrt{q} K \text{,}
\end{eqnarray}
where in the second line, we used (\ref{eq:MetricMomenta}, \ref{eq:ScalarFieldMomentum}).

\subsection{Gauge fixing and unfixing}
The content of this section strongly resembles the exposure in sections \ref{sec:GF} and \ref{sec:GU:SD}. We will first introduce a suitable gauge fixing condition for the Hamiltonian constraint, discuss its quality and in the end gauge unfix the theory by dropping $\mathcal{H}$. \\ 
\\
In analogy to the ADM case, the partial gauge fixing condition we want to introduce is the CMC condition
\be
	\mathcal{D}_{\delta} = \mathcal{D} - \sqrt{q} \delta
\ee
for some constant $\delta \in \mathbb{R}$. It is convenient to split the calculation of $\{\mathcal{H}[N], \mathcal{D}_{\delta}[\rho]\}$ into several parts. First, we calculate how the individual parts of $\mathcal{H}$ displayed in (\ref{eq:Hparts}) transform under conformal transformations generated by $\mathcal{D}$. We find that the matter part of $\mathcal{H}$ transforms with the negative conformal weight of the lapse function as
\be
\{\mathcal{H}_{\Phi}[N], \mathcal{D}[\rho]\} = \mathcal{H}_{\Phi}[- \frac{1}{2} \Delta^g \rho N] \text{,}
\label{eq:TrafoScalarFieldHamiltonian}
\ee
which, as we already argued, is due to the conformal invariance of $S_{\Phi}$. For the gravitational part $\mathcal{H}_{\text{Grav}}$, we find
\be
\{\kappa \mathcal{H}_{\text{Grav}}[N], \mathcal{D}[\rho]\} &=& \int_{\sigma} d^Dx  ~ \rho \left\{  \frac{\kappa^2}{\sqrt{q} a(\Phi)} P^{\text{tf}}_{ab} P_{\text{tf}}^{ab} \left[-\frac{D}{2}\Delta^g + \frac{1}{a(\Phi)} \frac{\kappa}{D \lambda}\frac{(\Delta^{\Phi})^2}{\Delta^g} \Phi^2 \right] \right. \nonumber \\ 
	& &\hspace{30mm} \left. - \sqrt{q} \Delta^g \left[\frac{D-2}{2} R^{(D)} - (D-1) D_a D^a \right] \right\} N \nonumber \\
	&=& \int_{\sigma} d^Dx  ~ \rho \left\{  \frac{\kappa^2}{\sqrt{q} a(\Phi)} P^{\text{tf}}_{ab} P_{\text{tf}}^{ab} \left[-\frac{D-1}{2}\Delta^g + \frac{1}{a(\Phi)} \frac{\kappa}{D \lambda}\frac{(\Delta^{\Phi})^2}{\Delta^g} \Phi^2 \right] \right. \nonumber \\ 
	& &\hspace{30mm} \left. - \sqrt{q} \Delta^g \left[\frac{D-1}{2} R^{(D)} - (D-1) D_a D^a \right] \right\} N \nonumber \\
	& & + ~\kappa \mathcal{H}_{\text{Grav}}[-\frac{1}{2} \Delta^g \rho N] \nonumber \\
		&=&\int_{\sigma} d^Dx  ~ (D-1) \sqrt{q} \rho \Delta^g \left[D_a D^a - \frac{\kappa^2}{2 q a(\Phi)^2} P^{\text{tf}}_{ab} P_{\text{tf}}^{ab} - \frac{1}{2} R^{(D)}  \right] N \nonumber \\
	 	& & + ~  \kappa \mathcal{H}_{\text{Grav}}[-\frac{1}{2} \Delta^g \rho N]  \text{,} \label{eq:HGravD}
\ee
and for the total Hamiltonian constraint
\be
\{\kappa \mathcal{H}[N], \mathcal{D}[\rho]\} &=&\kappa \mathcal{H}[-\frac{1}{2}\Delta^g \rho N] + \mathcal{D}[\frac{\kappa^2 N \rho}{2 \Delta^g D\sqrt{q}}\mathcal{D}]  \\ 
& & +\int_{\sigma}d^Dx ~ (D-1) \sqrt{q} \rho \Delta^g \left[D_a D^a - \frac{\kappa^2}{2q a(\Phi)^2} P^{\text{tf}}_{ab} P_{\text{tf}}^{ab} - \frac{1}{2} R^{(D)}  \right] N \nonumber  \text{.}
\ee
This result already strongly resembles (\ref{eq:DiracMatrix}, \ref{eq:Pvac}) for the case $\delta=0$. Now let us include $\delta \neq 0$ also in the matter coupled case. To this end, note that both, $\{\mathcal{H}_{\text{Grav}}[N], \sqrt{q}\} = \{\mathcal{H}_{\Phi}[N], \sqrt{q}\} = 0$. Therefore, like in the standard ADM case we find
\be
\{\kappa \mathcal{H}[N], \int_{\sigma} d^Dx ~ \rho \sqrt{q} \delta \} = \frac{\delta \kappa^2}{\Delta^g (D-1)}  \mathcal{D}[N \rho] \text{.}
\ee
Combining the above results, we obtain
\be
	\{\kappa \mathcal{H}[N], \mathcal{D}_{\delta}[\rho]\} &=& \kappa \mathcal{H}\left[-\frac{1}{2}\Delta^g N \rho\right] +\mathcal{D}_{\delta}\left[\frac{\kappa^2 N \rho}{\Delta^g}\left(\frac{1}{2 D\sqrt{q}}\mathcal{D}_{\delta} - \frac{\delta}{D(D-1)}\right)\right] \nonumber \\ & & + \int_{\sigma} d^Dx~ (D-1)\sqrt{q}\Delta^g \rho \left[ D_a D^a - \mathcal{P}(x)\right] N \text{,}
\ee
with
\be
	\mathcal{P}(x) := \frac{\kappa^2}{2qa(\Phi)^2}P^{ab}_{\text{tf}} P_{ab}^{\text{tf}} + \frac{1}{2} R^{(D)} + \delta^2 \frac{\kappa^2(D+1)}{2(\Delta^g)^2D(D-1)^2} \text{,}
\ee
which coincides with the vacuum result (\ref{eq:Pvac}) up to the factor of $a(\Phi)^2$ appearing in the denominator of the first term in $\mathcal{P}(x)$.\\ 
\\
Again, the constraint $\mathcal{D}_{\delta}$ locally is a good gauge fixing condition if and only if the above elliptic PDE for $N$,
\be
	[D_a D^a - \mathcal{P}(x)] N(x) = 0\text{,}
\ee
has the unique solution $N(x) = 0$, because of the (formal) self-adjointness of $D^a D_a$ (cf. \cite{ChoquetBruhatGeneralRelativityAnd}, Appendix II, Theorem 2.3, p.549). A sufficient condition for this to hold is that $\mathcal{P}(x) > 0$, at least in physically interesting regions of the phase space. 
In the matter free case, it was possible, using $\mathcal{H}$ and $\mathcal{D}_{\delta}$, to free $\mathcal{P}(x)$ completely from the dependence on $R^{(D)}$. In the matter coupled case, we have to proceed differently, since otherwise we would end up with indefinite scalar field terms appearing in $\mathcal{P}(x)$. The first and the last summand in $\mathcal{P}(x)$, being squares of real functions, obviously are non-negative and in the case $\delta \neq 0$, their sum is even strictly positive. Therefore, if $\delta \neq0$, we need to show that $R^{(D)}\geq 0$, and $R^{(D)}>0$ if $\delta = 0$. A lower bound for the second summand can be obtained using Einstein's equation,
\be
R^{(D)} + [K^2- K_{\mu \nu} K^{\mu \nu}] &=& q^{\mu \nu} q^{\rho \sigma} R^{(D+1)}_{\mu \rho \nu \sigma} = R^{(D+1)} + 2 R^{(D+1)}_{\mu \nu} n^{\mu} n^{\nu} \nonumber \\ &=& 2 G_{\mu \nu} n^{\mu} n^{\nu} = 16 \pi T_{\mu \nu} n^{\mu} n^{\nu} \text{.}
\ee
Demanding a usual energy condition, namely either the weak energy condition ($T_{\mu \nu} \zeta^{\mu} \zeta^{\nu} \geq 0$ for all future timelike vectors $\zeta$) in the case $\delta \neq 0$ or the dominant energy condition ($- T_{\mu}^{\nu} \zeta^{\mu}$ is a future causal vector for all future timelike vectors $\zeta$) in the case $\delta=0$, the right hand side of the above equation is either non-negative ($\delta \neq0$) or strictly greater than $0$ ($\delta=0$). \\
The conformally coupled scalar field is known to possibly violate all common variants of energy conditions, even the weak energy condition (and therefore necessarily the dominant condition). Nevertheless, it is also known that all experimentally observed forms of classical\footnote{A violation of the energy conditions can also be observed in the context of quantum fields on classical, possibly curved backgrounds (see \cite{EpsteinNonpositivityOfThe}), but we refrain from a discussion of this aspect since we are only concerned with classical objects at this stage.} matter fields satisfy at least the weak energy condition  (see, e.g. \cite{FordClassicalScalarFields} and references therein). Therefore, as there seem to be no results indicating that violations of the weak respectively dominant energy condition are generic, at least to the authors' knowledge, and as the physical status, apart from mathematical existence, of scalar field solutions violating all typical energy conditions is unsettled, we will restrict the solutions to those that satisfy the required assumptions. Additionally, we want to point out that it could be possible to extend our results to more general situations since the energy conditions are only utilised to give sufficient, not necessary, prerequisites to achieve the invertibility of the Dirac matrix. We conclude that
\be
	\frac{1}{2} R^{(D)}  \left\{\begin{array}{cl}  \geq \frac{1}{2} [K_{ab} K^{ab} - K^2] = \frac{\kappa^2}{2 q a(\Phi)^2}  P^{\text{tf}}_{ab} P_{\text{tf}}^{ab}  - \frac{\kappa^2}{2 (\Delta^g)^2 D(D-1)q} \mathcal{D}^2, & \mbox{if } \delta \neq 0 \\  > \frac{1}{2} [K_{ab} K^{ab} - K^2] = \frac{\kappa^2}{2 q a(\Phi)^2}  P^{\text{tf}}_{ab} P_{\text{tf}}^{ab}  - \frac{\kappa^2}{2 (\Delta^g)^2 D(D-1)q} \mathcal{D}^2, & \mbox{if } \delta = 0  \text{,} \end{array}\right.
\ee
which implies
\be
	\mathcal{P}(x) &=& \frac{\kappa^2}{2 q ~ a(\Phi)^2} P^{\text{tf}}_{ab} P_{\text{tf}}^{ab} + \frac{1}{2} R^{(D)} + \delta^2 \frac{\kappa^2(D+1)}{2(\Delta^g)^2D(D-1)^2} \nonumber \\ 
	&\geq& \frac{\kappa^2}{q~  a(\Phi)^2} P^{\text{tf}}_{ab} P_{\text{tf}}^{ab} - \frac{\kappa^2}{2 (\Delta^g)^2 D(D-1) q} \mathcal{D}^2 + \delta^2 \frac{\kappa^2(D+1)}{2(\Delta^g)^2D(D-1)^2} \nonumber \\ 
	&=& \frac{\kappa^2}{q~  a(\Phi)^2} P^{\text{tf}}_{ab} P_{\text{tf}}^{ab} - \frac{\kappa^2}{2 (\Delta^g)^2 D(D-1) q} \mathcal{D}_{\delta}\left(\mathcal{D}_{\delta} +2 \sqrt{q} \delta \right)  + \delta ^2 \frac{\kappa^2}{(\Delta^g)^2 D(D-1)^2} \nonumber \\ 
	&\approx&  \frac{\kappa^2}{q ~ a(\Phi)^2} P^{\text{tf}}_{ab} P_{\text{tf}}^{ab}  + \delta^2 \frac{\kappa^2}{(\Delta^g)^2 D(D-1)^2} > 0 
\ee
in the case $\delta \neq 0$ and
\be
	\mathcal{P}(x) &=& \frac{\kappa^2}{2 q ~ a(\Phi)^2} P^{\text{tf}}_{ab} P_{\text{tf}}^{ab} + \frac{1}{2} R^{(D)} \nonumber \\ 
	&>& \frac{\kappa^2}{q ~ a(\Phi)^2} P^{\text{tf}}_{ab} P_{\text{tf}}^{ab} - \frac{\kappa^2}{2 (\Delta^g)^2 D(D-1) q} \mathcal{D}^2 \nonumber \\ &\approx&  \frac{\kappa^2}{q~ a(\Phi)^2} P^{\text{tf}}_{ab} P_{\text{tf}}^{ab} \geq 0
\ee
for $\delta=0$. Therefore, under the above mentioned energy conditions, $\mathcal{D}_{\delta}$ locally is a good gauge fixing condition, at least in the spatially compact case (for the asymptotically flat case, cf. the discussion in section \ref{sec:AsympFlat}). As in the vacuum case, global aspects of the chosen gauge are left for further studies.\\
\\
The algebra of constraints in this case is the same as in (\ref{eq:SDA}), and gauge unfixing is straightforward and implies dropping of $\mathcal{H}$. Locally, $\mathcal{H}$ is again a good gauge fixing condition for $\mathcal{D}_{\delta}$, since the differential operator is self-adjoint. Globally, the situation is less clear than in the vacuum case. It is again possible to calculate the finite conformal transformations generated by $\mathcal{D}_{\delta}$ on $\mathcal{H}$. Using the fact that $\mathcal{H}_{\Phi}$ transforms with weight $-\frac{1}{2} \Delta^g$, the result can be read off from (\ref{eq:Global}). With $\Omega := e^{\frac{D-2}{4}\Delta^g \rho}$ and therefore $e^{-\frac{1}{2}\Delta^g \rho} = \Omega^{-\frac{2}{D-2}}$ and $e^{\Delta^{\Phi}\rho} = \Omega^{-\frac{D-1}{D-2}}$, we find
\be
	\alpha_{\rho}^{\mathcal{D}_{\delta}}(\kappa \mathcal{H}) &=& \Omega^{-\frac{2D}{D-2}} \frac{\kappa^2}{\sqrt{q} ~ a(\Omega^{-\frac{D-1}{D-2}} \Phi)} P^{ab}_{\text{tf}} P_{ab}^{\text{tf}}  - \Omega^2 \sqrt{q} \left[R^{(D)} - \frac{4(D-1)}{(D-2)} \Omega^{-1} D_a D^a \Omega \right] + \Omega^{-\frac{2}{D-2}} \mathcal{H}_{\Phi} \nonumber \\ & & - \frac{\kappa^2}{(\Delta^g)^2 D(D-1)} \left[\Omega^{- \frac{2D}{D-2}} \frac{1}{\sqrt{q}}\mathcal{D}_{\delta}^2 + 2\mathcal{D}_{\delta}\delta + \Omega^{\frac{2D}{D-2}} \sqrt{q}\delta^2 \right] \text{.}
\ee
For accessibility, we have to study existence of solutions to this equation on $S_{J(\mathcal{D}_{\delta})}$, while for uniqueness (up to large gauge transformations), it is sufficient to start directly from a point on the gauge cut $S_{S(\mathcal{D}_{\delta}, \mathcal{H})}$ and check if the unique positive solution to this equation is $\Omega = 1$. For simplicity, restrict again to $D=3$  
and multiply the above equation by $(\Omega \sqrt{q})^{-1}$ to obtain
\be
	0 &\approx_{S_{J(\mathcal{D}_{\delta})}}\hspace{2mm}& 8 D_a D^a \Omega - R^{(3)} \Omega + \frac{a(\Phi)^2}{a(\Omega^{-2} \Phi)} K_{ab}^{\text{tf}} K^{ab}_{\text{tf}} \Omega^{-7} + \frac{\kappa}{\sqrt{q}} \mathcal{H}_{\Phi}  \Omega^{-3} - \frac{\kappa^2 \delta^2}{(\Delta^g)^2 6} \Omega^5 \nonumber \\
	&\approx_{S_{J(\mathcal{D}_{\delta}, \mathcal{H})}}& 8 D_a D^a \Omega + \mathfrak{P}(\Omega)\text{,}
\ee
where
\be
	 \mathfrak{P}(\Omega) := \frac{\Omega^{-7} (1-\Omega^4)}{a(\Omega^{-2}\Phi)} \left[\Omega^4 a(\Omega^{-2}\Phi) \left(R^{(3)} + \frac{\kappa^2 \delta^2}{(\Delta^g)^2 6} (1+\Omega^4)\right) + a(\Phi)K_{ab}^{\text{tf}} K^{ab}_{\text{tf}} \right] \text{.}
\ee
Note that, unlike in the usual Lichnerowicz-York equation (\ref{eq:York}), $\mathfrak{P}$ in this case is non-polynomial and has a pole for positive $\Omega$ if $a(\Omega^{-2}\Phi) = 0$ (unless the last term in the square brackets vanishes), which corresponds to the conformal transformation which, at that point $(x)$, leads to a complete vanishing of the curvature terms in the action. Moreover, apart from the trivial root $\Omega =1$, it possibly has more real positive roots stemming from the term in square brackets. Due to the complicated form of $\mathfrak{P}$, the standard methods for elliptic PDEs of second order \cite{GilbargEllipticPartialDifferential} cannot be applied in a straightforward way. We leave the question on accessibility and existence of Gribov copies for further research. \\
\\
In the following, we will construct a connection formulation for the gauge unfixed system. As said above, we could equally well use the Dirac bracket associated with $\mathcal{H} =\mathcal{D}_{\delta}=0$, since we are going to construct a connection which Poisson commutes with $\mathcal{D}_{\delta}$.

\subsection{Non-compact spatial slices}

The generalisation to non-compact spatial slices works in the same way as in the matter-free case, i.e. as long as the spatial metric and its conjugate momentum satisfy the necessary fall-off behaviour at spatial infinity, which one has to check for the spacetime under consideration. In case of a different fall-off behaviour, e.g. induced by matter coupling \cite{HenneauxAsymptoticBevaiorAnd}, it might still be possible to define asymptotic Poincar\'e or anti-de Sitter charges and derive a, maybe less restrictive, fall-off behaviour for the lapse function which still leads to a vanishing of the respective boundary terms.  

\subsection{Connection formulation and solution to kinematical constraints}
The idea of how to obtain a connection formulation with nice transformation properties now is the following: First, we perform a canonical transformation to a new metric which is invariant under conformal transformations\footnote{In slight abuse of terminology, we will call a transformation generated by $\mathcal{D}_{\delta}$ also conformal. ``Conformally covariant" or ``conformally invariant" here is also meant w.r.t. $\mathcal{D}_{\delta}$}. Because of the non-trivial conformal weight of the scalar field, this can be done easily in the case at hand. Then, we can introduce a vielbein $E^{ai}$ related to the new metric and its conjugate variable $K_{ai}$ in the standard way, which, by construction, will also be invariant. Now we can straightforwardly either follow the Ashtekar-Barbero construction \cite{AshtekarNewVariablesFor, BarberoRealAshtekarVariables} or the one introduced in \cite{BTTI, BTTII} for $D \geq 2$ to obtain a connection formulation with nice transformation properties, namely, the connection and its conjugate momentum will be invariant under conformal transformations.

\subsubsection{Conformally covariant variables}
The first step consists in performing a canonical transformation to variables on which $\mathcal{D}_{\delta}$ acts as the generator of conformal transformations. This already holds true for all phase space variables except $P^{ab}$. It is easy to verify that
\be
	\{q_{ab}, P^{cd}\} \rightarrow \{ q^{\delta}_{ab} := q_{ab}, ~P^{cd}_{\delta} := P^{cd} + \frac{\delta}{\Delta^g D} \sqrt{q} q^{cd}\}
\ee
is a canonical transformation and $P^{cd}_{\delta}$ transforms as desired under the action of $\mathcal{D}_{\delta}$, which, in terms of the new variables, reads
\be
	\mathcal{D}_{\delta} := \Delta^g P_{\delta} + \Delta^{\Phi} \pi_{\Phi} \Phi \text{.}
\ee
Slightly abusing the notation, we will drop the index $\delta$ on $P^{ab}_{\delta}$, $q^{\delta}_{ab}$ and $\mathcal{D}_{\delta}$ in what follows.

\subsubsection{Dilaton - type field}
\label{sec:dilaton}
The obvious procedure to render the metric and its conjugate momentum conformally invariant is to multiply the metric by $\Phi^{\frac{4}{D-1}}$ and divide the momentum conjugate to $q_{ab}$ by the same power of $\Phi$. This, of course, is not well-defined unless $\Phi$ is strictly positive (negative). Therefore, we will restrict to $\Phi>0$ in all what follows. On the resulting open region in the phase space, we can perform the following canonical transformation
\be
	\label{eq:CanonicalTrafo1}
	\{\Phi, \pi_{\Phi} \} \rightarrow \{ \phi := \ln \Phi, \pi_{\phi} := \Phi \pi_{\Phi} \} \text{,}
\ee
to the convenient scalar field variable $\phi \in \mathbb{R}$ which now is unrestricted. To interpret this restriction on the original field $\Phi$, note that if we had started by coupling general relativity conformally to a dilaton field, i.e. replacing $\Phi$ in the action (\ref{eq:CombinedAction}) by $e^{\phi}$ (and thereby restricting to $\Phi>0$ from the beginning), the canonical analysis of the system would have yielded exactly the same result that we obtained by performing the above canonical transformation (\ref{eq:CanonicalTrafo1}). Therefore, the interpretation of $\phi$ will be that of a dilaton field in what follows.
In terms of the new scalar field variable $\phi$, the constraints read
\be
	\mathcal{H}_a[N^a] &=& \int_{\sigma} d^Dx ~ N \left(P^{ab} (\mathcal{L}_N q)_{ab} + \pi_{\phi} (\mathcal{L}_N \phi) \right) \text{,}\\
	\mathcal{D}[\rho] &=& \int_{\sigma} d^Dx~ \rho \left( \Delta^g P + \Delta^{\Phi} \pi_{\phi} \right)\text{.}
\ee
Of course, another interesting, possible interpretation of the scalar field $\Phi$ would be that of the Higgs field. However, $\Phi$ should not transform under some internal group to allow for the construction of invariant metric variables $\tilde{q}_{ab}$, $\tilde{P}^{ab}$. Therefore, a component of the su(2) Higgs fields cannot be used for that purpose, but an interpretation of $\Phi$ as the weak isospin singlet constructed from the Higgs field is possible. However, before jumping to conclusions, the experimental consequences of an identification of $\Phi$ with the Higgs field have to be investigated, e.g. is the proposed quantisation compatible with standard model calculations in which the Higgs appears in loop corrections to scattering amplitudes? A possible caveat is that the Higgs field would not be quantised as a usual scalar field, but as a part of the $3$-metric $q_{ab}$, which makes it very non-trivial to check for consistency with the standard model, see also the discussion in \cite{ThiemannReducedPhaseSpace}.

\subsubsection{Conformally invariant metric and connection variables}
\label{sec:ConformallyInvariantMetricAndConnectionVariables}

The above described transformation of rescaling the metric and its momentum by some power of the scalar field $\Phi$ is, of course, not canonical since the new metric and momentum will not Poisson commute with $\pi_{\Phi}$. But if we additionally change $\pi_{\Phi}$, we can render the transformation canonical. In terms of the dilaton variables,
\be
	\{q_{ab}, P^{ab}, \phi, \pi_{\phi}\} \rightarrow \{\tilde{q}_{ab} := e^{\frac{4}{D-1}\phi} q_{ab}, ~\tilde{P}^{ab} := e^{- \frac{4}{D-1}\phi} P^{ab}, ~\tilde{\phi} := \phi, ~\tilde{\pi}_{\tilde{\phi}} := \frac{1}{\Delta^{\Phi}} \mathcal{D} \}
\ee
can be easily verified to be a canonical transformation. Moreover, $\mathcal{D}$ now appears as one of the canonical variables, i.e. it generates translations on its conjugate field $\phi$ while all other fields are invariant, and can be easily solved classically by dropping the scalar field variables.\footnote{Note that we could as well quantise the system including the scalar field along the lines of \cite{ThiemannKinematicalHilbertSpaces, ThiemannQSD5} and solve the constraint $\mathcal{D}$ after quantisation. In this simple case, it is easy to show that, like for the Gau{\ss} constraint in usual LQG, quantisation and solving the constraint actually do commute.}\\ 
\\
The freely specifiable and independent initial data now is a Riemannian metric $\tilde{q}_{ab}$, a transversal, symmetric, not necessarily trace free tensor field $\tilde{P}^{ab}$ and an unphysical scalar field $\phi$. This should be compared to the case of (vacuum) general relativity, where we had a conformal equivalence class of Riemannian metrics and a transversal trace free symmetric second rank tensor field as freely specifiable initial data. Due to the canonical transformations applied, the conformal mode of the metric is combined with the scalar field in such a way that any metric becomes good initial data.\\
\\
To obtain a connection formulation, one now can either follow the original route of Ashtekar and Barbero \cite{AshtekarNewVariablesFor, BarberoRealAshtekarVariables} for $D=3$  or the higher dimensional ($D \geq 2$) variant introduced in \cite{BTTI, BTTII}. All of them can be straightforwardly applied to the theory at hand and result in a connection invariant under conformal transformations. Since, in view of the results of section \ref{sec:ExtensionToStandardModelMatterFields}, the focus of this paper is on $3+1$ dimensions, like in the vacuum case we will stick to the original Ashtekar-Barbero approach. The resulting phase space is coordinatised by a real SU$(2)$ connection $\m^{(\gamma)}A_{ai}$ and a densitised triad $E^{ai}$ conjugated to the connection, 
\be
	\{\m^{(\gamma)}A_{a}^i(x), E^{bj}(y)\} = \gamma \delta_a^b \delta^{ij} \delta^3(x,y) \text{,}
\ee
where $\gamma$ denotes the Barbero-Immirzi parameter, subject to a su$(2)$ Gau{\ss} law constraint
\be
	G^{i} := D_a(\m^{(\gamma)}A) E^{ai}
\ee
in order that its symplectic reduction leads back to the ADM type phase space we considered above. In terms of the connection variables, the remaining spatial diffeomorphism constraint reads
\be
	\mathcal{H}_a[N^a] = \int_{\sigma} d^3x E^{ai} (\mathcal{L}_N \m^{(\gamma)}A)_{ai}  \text{.}
\ee
One more remark is in order: Note that the variables $ \{\tilde{q}_{ab}, \tilde{P}^{ab}, \tilde{\phi}, \tilde{\pi}_{\tilde{\phi}} \}$ would also be good coordinates in the original theory, i.e. before gauge unfixing, since up to this point we only performed canonical transformations. Using the observable projector, we could have constructed the $\mathcal{H}$ observables $\mathbb{P}^{\mathcal{D}_{\delta}}_{\mathcal{H}}(\tilde{q}_{ab}), \mathbb{P}^{\mathcal{D}_{\delta}}_{\mathcal{H}}(\tilde{P}^{cd})$ with Poisson brackets
\be\label{eq:diracobsalg}
	\{ \mathbb{P}^{\mathcal{D}_{\delta}}_{\mathcal{H}}(\tilde{q}_{ab}), \mathbb{P}^{\mathcal{D}_{\delta}}_{\mathcal{H}}(\tilde{P}^{cd})\} &\approx& \mathbb{P}^{\mathcal{D}_{\delta}}_{\mathcal{H}}(\{ \tilde{q}_{ab}, \tilde{P}^{cd} \}_{\text{DB}(\mathcal{D}_{\delta},\mathcal{H})}) \nonumber \\ 
	&=& \mathbb{P}^{\mathcal{D}_{\delta}}_{\mathcal{H}}(\{ \tilde{q}_{ab}, \tilde{P}^{cd} \})  \nonumber \\ 
	&=&  \mathbb{P}^{\mathcal{D}_{\delta}}_{\mathcal{H}}(\delta_{(a}^c \delta_{b)}^d) \nonumber \\ 
	& =& \delta_{(a}^c \delta_{b)}^d \text{,}
\ee
where in the first step, we used that $\mathbb{P}^{\mathcal{D}_{\delta}}_{\mathcal{H}}$ defines a weak Poisson homomorphism (w.r.t. the Dirac bracket), and in the second that the Dirac bracket and the Poisson bracket weakly coincide for observables w.r.t. $D_{\delta}$, which holds even strongly in our case. Similarly, we find $\{ \mathbb{P}^{\mathcal{D}_{\delta}}_{\mathcal{H}}(\tilde{q}_{ab}), \mathbb{P}^{\mathcal{D}_{\delta}}_{\mathcal{H}}(\tilde{q}_{cd})\} \approx 0 \approx \{ \mathbb{P}^{\mathcal{D}_{\delta}}_{\mathcal{H}}(\tilde{P}^{ab}), \mathbb{P}^{\mathcal{D}_{\delta}}_{\mathcal{H}}(\tilde{P}^{cd})\}$, i.e. simple brackets. The projections of $\tilde{\phi},\ \tilde{\pi}_{\tilde{\phi}}$ can be ignored since $\tilde{q}_{ab}.\ \tilde{P}^{ab}$ already constitute a set of phase space functions that is projected onto $\mathcal{O}^{w}_{J\left(\mathcal{D}\right)}$ (cf. proposition \ref{prop:comptaylorprop}).\\
This treatment would be more alike the one in deparametrised models, but the algebra obtained coincides with the one we found via gauge unfixing. This opens the possibility of interpreting the quantum theory which we describe in the following section also from this point of view. However, in contrast to the deparametrised models, the expressions for $\mathbb{P}^{\mathcal{D}_{\delta}}_{\mathcal{H}}(\tilde{q}_{ab}), \mathbb{P}^{\mathcal{D}_{\delta}}_{\mathcal{H}}(\tilde{P}^{cd})$ are problematic, since $\mathcal{H}$ is not first class with itself. Therefore, the observable projectors $\mathbb{P}_{\mathcal{H}}^{\mathcal{D}_{\delta}}$ can only be constructed explicitly using the methods described in section \ref{sec:PartialComplete}, if we additionally fix the diffeomorphism constraints (complete gauge fixing as e.g. in \cite{GieselAQG4}). Still, if we are interested in a reduced phase space quantisation, we will not need explicit knowledge of the observable projector, but only of the algebra of Dirac observables given by (\ref{eq:diracobsalg}).

\subsection{Remarks on quantisation and observables}
\label{sec:RemarkOnQuantisationAndObservables}

The above system can straightforwardly be quantised using loop quantum gravity methods (cf. e.g. \cite{RovelliQuantumGravity, ThiemannModernCanonicalQuantum}). In particular, one can represent and solve the Gau{\ss} and spatial diffeomorphism constraints. This means that, like in the deparametrised models, we have direct access to the {\it{physical}} Hilbert space of the model. However, we have to be careful in how we interpret the elements of this Hilbert space as well as operators acting on it. Classically, the gauge fixing $\mathcal{D}_\delta=0$ corresponds to fixing a certain CMC spatial slice, or, in other words, looking at the universe at the ``time'' $\delta \propto K$. The missing of the Hamiltonian constraint can thus be attributed to the fact that we are just not considering time evolution in our framework. Still, simply neglecting the Hamiltonian constraint is not a valid option: Due to the gauge fixing $\mathcal{D}_{\delta}=0$, we have to employ a Dirac bracket which only equals the Poisson bracket for phase space functions which Poisson commute with $\mathcal{D}_\delta$ (or with $\mathcal{H}$, however, these functions are not known explicitly). Since the derivation of the Ashtekar-Barbero-type variables heavily relies on the exact form of the ADM Poisson bracket, we have to restrict to phase space functions which have a similar Dirac bracket.  Counting degrees of freedom, one directly sees that considering the functions Poisson commuting with either $\mathcal{D}_\delta$ or $\mathcal{H}$ is already enough to coordinatise the reduced phase space. On the other hand, gauge unfixing the second class system $\mathcal{D}_\delta=\mathcal{H}=0$ and considering Dirac observables in those systems yields the same result. \\
The result of the above discussion may appear a bit puzzling at this point, since the quantisation problem seems to have been trivialized by choosing a certain gauge at the classical level. From the discussion in section \ref{sec:PartialComplete}, we know that the observable algebras $\mathcal{O}^w_\mathcal{H}$ and $\mathcal{O}^w_{\mathcal{D}_\delta}$ are indeed isomorphic, meaning that they both coordinatise the reduced (w.r.t. $\mathcal{D}_\delta, \: \mathcal{H}$) phase space. On the other hand, finding a representation of the reduced phase space functions according to their Poisson structure (i.e. the Dirac bracket for the second class system  $\mathcal{D}_\delta=\mathcal{H}=0$) is the very definition of a (reduced phase space) quantisation. However, as also remarked in section \ref{sec:PartialComplete}, the classical Poisson isomorphism between $\mathcal{O}^w_\mathcal{H}$ and $\mathcal{O}^w_{\mathcal{D}_\delta}$ ceases to have a quantum analogue for general situations. Thus, both, the presented quantisation and a hypothetical reduced phase space quantisation using phase space functions Poisson commuting with $\mathcal{H}$, are genuine reduced phase space quantisations, however, the Poisson isomorphism between $\mathcal{O}^w_\mathcal{H}$ and $\mathcal{O}^w_{\mathcal{D}_\delta}$ does not have to be implementable as a unitary transformation, or even an algebra automorphism, at the quantum level. Therefore, the results of the two quantum theories can be different and, as always, an experiment has to decide which quantisation is the correct one. In this sense, we cannot claim to have solved the problem of quantum dynamics, especially since it is even more unclear\footnote{E.g. in the view of Haag's theorem (\cite{HaagOnQuantumField, HaagLocalQuantumPhysics}).} if we can relate the proposed quantisation to a Dirac type quantisation which implements the Hamiltonian constraint on the quantum level. \\
In order to implement a time evolution in our framework and thus making contact with other reduced phase space quantisations, one would have to use an undensitised version of $\mathcal{D}$ as a time variable\footnote{Here, we would have to assume the existence of a CMC-time function generalised to the conformally coupled scalar field, meaning that, at least the region of spacetime which we are interested in should foliate into $\mathcal{D}_{\delta=t}=0$ hypersurfaces, where $t$ is a monotonically increasing function along all future directed causal curves.} and calculate the physical Hamiltonian which evolves the spatial slice $\mathcal{D}_{\delta=t_1}=0$ to some later time $\mathcal{D}_{\delta=t_2}=0$. However, the explicit construction of this Hamiltonian is very difficult already at the classical level, since one has to invert the partial differential operator in (\ref{eq:DiracMatrix}). Also, an exact quantisation of this Hamiltonian seems hopeless. We are thus led to prefer the other reduced phase space quantisations \cite{GieselAQG4, DomagalaGravityQuantizedLoop, HusainTimeAndA} in order to implement a time evolution and continue to investigate the properties of our formulation as a quantisation of general relativity conformally coupled to a scalar field on a fixed spatial slice. Since, up to the spatial diffeomorphism constraint\footnote{The spatial diffeomorphism constraint has to be solved on the quantum level, see \cite{ThiemannModernCanonicalQuantum} for an exposition of the current state of the art. While the construction of a Hilbert space of diffeomorphism invariant states (more precisely, spatially diffeomorphism invariant distributions in the dual of the kinematical Hilbert space) is well understood, the construction of spatially diffeomorphism invariant operators is more subtle. In principle, one could solve the constraint by using matter fields as physical coordinate system as in \cite{GieselAQG4}.}, there are no constraints left when restricting to $\mathcal{D}_{\delta}=\mathcal{H}=0$ and passing to the Dirac bracket, the interpretation of the $\mathcal{D}_{\delta}$-invariant observables $\mathcal{O}^w_{\mathcal{D}_{\delta}}$ is that of allowed data on the spatial slice, or in a spacetime context, the allowed initial data which can be specified on the spatial slice and then be evolved to a complete spacetime using the Hamiltonian constraint. After quantisation, the spectral theory of the operator representation derived from  $\mathcal{O}^w_{\mathcal{D}_{\delta}}$ tells us what the possible measurement outcomes for the phase space functions  $\mathcal{O}^w_{\mathcal{D}_{\delta}}$ are. Therefore, one can directly extract a prediction from this theory: Measuring $f \in \mathcal{O}^w_{\mathcal{D}_{\delta}}$ by an observer living on the spatial slice $\mathcal{D}_{\delta}=0$ (or  $\mathcal{D}_{\delta=t})$, the possible measurement values are determined by the spectrum of the operator $\hat{f}$ corresponding to $f$.  \\
One of the main differences to other (deparametrised) reduced phase space quantisations of general relativity coupled to suitable matter is the form of our clock $\mathcal{D}$. While in other approaches, one uses configuration space variables as clocks and rods, we use a combination of momenta as a clock. Especially for static spacetimes, which are of great interest in black hole physics, accessibility of the gauge $\mathcal{D}=0$ immediately follows when using a $t=\text{const.}$ foliation. The CMC gauge and its global properties have been subject of intensive studies (cf. e.g. \cite{RendallConstantMeanCurvature} and references therein). 
An application of the present framework to black hole entropy calculations, which exemplifies the accessibility properties of the gauge $\mathcal{D}=0$ for static situations, has been sketched in our companion paper \cite{BSTI}. Also, when interpreting Dirac observables $\mathcal{O}^w_\mathcal{D}$, e.g. the invariant  metric and its conjugate momentum, we have to keep in mind that the invariant metric is a product of the spatial metric and a suitable power of the scalar field. Thus, the invariant geometric operators constructed in \cite{BSTI} do not predict a discrete spectrum for the area and volume, but a discrete spectrum for the products of area and volume with a suitable power of the scalar field. \\
A last remark concerns gauge unfixing and how to recover the original ADM formulation from the theory which has first been gauge fixed to the second class system $\mathcal{D}_{\delta}=\mathcal{H}=0$ and then gauge unfixed to the first class system $\mathcal{D}_{\delta}=0$. At first sight, it seems false to simply drop the Hamilton constraint in the second class system $(\mathcal{H}, \mathcal{D}_{\delta})$ and work with $\mathcal{D}_{\delta}$. Of course, from the point of view of the first class $\mathcal{D}_{\delta}$ system, all gauge choices are (up to technical details) equivalent, and therefore we can forget about the special choice $\mathcal{H} = 0$. However, we did not start from the first class $\mathcal{D}_{\delta}$ system, but from the ADM formulation and therefore there definitely is a special role of the constraint $\mathcal{H}$, dropping it feels like forgetting about information that is forced upon us by the theory we are considering. The point is that the first class theory obviously reduces to the second class system in the gauge $\mathcal{H} = 0$, and that any other gauge fixing condition for $\mathcal{D}_{\delta}$ is related to this gauge choice by a gauge transformation, and the observable algebras for different gauge fixings, constructed via the corresponding observable projectors, are isomorphic, as was discussed in section \ref{sec:PartialComplete}. \\
Moreover, in the case at hand, there actually is a guiding principle which leads us back to ADM, and in this sense a preferred gauge choice for the constraint $\mathcal{D}_{\delta}$: Remember that the Hamiltonian constraint has a clear geometric interpretation in the spacetime picture: It generates time translations. Furthermore, since we still have the spatial diffeomorphism constraint, we can ask what properties we can demand from a gauge fixing condition for $\mathcal{D}_{\delta}$ in order to interpret it as generating time translations. The answer to this question, which is that the constraint algebra should be identical to the hypersurface deformation algebra, has been investigated by Hojman, Kucha{\v r} and Teitelboim \cite{HojmanGeometrodynamicsRegained}, with the result, that, for the ADM phase space varibles $q_{ab}$ and $P^{ab}$, the only constraints satisfying the hypersurface deformation algebra are the ADM constraints $\mathcal{H}$ and $\mathcal{H}_a$. While this settles the question for gravity (and also standard matter \cite{HojmanGeometrodynamicsRegained}), analogous results are, to the best of the authors knowledge, not available for derivative coupling of matter to general relativity like in the case considered here. Still, it transpires that the demand for a spacetime interpretation should restrict the allowed gauge choices also for that case. 

\section{Extension to standard model matter fields}
\label{sec:ExtensionToStandardModelMatterFields}
Finally, we will show that we can include ($D=3$) the complete matter content of the Standard Model into the framework presented in this work. The key observation is that the actions for Dirac fermions (in arbitrary dimension) and Yang-Mills fields (only for $D=3$) are conformally invariant for certain choices of conformal weights for the fields, and therefore can be treated on equal footing. The considerations in this section will be purely classical. A quantisation thereof can be done when using the methods developed in \cite{ThiemannKinematicalHilbertSpaces, ThiemannQSD5}. We will also comment on inclusion of a cosmological constant.
\subsection{Conformally invariant matter field actions}
\label{sec:ConformallyInvariantMatterFieldActions}
Consider an action $S[g, \partial g, \phi_I, \partial \phi_I]$ with $I$ running in some index range, which is conformally invariant if we specify the conformal weight of the fields $\phi_I$ relative to the conformal weight of the metric $g$, $\Delta^{\phi_I} = \beta^I \Delta^{g}$. After the $D+1$ split, the action takes the form
\be
	S = \int_{\mathbb{R}} dt \int_{\sigma}d^Dx~ \left(P^{ab} \dot{q}_{ab} + \pi_{\phi}^I \dot{\phi}_I - N \mathcal{H}_{\phi} - N^a \mathcal{H}_a - \lambda_i \mathcal{C}^i \right)\text{,}
\ee
where $\mathcal{C}^i$ are gauge constraints for the fields $\phi_I$ and $\lambda_i$ corresponding Lagrange multipliers. The conformal weights of the fields on the spatial slice $\sigma$ are inherited from the corresponding spacetime fields. $q_{ab}$ for example transforms with the same weight as $g_{\mu \nu}$ and it follows that $N$ has weight $\frac{1}{2} \Delta^g$. Note that under a conformal transformation, the kinetic terms and the three different constraint terms cannot mix, therefore each of them has to be invariant by itself. An immediate consequence is that the Hamiltonian constraint $\mathcal{H}_{\phi}$ has to transform with weight $-\frac{1}{2}\Delta^g$, which already has been confirmed for the conformally coupled scalar field (\ref{eq:TrafoScalarFieldHamiltonian}) and will be furthermore explicitly shown for Dirac fermions in appendix \ref{app:2}. Therefore, coupling the fields $\phi_I$ to (\ref{eq:CombinedAction}) has the following effects:\\
\\
1. The generator of conformal transformations becomes 
\be
\mathcal{D}^{\text{new}} = \mathcal{D}^{\text{old}} + \beta^I \Delta^g \pi_{\phi}^I \phi_I 
\ee
2. The form of the commutator of the gauge fixing condition $\mathcal{D}^{\text{new}}_{\delta}$with the new Hamiltonian constraint is unchanged
\be
\{\mathcal{H}^{\text{new}}[N], \mathcal{D}^{\text{new}}_{\delta}[\rho]\} = \mathcal{H}^{\text{new}}[-\frac{1}{2}\Delta^g \rho N] + \mathcal{D}^{\text{new}}_{\delta}[...] + \int_{\sigma} d^Dx ~ \sqrt{q} \frac{(D-1) \rho}{\kappa} \Delta^g [D^a D_a - \mathcal{P}^{\text{new}}] N
\ee
3. The form of the $\mathcal{P}^{\text{new}}(x)$ is similar
\be
	\mathcal{P}^{\text{new}}(x) = \frac{\kappa^2}{2 q} f(\Phi)^2 P^{\text{tf}}_{ab} P_{\text{tf}}^{ab} + \frac{1}{2} R^{(D)} + \delta^2 \frac{\kappa^2 (D+1)}{2(\Delta^g)^2 D(D-1)^2}
\ee
4. The positivity of $\mathcal{P}^{\text{new}}(x)$ can be shown in exactly the same way.\\
\\
It is a standard result that the free Dirac action (weight $\Delta^{\Psi} = -\frac{D}{4} \Delta^g$) 
\be
	S_{\Psi}[e, \Psi, \Psi^*] := \frac{i}{2} \int_M d^{D+1}X \sqrt{g} \left( \overline{\Psi} \,\slashed{e}^{\mu} \nabla_{\mu} \Psi -\overline{ \nabla_{\mu} \Psi}\, \slashed{e}^{\mu} \Psi \right) \text{,}
\ee
(see appendix \ref{app:2} for notation) as well as the free Yang-Mills action in $D=3$ (weight $\Delta^A = 0$) 
\be
	S_{A}[g, A] := - \frac{1}{4g^2_{\text{YM}}} \int_M d^{D+1}X \sqrt{g} g^{\mu \rho} g^{\nu \sigma} \text{Tr}(F_{\mu \nu} F_{\rho \sigma})\text{,}
\ee
where $g_{\text{YM}}$ denotes the Yang-Mills coupling and $F_{\mu \nu}$ is the field strength of $A_{\mu}$, are conformally invariant (cf., e.g., \cite{WaldGeneralRelativity}). Moreover, for loop quantisation it is necessary to work with fermionic spatial half densities at the Hamiltonian level \cite{ThiemannKinematicalHilbertSpaces}, which then transform with weight zero. Therefore, $\mathcal{D}$ is unchanged for these fields. As an example, the Dirac field is discussed in appendix \ref{app:2}. We leave the easier case of the Yang-Mills field to the interested reader. \\
\\
Finally, we also want to discuss minimal coupling of matter fields to the Yang-Mills gauge field. The corresponding coupling terms for Dirac fields ($\frac{i}{2} \sqrt{g} \overline{\Psi} \slashed{e}^{\mu} A_{\mu}^{IJ} \Sigma_{IJ} \Psi$ + c.c.) can easily be shown to be conformally invariant. For additional, conformally coupled scalar fields transforming in the adjoint of the Yang-Mills gauge group\footnote{These scalar fields could e.g. correspond to the Higgs sector of the standard model.}, the conformal invariance of the coupling term \mbox{$\sqrt{g} g^{\mu \nu} (\text{Tr}\{(\partial_{\mu} \Phi) g_{\text{YM}} [A_{\nu}, \Phi] + g^2_{\text{YM}} [A_{\mu}, \Phi][ A_{\nu}, \Phi]\})$} follows from trace properties.

\subsection{Scalar field potentials}
As just mentioned, we can, in addition to the already present dilaton field, also couple more scalar fields. In order to preserve the form of $\mathcal{P}(x)$ they have to be coupled conformally. For the standard model, we furthermore need to include potential terms for these scalar fields into the action. In this section, we will discuss restrictions on this potential needed in order that the estimation $\mathcal{P}(x) > 0$ still holds (weakly). In particular, we will see that Higgs potentials are allowed in dimensions $D\geq 3$. \\
\\
Adding to the action (\ref{eq:CombinedAction}) the term
\be
	+ \frac{1}{2\lambda} \int_{M}d^{D+1}X \sqrt{g} V(\Phi) \text{,}
\ee
for some potential $V(\Phi)$, the Hamiltonian constraint is modified to
\be
	\kappa \mathcal{H} \rightarrow \kappa \mathcal{H} - \frac{\kappa}{2\lambda} \sqrt{q} V(\Phi) \text{.}
\ee
Calculating the Poisson brackets with the conformal transformation generator $\mathcal{D}$, we conclude that $\mathcal{P}(x)$ obtains an extra contribution
\be
	\mathcal{P} \rightarrow \mathcal{P} + \frac{\kappa}{4 \lambda} \left[ \frac{D+1}{D-1} V(\Phi) - \frac{1}{2} \Phi V'(\Phi)\right] \text{.}
\ee
If the term in square brackets is negative, we still have $\mathcal{P}(x) > 0$ (remember that $\frac{\kappa}{\lambda} < 0$ in our conventions). In particular, for the Higgs potential
\be
	V(\Phi) = \tilde{\lambda} [(\Phi^2 -  v^2)^2 - v^4] = \tilde{\lambda} (\Phi^4 - 2 v^2 \Phi^2) 
\ee
with $\tilde{\lambda} \geq 0$ and $v$ the classical expectation value of the field\footnote{Note that the constant term $\tilde{\lambda} v^4$ sometimes included into the Higgs potential is subtracted here since it can be absorbed into the cosmological constant, which will be discussed in the next section.}, this bracket is given by
\be
	\left[ \frac{D+1}{D-1} V(\Phi) - \frac{1}{2} \Phi V'(\Phi)\right] = \frac{\tilde{\lambda}}{D-1} \Phi^2 \left[(3-D) \Phi^4 - 2v^2 \right] \leq 0 \text{ for } D\geq 3 \text{.}
\ee
\subsection{Cosmological constant}
The cosmological constant term\footnote{The constants in front of $S_{\Lambda}$ are chosen such that the equations of motion will read $G_{\mu \nu} + \Lambda g_{\mu \nu} = 8 \pi T_{\mu \nu}$.} 
\be
	S_{\Lambda} := -\frac{2}{\kappa} \int_M d^{D+1}X ~ \sqrt{g} \Lambda
\ee
leads to an additional term
\be
	\mathcal{H}_{\Lambda} :=  \frac{2}{\kappa} \sqrt{q} \Lambda
\ee
in the Hamilton constraint and changes $\mathcal{P}(x)$ to
\be
	\mathcal{P} &\rightarrow& \mathcal{P} - \frac{D+1}{D-1} \Lambda \nonumber \\
	 &=& \frac{\kappa^2}{2qa(\Phi)^2}P^{ab}_{\text{tf}} P_{ab}^{\text{tf}} + \frac{1}{2} R^{(D)} + \delta^2 \frac{\kappa^2(D+1)}{2(\Delta^g)^2D(D-1)^2} - \frac{D+1}{D-1} \Lambda  \text{.}
\ee
Demanding the weak energy condition, we find
\be
 R^{(D)} - \frac{\kappa^2}{ q a(\Phi)^2} P^{ab}_{\text{tf}} P_{ab}^{\text{tf}} + \frac{\kappa^2}{(\Delta^g)^2 D(D-1)q} \mathcal{D}^2 - 2 \Lambda \geq 0 \text{,}
\ee
and therefore
\be
	\mathcal{P} &\geq&  \frac{\kappa^2}{q a(\Phi)^2} P^{ab}_{\text{tf}} P_{ab}^{\text{tf}} + \delta^2 \frac{\kappa^2}{(\Delta^g)^2D(D-1)^2} - \frac{\kappa^2}{2 (\Delta^g)^2 D(D-1)q} (\mathcal{D}_{\delta}^2 + 2 \mathcal{D}_{\delta} \sqrt{q} \delta) - \frac{2}{D-1} \Lambda \nonumber \\
	 &\approx&  \frac{\kappa^2}{q a(\Phi)^2} P^{ab}_{\text{tf}} P_{ab}^{\text{tf}} + \delta^2 \frac{\kappa^2}{(\Delta^g)^2D(D-1)^2}- \frac{2}{D-1} \Lambda  \nonumber \\
	 &\geq& \delta^2 \frac{\kappa^2}{(\Delta^g)^2D(D-1)^2} - \frac{2}{D-1} \Lambda  \text{.}
\ee
A sufficient condition that (\ref{eq:Estimation}) still holds is therefore 
\be
	\delta^2 > \frac{2 (\Delta^g)^2 D(D-1)}{\kappa^2} \Lambda \text{,}
\ee
which is always satisfied for $\Lambda < 0$, and imposes a restriction on the allowed values of $\delta$ if $\Lambda > 0$.

\section{Concluding remarks}
\label{sec:ConcludingRemarks}

Let us finish this paper with some remarks. In the following, we will neglect the spatial diffeomorphism constraint, since it is treated at the quantum level and for the sake of readability, we will use the words observable and physical state with respect to the Hamiltonian constraint only.\\
The presented formalism exemplifies the canonical structure of general relativity and to what extend it can be modified. In the canonical treatment, it is important to distinguish between equations which determine good initial data and time evolution: $\mathcal{H}=0$ selects good initial data which can then be evolved, using the Hamiltonian flow of $\mathcal{H}$, to a complete spacetime. On the reduced phase space however, we do not see time evolution, since we have restricted ourselves to a CMC spatial slice, given by $\mathcal{D}=0$. On the other hand, the complete time evolution is already determined by a set of good initial data, and it is therefore classically equivalent just to work, i.e. to specify initial data, on a single slice. Regarding quantisation, we have to find an operator representation of the phase space functions, where the Dirac bracket is reproduced by commutators. The spectral analysis of these operators will then yield the allowed values, which these phase space functions can take in the quantum theory. Thus, the operator representation derived in this paper corresponds to good initial data and the theory predicts a certain spectrum for the initial values, exemplified by the invariant geometric operators \cite{BSTI}. In order to define a time evolution in our picture, one could try to define a physical Hamiltonian, whose associated time function corresponds to CMC time. However, already at the classical level, this operator is very difficult to define, because it involves inverting the differential operator $D_a D^a - \mathcal{P}(x)$. Nevertheless, by restricting to a spatial slice where $\mathcal{D}=0$ is a good gauge fixing, we can access the physical Hilbert space by trading the equation $\mathcal{H}=0$ for $\mathcal{D}$-invariance, which is easily solvable, and attack problems for which this ``Hilbert space of initial values'' suffices. Of course, a similar treatment of general relativity coupled to other types of matter, e.g. a minimally coupled scalar field as a clock as in \cite{DomagalaGravityQuantizedLoop}, is also possible, but the upsides of our treatment are that the gauge fixing $\mathcal{D}=0$ is locally generically valid, at least if we assume suitable energy conditions for the matter fields, and accessible for the proposed application to black holes \cite{BSTI}. The (global) properties of the CMC-gauge have been extensively studied in general relativity, see e.g. \cite{RendallConstantMeanCurvature} and references therein, and strong results are available. \\
A count of the degrees of freedom exemplifies that no obvious error lies in this train of thought: For pure gravity, the conformally invariant degrees of freedom are reduced by 1, as they would have been when restricting to $\mathcal{H}=0$. In the case of interest for this paper, general relativity conformally coupled to a scalar field, the count of degrees of freedom also provides a good check for the consistency of the framework. In the end, we have the phase space on which loop quantum gravity is based, however without the Hamiltonian constraint. On the other hand, the scalar field degree of freedom seems to have vanished, i.e. it has been explicitly rendered unphysical by a suitable canonical transformation after which the scalar field does not appear anywhere and has trivial Dirac brackets (or is killed at the quantum level when performing a Dirac type quantisation of the gauge unfixed theory). Still, the corresponding degree of freedom is not lost, but has been absorbed into the invariant metric $\tilde{q}_{ab}$ (cf. sections \ref{sec:dilaton}, \ref{sec:RemarkOnQuantisationAndObservables}). Morally speaking, by specifying initial data via specifying the invariant metric $\tilde{q}_{ab}$ and its conjugate momentum $\tilde{P}^{ab}$, we do not make any assertions about the unphysical scalar field, which then takes the role of an auxiliary field which can thought of to be tunable to a specific value which enforces $\mathcal{H}=0$. Technically however, this idea is implemented via the Dirac bracket, and a quantisation thereof (or considering $\mathcal{D}$-invariant functions in the gauge unfixed theory), which means that a quantum equation like $\hat{\mathcal{H}} \ket{\Psi} = 0$ is never even defined. \\
For future research, it will be very interesting to think about physical consequences of the presented quantisation of general relativity conformally coupled to a scalar field. For this, it is important to remember that the invariant geometric operators constructed in \cite{BSTI} exemplify that the choice of a time function has great impact on what the observables of a theory are. While the area operator in, e.g. \cite{GieselAQG4}, is physical only for surfaces which are embedded into a single constant time hypersurface labeled by a certain value of a dust field, the invariant area operators based on the present formalism are only valid for areas living on a CMC hypersurface labeled by $\mathcal{D}_{\delta} = 0$. While the matter content of the two theories is different form the beginning and it is thus hard to compare them, there is no indication that the above hypersurfaces should agree if we would formulate a theory with both matter fields. Regarding an experiment, one would therefore first have to ensure that the process measured is associated to a hypersurface which is compatible with the chosen time function. While an intuitive understanding of such a choice might be straight forward for measuring the area of a given surface, it is, at least to the authors, unclear for the situation of scattering in quantum field theory. Since, however, it has to be the goal of loop quantum gravity to reproduce quantum field theory in a certain regime, more research in this direction is necessary. Focussing on the invariant geometric operators, one can ask about the physical nature of the conformally coupled scalar field. In analogy to the Higgs field, one might consider it as a (heavy) field with a certain non-zero vacuum expectation value (cf. section \ref{sec:dilaton}). A naive insertion of the vacuum expectation value into the invariant geometric operators would increase the fundamental geometric scale of the theory, basically the lowest non-zero eigenvalues of the geometric operators, at least for vacuum expectations values below the Planck scale. While it might be tempting to conclude that this presents a mechanism to significantly raise the fundamental geometric scale of loop quantum gravity, much more research in this direction is needed before jumping to conclusions. Especially the question of the dynamics in the present formalism has to be taken into account when speculating about possible signatures of a raised fundamental scale which might be seen in, e.g. collider experiments. Also, applicability of the chosen time function has to be ensured for the experiment under consideration. This also means that when increasing the energy scale of the experiment far enough, one would expect that {\it any} time function would lose its applicability (as soon as quantum effects become important for the time function) and one would necessarily have to resort to a Dirac-type quantisation, as argued in \cite{ThiemannReducedPhaseSpace}.\\
To conclude, we have presented a new reduced phase space quantisation of general relativity conformally coupled to a scalar field, which is in several aspects different from the other available reduced phase space quantisations. While all of the so far proposed models seem to have their merits, their range of applicability is different and has to be checked for the specific problem at hand.

\section*{Acknowledgments} 
NB and AT thank the German National Merit Foundation for financial support. AS thanks the Evangelisches Studienwerk Villigst for financial support. We thank Kristina Giesel, Henrique Gomes, Sean Gryb, Tim Koslowski, and Thomas Thiemann for helpful and stimulating discussions. Also, we thank Kristina Giesel and Thomas Thiemann for carefully reading a draft of this paper and proposing numerous improvements. The idea for this work was born at the ESF-funded Quantum Gravity Colloqium 6, where Sean Gryb urged us to investigate the possibility of a connection formulation of shape dynamics with nice transformation properties under local conformal transformations.

\begin{appendix}

\newpage

\section{Conformally invariant Klein Gordon equation}
\label{app:1}
\label{app:A}
In this appendix, we summarise some formulas helpful to fill in the gaps in various calculations related with conformal transformations presented in the main text. These formulas can also be found in \cite{WaldGeneralRelativity}, with slightly different notation and $\Delta_g = 2$. Throughout this paper, if $\Omega$ is a striclty positive, smooth function, then we will call a transformation {\it{conformal}} if the metric is changed by
\be
	\tilde{g}_{\mu \nu} = \Omega^{\Delta^g} g_{\mu \nu}\text{,}
\ee
where $\Delta^g$ is called {\it{conformal weight}} of $g_{\mu\nu}$. The metric-compatible torsion-free covariant derivative is changed by
\be
	\tilde{\Gamma}_{\mu \nu}^{\rho} &=& \Gamma_{\mu \nu}^{\rho} - C^{\rho}_{\mu \nu}\text{,} \\
	C^{\rho}_{\mu \nu} &=& \frac{\Delta^g}{2} \Omega^{-1}\left( 2 \delta_{(\mu}\m^{\rho} \nabla_{\nu)}\Omega - g_{\mu\nu} \nabla^{\rho} \Omega \right) \text{.}
\ee
For the transformed Riemann tensor, Ricci tensor and Ricci scalar we find
\be
	\tilde{R}^{(D+1)}_{\mu \nu \rho}\m^{\sigma} &=& R^{(D+1)}_{\mu \nu \rho}\m^{\sigma} + \Omega^{-2} \Delta^g \left(\frac{\Delta^g}{2} + 1\right) (\nabla_{[\mu}\Omega) \left[ \delta_{\nu]}^{\sigma} (\nabla_{\rho} \Omega) - g_{\nu]\rho} \nabla^{\sigma} \Omega \right]  \nonumber \\ 
	& &- \Omega^{-2} \frac{(\Delta^g)^2}{2} g_{\rho [\mu}\delta_{\nu}^{\sigma} (\nabla_{\alpha} \Omega)(\nabla^{\alpha} \Omega) - \Delta^g \Omega^{-1} \left[ \delta_{[\nu}^{\sigma}\nabla_{\mu]}\nabla_{\rho} \Omega- g_{\rho [\nu} \nabla_{\mu]}\nabla^{\sigma}\Omega \right] \text{,} \\
	\tilde{R}^{(D+1)}_{\mu \nu} &=& R^{(D+1)}_{\mu \nu} + \Omega^{-2} \frac{\Delta^g }{2}  \left(\frac{\Delta^g}{2} + 1\right) \left[ (D-1) (\nabla_{\mu} \Omega)(\nabla_{\nu} \Omega) + g_{\mu \nu} (\nabla_{\rho} \Omega)(\nabla^{\rho} \Omega)\right]  \nonumber \\ 
	& &- \Omega^{-2} \frac{(\Delta^g)^2}{4} D g_{\mu \nu}(\nabla_{\rho} \Omega)(\nabla^{\rho} \Omega) -\Omega^{-1} \frac{\Delta^g }{2}  \left[ (D-1) \nabla_{\mu}\nabla_{\nu}\Omega + g_{\mu \nu} \nabla_{\rho}\nabla^{\rho}\Omega \right]\text{,}  \\
	\tilde{R}^{(D+1)}&=& \Omega^{-\Delta^g} \left[ R^{(D+1)} + \Omega^{-2} D \Delta^g  \left(- \frac{(D+1)\Delta^g}{4}+ \frac{\Delta^g}{2} + 1\right) (\nabla_{\rho} \Omega)(\nabla^{\rho} \Omega) \right. \nonumber \\
	& &\hspace{12mm}\left. -\Omega^{-1} D \Delta^g \nabla_{\rho}\nabla^{\rho}\Omega \right. \bigg]\text{.} \label{eq:R}
\ee
For the Laplacian of a scalar field $\Phi$ of conformal weight $\Delta^{\Phi}$, we find that
\begin{alignat}{4}
	\tilde{g}^{\mu \nu} \tilde{\nabla}_{\mu} \tilde{\nabla}_{\nu} \tilde{\Phi} &=& \Omega^{-\Delta^{g}} & \left[  \Omega^{\Delta^{\Phi}} \nabla_{\mu} \nabla^{\mu} \Phi + \Omega^{\Delta^{\Phi}-1} \left(\Phi \Delta^{\Phi}  \nabla_{\mu} \nabla^{\mu} \Omega + \left(2 \Delta^{\Phi} + \frac{D-1}{2} \Delta^{g} \right)  (\nabla_{\mu} \Phi) (\nabla^{\mu} \Omega)\right) \nonumber \right. \\
	& &\hspace{12mm} & \left. + \Omega^{\Delta^{\Phi} - 2} \Delta^{\Phi}\left(\Delta^{\Phi} - 1 + \frac{D-1}{2}\Delta^g\right) \Phi  (\nabla_{\mu} \Omega )(\nabla^{\mu} \Omega) \right] \text{.} 
\end{alignat}
Choosing 
\be
 \Delta^{\Phi} = -\frac{D-1}{4} \Delta^g \text{,}
\ee
we find that the terms $\propto(\nabla \Phi)(\nabla \Omega )$ vanish. Moreover, comparing with (\ref{eq:R}), we find that in the equation $\nabla\nabla \Phi + \alpha R^{(D+1)} \Phi$ both the terms $\propto \nabla\nabla \Omega$ and $\propto (\nabla \Omega)(\nabla \Omega)$ will cancel provided $\alpha = \frac{\Delta^{\Phi}}{D \Delta^g}$.

\section{Constraint algebra of general relativity conformally coupled to a scalar field}
This appendix provides calculational details showing that the algebra of constraints for general relativity conformally coupled to a scalar field is given by the standard hypersurface deformation algebra (\ref{eq:HDA}). Obviously, the Poisson brackets involving the spatial diffeomorphism constraint are the same as in the matter free case, since it generates spatial diffeomorphisms on all field variables. We therefore concentrate on the calculation of the Poisson bracket between two Hamiltonian constraints. Using the notation of section \ref{sec:ConformallyCoupledScalarField}, we split the calculation in the following parts:
\be
	\left\{\mathcal{H}[N], \mathcal{H}[M] \right\}  &=& \hspace{4mm} \underbrace{\left\{\mathcal{H}_{\text{Grav}}[N], \mathcal{H}_{\text{Grav}}[M] \right\}  }_{I} +  \underbrace{\left\{\mathcal{H}_{\Phi}[N], \mathcal{H}_{\Phi}[M] \right\}  }_{II}  \nonumber \\
	& &+  \underbrace{\left\{ \frac{\kappa}{(\Delta^g)^2 D(D-1) \sqrt{q}} \mathcal{D}^2[N], \frac{\kappa}{(\Delta^g)^2 D(D-1)\sqrt{q}} \mathcal{D}^2 [M] \right\}  }_{III}  \nonumber \\
	& & + \underbrace{ \left[ \left\{\mathcal{H}_{\text{Grav}}[N], \mathcal{H}_{\Phi}[M] \right\}  - (N \leftrightarrow M)\right]}_{IV} \nonumber \\
	& & -  \underbrace{ \left[  \left\{\mathcal{H}_{\text{Grav}}[N], \frac{\kappa}{(\Delta^g)^2 D(D-1) \sqrt{q}} \mathcal{D}^2[M] \right\}  - (N \leftrightarrow M) \right]}_{V} \nonumber \\
	& & -  \underbrace{ \left[ \left\{\mathcal{H}_{\Phi}[N], \frac{\kappa}{(\Delta^g)^2 D(D-1) \sqrt{q}} \mathcal{D}^2[M] \right\}  - (N \leftrightarrow M) \right]}_{VI} \text{.}
\ee
We will drop all surface terms, which is justified if the multipliers $N,M$ are compactly supported. If we allow for more general multipliers (like e.g. supertranslations and asymptotical Poincar{\'e} transformations in the asymptotically flat case), we might have to pass to improved generators, cf. sec. \ref{sec:Invertibility}. The calculation is simplified noting that terms without derivatives acting on the multipliers $N, M$ vanish due to the antisymmetry in $N, M$, which we will use repeatedly. First, we give some variational equations which will be useful for the calculation of the Poisson brackets,
\be
	\delta R^{(D)} &=& -R_{(D)}^{ab} \delta q_{ab} - q^{ab} (D_c D^c \delta q_{ab}) + (D^a D^b \delta q_{ab})\text{,} \label{eq:VarR}\\
	\delta (D_c D^c  \Phi) &=&  - (D^a D^b \Phi) \delta q_{ab} +(D_c D^c \delta \Phi)- (D^b \Phi) (D^a \delta q_{ab}) + \frac{1}{2} q^{ab} (D^c \Phi) (D_c \delta q_{ab}) \text{.} \label{eq:VarLaplace}
\ee
Note that the first summands on the right hand sides of (\ref{eq:VarR}) and (\ref{eq:VarLaplace}) do not contain derivates acting on $\delta q_{ab}$ and since $\mathcal{H}$ does not contain terms with derivatives acting on $P^{ab}$, these summands cannot contribute due to the afore mentioned antisymmetry. The Poisson brackets  for $I$ and $IV$ can be further simplified by noting that the terms containing $q^{ab} ... D \delta q_{ab}$ in (\ref{eq:VarR}) and (\ref{eq:VarLaplace}) can be neglected when Poisson commuting with $P^{ab}_{\text{tf}}$ due to its tracefreeness. \\
\\
Since $\mathcal{D}$ does not contain derivatives, $III$ vanishes. Since there are no derivates acting on the multipliers in (\ref{eq:TrafoScalarFieldHamiltonian}), we conclude that $VI$ gives zero as well. Using (\ref{eq:VarR}), only the third summand of which can contribute as described above, we find for $I$
\be
	I &=& \left\{\frac{\kappa}{\sqrt{q} a(\Phi)} P^{ab}_{\text{tf}} P_{ab}^{\text{tf}}[N], - \frac{1}{\kappa}\sqrt{q} R^{(D)}[M]  \right\}  - (N \leftrightarrow M) \nonumber \\
	&=&  \int_{\sigma} d^Dx ~ \frac{2 N}{a(\Phi)} P^{ab}_{\text{tf}} (D_a D_b M)  - (N \leftrightarrow M) \text{.}
\ee
A similar calculation using (\ref{eq:VarR}) and (\ref{eq:VarLaplace}) (only the third summands in both of these equations are relevant for the calculation, as was indicated above) yields for $IV$
\be
	IV &=& \left\{\frac{\kappa}{\sqrt{q} a(\Phi)} P^{ab}_{\text{tf}} P_{ab}^{\text{tf}}[N], - \frac{4 \alpha}{\kappa}\sqrt{q} \Phi D_a D^a \Phi + \frac{\alpha}{\kappa} \sqrt{q} R^{(D)} \Phi^2 \right\}  - (N \leftrightarrow M)\nonumber \\
	 &=& \int_{\sigma} d^Dx ~ \frac{8N\alpha}{a(\Phi)} P^{ab}_{\text{tf}} \Phi (D_a \Phi) (D_b M) - (N \leftrightarrow M) \nonumber \\
	 & & - \int_{\sigma} d^Dx ~ \frac{2 N \alpha}{a(\Phi)} P^{ab}_{\text{tf}} (D_a D_b \Phi^2 M)  - (N \leftrightarrow M) \text{.}
\ee
Combining these two terms, we find
\be
	I+IV &=&\int_{\sigma} d^Dx ~ \frac{2N}{a(\Phi)} P^{ab}_{\text{tf}} \left[ (D_a D_b a(\Phi) M) + 4 \alpha \Phi (D_a \Phi) (D_b M) \right]  - (N \leftrightarrow M)\nonumber \\
	&=& \int_{\sigma} d^Dx ~ \frac{2N}{a(\Phi)} P^{ab}_{\text{tf}} \left[ a(\Phi) (D_a D_b M) + 2 (D_a a(\Phi)) (D_b M) + 4 \alpha \Phi (D_a \Phi) (D_b M) \right]  - (N \leftrightarrow M) \nonumber \\
	&=&\int_{\sigma} d^Dx ~ 2N P^{ab}_{\text{tf}} (D_a D_b M) - (N \leftrightarrow M) \nonumber \\
	&=&- \int_{\sigma} d^Dx ~ 2N (D_a P^{ab}_{\text{tf}}) (D_b M) - (N \leftrightarrow M) \text{.}
\ee
The first three steps are straightforward. In the last one, we partially integrated and dropped the term $\propto (D_aN) P^{ab}_{\text{tf}} (D_b M)$ due to antisymmetry in $N,M$. For $II$, we find
\be
	II &=& \left\{-\frac{\lambda}{2\sqrt{q}} \pi_{\Phi}^2[N], \frac{1}{2\lambda} \sqrt{q} \left[-\frac{1}{D} (D^a \Phi)(D_a \Phi) + \frac{D-1}{D} \Phi (D_a D^a \Phi)\right][M] \right\}  - (N \leftrightarrow M)\nonumber \\
	&=& \int_{\sigma} d^Dx ~ \frac{N}{2} \pi_{\Phi} \left[ \frac{2}{D} (D_a \Phi)(D^a M) + \frac{D-1}{D} (D_a D^a \Phi M)\right] - (N \leftrightarrow M)\nonumber \\
	&=& \int_{\sigma} d^Dx ~ \frac{N}{2} \pi_{\Phi} \left[ 2 (D_a \Phi)(D^a M) + \frac{D-1}{D} \Phi (D_a D^a M)\right] - (N \leftrightarrow M)\nonumber \\
	&=& \int_{\sigma} d^Dx ~ \left[ N \pi_{\Phi} (D_a \Phi)(D^a M) - \frac{(D-1)N}{2D} (D_a \pi_{\Phi}  \Phi) (D^a M) \right] - (N \leftrightarrow M) \text{.}
\ee
Due to the antisymmetry in $N,M$, the only contributing terms are those with derivatives acting on $\Phi$, which do not Poisson commute only with the $\pi_{\Phi}^2$ term of $\mathcal{H}_{\Phi}$. For the second line, we used (\ref{eq:VarLaplace}) (only the second summand is relevant). In the third line, we computed the action of the Laplacian on $\Phi M$, in the fourth line we integrated by parts and dropped the terms $(D_aN) (D^aM)$. Finally, we have to calculate $V$, which, however, can be read off from (\ref{eq:HGravD}) keeping only the terms containing derivatives acting on the multipliers,
\be
	V &=& \int_{\sigma} d^Dx  ~  \frac{- 2M}{\Delta^g D} \mathcal{D} (D_a D^a N)  - (N \leftrightarrow M) \nonumber \\
	&=& \int_{\sigma} d^Dx  ~  \left[ \frac{2M}{D} (D_a  P)(D^a N)  - \frac{(D-1)M}{2D} (D_a \pi_{\Phi} \Phi)(D^a N)  \right]  - (N \leftrightarrow M) \text{.}
\ee
The factor of 2 in the first line stems from the $\mathcal{D}^2$ term in $V$, and in the second line we performed a partial integration. The second summands in the square brackets of $II$ and $V$ cancel each other and we finally obtain
\be
	\left\{\mathcal{H}[N], \mathcal{H}[M] \right\}  &=& I + ...+ VI \nonumber \\
	&=& \int_{\sigma} d^Dx ~ \left[ - 2N (D_a (P^{ab}_{\text{tf}} + \frac{1}{D} q^{ab} P)) (D_b M) + N \pi_{\Phi} (D_a\Phi) (D^a M) \right] - (N \leftrightarrow M) \nonumber \\
	&=& \mathcal{H}_a[ q^{ab} (N D_b M - M D_b N)] \text{,}
\ee
which coincides with the corresponding bracket in the matter free case (\ref{eq:HDA}). In the last step, we used $\int_{\sigma} d^Dx ~ (-2N^a q_{ac} D_b P^{bc}) = \int_{\sigma} d^Dx ~ (\mathcal{L}_{N}q)_{ab} P^{ab}$ in order to find the expression for $\mathcal{H}_a$ as defined in (\ref{eq:Ha}).

\label{app:3}

\section{Inclusion of Dirac fermions}
\label{app:2}
The (second order) action\footnote{Note that the standard first order action for Dirac fermions differs from the second order action by a term quadratic in torsion (up to boundary terms). This torsion term is indefinite and therefore spoils the estimation (\ref{eq:Estimation}).} for Dirac fermions is given by
\be
	S_{\Psi}[e, \Psi, \Psi^*] := \frac{i}{2} \int_M d^{D+1}X \sqrt{g} \left( \overline{\Psi} \,\slashed{e}^{\mu} \nabla_{\mu} \Psi -\overline{ \nabla_{\mu} \Psi}\, \slashed{e}^{\mu} \Psi \right) \text{,}
\ee
where $\overline{\Psi} := \Psi^* \gamma^0$, $\nabla$ denotes the vielbein compatible connection, $\nabla_{\mu} \Psi = \partial_{\mu} \Psi + \frac{i}{2} \Gamma_{\mu IJ}(e) \Sigma^{IJ} \Psi$ and $\Sigma^{IJ} := - \frac{i}{4} [\gamma^I , \gamma^J]$. This action is conformally invariant if we choose
$\Delta^{\Psi} = -\frac{D}{4} \Delta^g$. Performing the $(D+1)$ - split and introducing half densitised (and therefore conformally invariant) fermionic variables $\psi := \sqrt[4]{q} \Psi$, the action reads
\be
	S_{\Psi} &=&  \int_{\mathbb{R}} dt \int_{\sigma} d^{D}x ~ \left\{ N \frac{i}{2} \overline{\psi} \left[\slashed{e}^{a} D^{\text{hyb}}_{a} \psi + \frac{2}{N} \, \slashed{n} \, e^{b}\m_{[I} n_{J]} (\partial_b N) i \Sigma^{IJ} \psi + \text{c.c.} \right]  \right.\nonumber \\
	&& \left. \hspace{2.5cm} - ~ \pi_{\psi} (\partial_T \psi - \partial_N \psi) + M^{b}_L (\mathcal{L}_n e)_{b}^L \right\} \text{,}
\ee 
where
\be
	\pi_{\psi} &:=&  - i \overline{\psi} \, \slashed{n} \text{,} \\
	M^{b}_L &:=& \frac{1}{4} N\overline{\psi} \left[2 \slashed{e}^{(a} e^{b)}\m_{[I} n_{J]} e_{aL} + \slashed{n} \, e^{b}\m_{[I} \left(\bar{\eta}_{J]L} - 2 n_{J]} n_L\right) \right] \Sigma^{IJ} \psi + \text{c.c.}
\ee
and $e_a^I$ denotes the pullback of the spacetime vielbein $e_{\mu}^I$. Moreover, $n^I$ is the unique unit normal (up to sign) orthogonal to $e_a^I$, $D_a^{\text{hyb}}$ denotes the hybrid spin connection introduced by Peldan \cite{PeldanActionsForGravity} which annihilates $e_a^I$ and $\bar{\eta}_{IJ}$ denotes the projector on the subspace orthogonal to $n^I$, $\bar{\eta}_{IJ} := \eta_{IJ} + n_I n_J$. Explicitly writing out the complex conjugate terms, one finds that the terms $\propto (\partial N)$ as well as the $\slashed{e}$ and the $\slashed{n} \, n_L$ terms in $M^b_L$ drop out, 
\be
S_{\Psi} &=&  \int_{\mathbb{R}} dt \int_{\sigma} d^{D}x ~ \left\{ N \frac{i}{2} \overline{\psi} \, \slashed{e}^{a} D^{\text{hyb}}_{a} \psi -  \pi_{\psi} (\partial_T \psi - \partial_N \psi) + M^{b}_L (\mathcal{L}_n e)_{b}^L \right\} \text{,} \\
M^{b}_L &=& \frac{1}{4} N\overline{\psi} (\slashed{n} \Sigma^{IJ} + \Sigma^{IJ} \slashed{n}) \psi e^{b}\m_{[I} \bar{\eta}_{J]L}  \text{.}
\ee
Combining this action with (\ref{eq:CombinedAction}), we obtain
\be
	S &=& \int_{\mathbb{R}} dt \int_{\sigma} d^Dx ~ \left\{ N [...] - \pi_{\psi} ( \partial_T \psi - \partial_N \psi) + M^{b}_L (\mathcal{L}_n e)_{b}^L + \frac{1}{2\kappa} M^{ab,cd} (\mathcal{L}_n q)_{ab}(\mathcal{L}_n q)_{cd} \right.  \nonumber \\
	& &\left. \hspace{2.5cm}+~  \frac{1}{\kappa} M^{ab} (\mathcal{L}_n q)_{ab} (\mathcal{L}_n \Phi)+\frac{1}{2\kappa} M (\mathcal{L}_n \Phi)^2 \right\} \text{,}	
\ee
where we dropped surface terms and $M^{ab,cd}$, $M^{ab}$ and $M$ were defined in section \ref{sec:ConformallyCoupledScalarField}. For convenience, we will use the co-vielbein as canonical coordinate, not the usual densitised vielbein. Therefore, its conjugate variable $k^{aI}$ will be a $\mathbb{R}^{1,D}$-valued vector density\footnote{To obtain the standard variables, note that the transformation $\{e_a\m^I, k^b\m_J\} \rightarrow \{K_{aI} := \frac{1}{\sqrt{q}} (e_{bI}e_{aJ} - \frac{1}{D-1} e_{aI}e_{bJ} - s q_{ab} n_I n_J) k^{bJ} , E^{aI} := \sqrt{q} q^{ab} e_b\m^I\}$ is canonical, as can be easily verified. Also, in order to match these variables with those used in section \ref{sec:ConformallyInvariantMetricAndConnectionVariables}, we need to impose time gauge, i.e. fix an internal timelike normal $n^I = \delta^I_0$.},
\be
	k^{aI} := \frac{\delta L}{\delta \dot{e}_{aI}} = \frac{2}{N \kappa} \left[ \frac{\kappa}{2} M^{aI} + M^{ab,cd} (\mathcal{L}_n q)_{cd} ~ e_b^I + M^{ab} (\mathcal{L}_n \Phi) e_b^I \right] = \frac{1}{N} M^{aI} + 2 P^{ab} e_b^I \text{,}
\ee
where $P^{ab}$ is the function of $q, \mathcal{L}_T q$ given in (\ref{eq:MetricMomenta}). Therefore, this equation can obviously not be solved for all velocities $(\mathcal{L}_T e)_{a}^I$ but only for $(\mathcal{L}_T q)_{ab}$. Contracting with $e^{aJ}$ and symmetrising / anti - symmetrising in the indices $I \leftrightarrow J$, we obtain
\be
	2 P^{ab}(q, \mathcal{L}_T q) &=& k^{(a| I} e^{b)}\m_I  \text{,} \\
	0 = \left(k^{a[I} - \frac{1}{N} M^{a[I} \right) e_a\m^{J]} &=& k^{a[I} e_a\m^{J]} - \frac{1}{4} \overline{\psi} \{\slashed{n}, \Sigma^{IJ} \} \psi =: G^{IJ} \text{.}
\ee
The second equation, i.e. the momenta which cannot be solved for the velocities, gives the usual so$(1,D)$ Gau{\ss} constraint. The first equation can be solved for $(\mathcal{L}_T q)_{ab}$ to obtain the same result as in (\ref{eq:SolvedForMetric}) if we replace $P^{ab}$ by $\frac{1}{2}k^{(a| I} e^{b)}_I$, and a minute's thought reveals that the same replacement gives the (gravitational and scalar field part of the) Hamiltonian and spatial diffeomorphism constraint, i.e. 
\be
	\mathcal{H}_a[N^a] &:=& \int_{\sigma} d^Dx N^a \mathcal{H}_a = \int_{\sigma} d^Dx \left[ k^{aI} (\mathcal{L}_N e)_{aI} + \pi_{\Phi} (\mathcal{L}_N \Phi) - \pi_{\psi} (\mathcal{L}_N \psi)\right]\text{,} \\
	\mathcal{H}[N] &:=& \int_{\sigma} d^Dx N \mathcal{H} =  \int_{\sigma} d^Dx N \left[ \mathcal{H}_{\text{Grav}} + \mathcal{H}_{\Phi} + \mathcal{H}_{\psi} - \frac{\kappa}{4 D(D-1)\sqrt{q}} \mathcal{D}^2 \right]\text{,} \\
	\kappa \mathcal{H}_{\text{Grav}} &:=&  \frac{\kappa^2}{4 \sqrt{q} a(\Phi)}  (k^{(a|I} e^{b)}\m_I)^{\text{tf}}(k_{(a}\m^{I} e_{b)I})^{\text{tf}} - \sqrt{q} R^{(D)} \text{,}\\
	\kappa \mathcal{H}_{\Phi} &:=& \frac{\kappa}{2\lambda} \sqrt{q} \left[ - \frac{\lambda^2}{q} \pi_{\Phi}^2 - \frac{1}{D} q^{ab} (D_a \Phi) (D_b \Phi) + \frac{D-1}{D} \Phi D_a D^a \Phi + \frac{1}{D} \frac{\Delta^{\Phi}}{\Delta^g} R^{(D)} \Phi^2\right]\text{,}~~~~~~\\
	\kappa \mathcal{H}_{\psi} &:=& \frac{\kappa i}{2} \left( \overline{\psi} \slashed{e}^a D_a^{\text{hyb}} \psi \right) + \text{c.c.} \text{,}\\
	\mathcal{D} &:=& \frac{1}{2} \Delta^g k^{aI} e_{aI} + \Delta^{\Phi} \pi_{\Phi} \Phi\text{.}
\ee
It is obvious that the half densitised fermionic variables do not transform under conformal transformations generated by the chosen $\mathcal{D}$ and it is easy to verify that $\mathcal{H}_{\psi}$ transforms with conformal weight $- \frac{1}{2} \Delta^g$, in agreement with the results we obtained by general arguments in section \ref{sec:ExtensionToStandardModelMatterFields}.

\end{appendix}

\newpage


\bibliography{pa10.bbl}

\providecommand{\href}[2]{#2}\begingroup\raggedright\begin{thebibliography}{10}

\bibitem{BSTI}
N.~Bodendorfer, A.~Stottmeister, and A.~Thurn, ``{Loop quantum gravity without
  the Hamiltonian contraint},'' {\tt arXiv:1203.6525 [gr-qc]}.

\bibitem{GieselAQG4}
K.~Giesel and T.~Thiemann, ``{Algebraic quantum gravity (AQG): IV. Reduced
  phase space quantization of loop quantum gravity},'' {\em Classical and
  Quantum Gravity} {\bf 27} (2010) 175009, {\tt arXiv:0711.0119 [gr-qc]}.

\bibitem{DomagalaGravityQuantizedLoop}
M.~Domagala, K.~Giesel, W.~Kaminski, and J.~Lewandowski, ``{Gravity quantized:
  Loop quantum gravity with a scalar field},'' {\em Physical Review D} {\bf 82}
  (2010) 104038, {\tt arXiv:1009.2445 [gr-qc]}.

\bibitem{HusainTimeAndA}
V.~Husain and T.~Pawlowski, ``{Time and a physical Hamiltonian for quantum
  gravity},'' {\em Physical Review Letters} {\bf 108} (2012) 141301, {\tt
  arXiv:1108.1145 [gr-qc]}.

\bibitem{WaldGeneralRelativity}
R.~M. Wald, {\em {General Relativity}}.
\newblock University Of Chicago Press, 1984.

\bibitem{MitraGaugeInvariantReformulationAnomalous}
P.~Mitra and R.~Rajaraman, ``{Gauge-invariant reformulation of an anomalous
  gauge theory},'' {\em Physics Letters B} {\bf 225} (1989) 267--271.

\bibitem{AnishettyGaugeInvarianceIn}
R.~Anishetty and A.~S. Vytheeswaran, ``{Gauge invariance in second-class
  constrained systems},'' {\em Journal of Physics A: Mathematical and General}
  {\bf 26} (1993) 5613--5619.

\bibitem{VytheeswaranGaugeUnfixingIn}
A.~S. Vytheeswaran, ``{Gauge unfixing in second-class constrained systems},''
  {\em Annals of Physics} {\bf 236} (1994) 297--324.

\bibitem{HenneauxQuantizationOfGauge}
M.~Henneaux and C.~Teitelboim, {\em {Quantization of Gauge Systems}}.
\newblock Princeton University Press, 1994.

\bibitem{DittrichPartialAndComplete}
B.~Dittrich, ``{Partial and complete observables for Hamiltonian constrained
  systems},'' {\em General Relativity and Gravitation} {\bf 39} (2007)
  1891--1927, {\tt arXiv:gr-qc/0411013}.

\bibitem{RovelliWhatIsObservable}
C.~Rovelli, ``{What is observable in classical and quantum gravity?},'' {\em
  Classical and Quantum Gravity} {\bf 8} (1991) 297--316.

\bibitem{RovelliPartialObservables}
C.~Rovelli, ``{Partial observables},'' {\em Physical Review D} {\bf 65} (2002)
  124013, {\tt arXiv:gr-qc/0110035}.

\bibitem{ThiemannReducedPhaseSpace}
T.~Thiemann, ``{Reduced phase space quantization and Dirac observables},'' {\em
  Classical and Quantum Gravity} {\bf 23} (2006) 1163--1180, {\tt
  arXiv:gr-qc/0411031}.

\bibitem{GomesTheLinkBetween}
H.~Gomes and T.~Koslowski, ``{The link between general relativity and shape
  dynamics},'' {\em Classical and Quantum Gravity} {\bf 29} (2012) 075009, {\tt
  arXiv:1101.5974 [gr-qc]}.

\bibitem{GomesEinsteinGravityAs}
H.~Gomes, S.~Gryb, and T.~Koslowski, ``{Einstein gravity as a 3D conformally
  invariant theory},'' {\em Classical and Quantum Gravity} {\bf 28} (2011)
  045005, {\tt arXiv:1010.2481 [gr-qc]}.

\bibitem{ThiemannModernCanonicalQuantum}
T.~Thiemann, {\em {Modern Canonical Quantum General Relativity}}.
\newblock Cambridge University Press, Cambridge, 2007.

\bibitem{BinzGeometryOfClassical}
E.~Binz, J.~Sniatycki, and H.~Fischer, {\em {Geometry of Classical Fields}}.
\newblock Dover Publications, 2006.

\bibitem{ChoquetBruhatGeneralRelativityAnd}
Y.~Choquet-Bruhat, {\em {General Relativity and the Einstein Equations}}.
\newblock Oxford University Press, USA, 2009.

\bibitem{GribovQuantizationOfNon}
V.~Gribov, ``{Quantization of non-Abelian gauge theories},'' {\em Nuclear
  Physics B} {\bf 139} (1978) 1--19.

\bibitem{BTTII}
N.~Bodendorfer, T.~Thiemann, and A.~Thurn, ``{New variables for classical and
  quantum gravity in all dimensions: II. Lagrangian analysis},'' {\em Classical
  and Quantum Gravity} {\bf 30} (2013) 045002, {\tt arXiv:1105.3704 [gr-qc]}.

\bibitem{GieselManifestlyGaugeInvariantI}
K.~Giesel, S.~Hofmann, T.~Thiemann, and O.~Winkler, ``{Manifestly
  gauge-invariant general relativistic perturbation theory: I. Foundations},''
  {\em Classical and Quantum Gravity} {\bf 27} (2010) 055005, {\tt
  arXiv:0711.0115 [gr-qc]}.

\bibitem{GieselManifestlyGaugeInvariantII}
K.~Giesel, S.~Hofmann, T.~Thiemann, and O.~Winkler, ``{Manifestly
  gauge-invariant general relativistic perturbation theory: II. FRW background
  and first order},'' {\em Classical and Quantum Gravity} {\bf 27} (2010)
  055006, {\tt arXiv:0711.0117 [gr-qc]}.

\bibitem{BarberoRealAshtekarVariables}
J.~Barbero, ``{Real Ashtekar variables for Lorentzian signature space-times},''
  {\em Physical Review D} {\bf 51} (1995) 5507--5510, {\tt
  arXiv:gr-qc/9410014}.

\bibitem{ImmirziQuantumGravityAnd}
G.~Immirzi, ``{Quantum gravity and Regge calculus},'' {\em Nuclear Physics B -
  Proceedings Supplements} {\bf 57} (1997) 65--72, {\tt arXiv:gr-qc/9701052}.

\bibitem{RovelliTheImmirziParameter}
C.~Rovelli and T.~Thiemann, ``{The Immirzi parameter in quantum general
  relativity},'' {\em Physical Review D} {\bf 57} (1998) 1009--1014, {\tt
  arXiv:gr-qc/9705059}.

\bibitem{HaagOnQuantumField}
R.~Haag, ``{On quantum field theories},'' {\em Danske Videnskabernes Selskab
  Matematisk-Fysiske Meddelelser} {\bf 29} (1955), no.~12 1--37.

\bibitem{HaagLocalQuantumPhysics}
R.~Haag, {\em {Local Quantum Physics: Fields, Particles, Algebras}}.
\newblock Springer, 2nd~ed., 1996.

\bibitem{ArnoldMathematicalMethodsOf}
V.~I. Arnold, A.~Weinstein, and K.~Vogtmann, {\em {Mathematical Methods of
  Classical Mechanics}}.
\newblock Springer, 1989.

\bibitem{WangConformalGeometrodynamicsTrue}
C.~Wang, ``{Conformal geometrodynamics: True degrees of freedom in a truly
  canonical structure},'' {\em Physical Review D} {\bf 71} (2005) 124026, {\tt
  arXiv:gr-qc/0501024}.

\bibitem{WangTowardsConformalLoop}
C.~Wang, ``{Towards conformal loop quantum gravity},'' {\em Journal of Physics:
  Conference Series} {\bf 33} (2006) 285--290, {\tt arXiv:gr-qc/0512023}.

\bibitem{BernalSmoothnessOfTime}
A.~N. Bernal and M.~S\'{a}nchez, ``{Smoothness of Time Functions and the Metric
  Splitting of Globally Hyperbolic Spacetimes},'' {\em Communications in
  Mathematical Physics} {\bf 257} (2005) 43--50, {\tt arXiv:gr-qc/0401112}.

\bibitem{GerochDomainOfDependence}
R.~Geroch, ``{Domain of Dependence},'' {\em Journal of Mathematical Physics}
  {\bf 11} (1970) 437.

\bibitem{BaerWaveEquationsOn}
C.~Baer, N.~Ginoux, and F.~Pfaeffle, ``{Wave Equations on Lorentzian Manifolds
  and Quantization},'' in {\em ESI Lectures in Mathematics and Physics},
  p.~199, European Mathematical Society Publishing House2007.
\newblock {\tt arXiv:0806.1036 [math.DG]}.

\bibitem{DiracLecturesOnQuantum}
P.~A.~M. Dirac, {\em {Lectures on Quantum Mechanics}}.
\newblock Belfer Graduate School of Science, Yeshiva University Press, New
  York, 1964.

\bibitem{ArnowittTheDynamicsOf}
R.~Arnowitt, S.~Deser, and C.~W. Misner, ``{The dynamics of general
  relativity},'' in {\em Gravitation: An introduction to current research}
  (L.~Witten, ed.), (New York), pp.~227--265, Wiley, 1962.
\newblock {\tt arXiv:gr-qc/0405109}.

\bibitem{DiracFixationOfCoordinates}
P.~Dirac, ``{Fixation of Coordinates in the Hamiltonian Theory of
  Gravitation},'' {\em Physical Review} {\bf 114} (1959) 924--930.

\bibitem{AshtekarIsolatedHorizonsThe}
A.~Ashtekar, A.~Corichi, and K.~Krasnov, ``{Isolated Horizons: the Classical
  Phase Space},'' {\em Advances in Theoretical and Mathematical Physics} {\bf
  3} (2000) 419--478, {\tt gr-qc/9905089}.

\bibitem{LeeTheYamabeProblem}
J.~M. Lee and T.~H. Parker, ``{The Yamabe problem},'' {\em Bulletin (New
  Series) of the American Mathematical Society} {\bf 17} (1987) 37--91.

\bibitem{GromovTheClassificationOf}
M.~Gromov and H.~{Blaine Lawson Jr.}, ``{The Classification of Simply Connected
  Manifolds of Positive Scalar Curvature},'' {\em The Annals of Mathematics}
  {\bf 111} (1980) 423--434.

\bibitem{AshtekarAUnifiedTreatment}
A.~Ashtekar and R.~O. Hansen, ``{A unified treatment of null and spatial
  infinity in general relativity. I. Universal structure, asymptotic
  symmetries, and conserved quantities at spatial infinity},'' {\em Journal of
  Mathematical Physics} {\bf 19} (1978) 1542--1566.

\bibitem{BeigEinsteinsEquationsNear}
R.~Beig and B.~G. Schmidt, ``{Einstein's equations near spatial infinity},''
  {\em Communications in Mathematical Physics} {\bf 87} (1982) 65--80.

\bibitem{MurchadhaExistenceAndUniqueness}
N.~O'Murchadha, ``{Existence and uniqueness of solutions of the Hamiltonian
  constraint of general relativity on compact manifolds},'' {\em Journal of
  Mathematical Physics} {\bf 14} (1973) 1551--1557.

\bibitem{HenneuaxHamiltonianTreatmentOf}
M.~Henneaux, ``{Hamiltonian treatment of asymptotically anti-de Sitter
  spaces},'' {\em Physics Letters B} {\bf 142} (1984) 355--358.

\bibitem{HenneuaxAsymptoticallyAntiDe}
M.~Henneaux and C.~Teitelboim, ``{Asymptotically anti-de Sitter spaces},'' {\em
  Communications in Mathematical Physics} {\bf 98} (1985) 391--424.

\bibitem{AshtekarAsymptoticallyAntiDe}
A.~Ashtekar and A.~Magnon, ``{Asymptotically anti-de Sitter space-times},''
  {\em Classical and Quantum Gravity} {\bf 1} (1984) L39--L44.

\bibitem{AshtekarAsymptoticallyAntiDeConserved}
A.~Ashtekar and S.~Das, ``{Asymptotically anti-de Sitter spacetimes: conserved
  quantities},'' {\em Classical and Quantum Gravity} {\bf 17} (2000) L17--L30,
  {\tt hep-th/9911230}.

\bibitem{BartnikRemarksOnCosmological}
R.~Bartnik, ``{Remarks on cosmological spacetimes and constant mean curvature
  surfaces},'' {\em Communications in Mathematical Physics} {\bf 117} (1988)
  615--624.

\bibitem{AshtekarNewVariablesFor}
A.~Ashtekar, ``{New Variables for Classical and Quantum Gravity},'' {\em
  Physical Review Letters} {\bf 57} (1986) 2244--2247.

\bibitem{BTTI}
N.~Bodendorfer, T.~Thiemann, and A.~Thurn, ``{New variables for classical and
  quantum gravity in all dimensions: I. Hamiltonian analysis},'' {\em Classical
  and Quantum Gravity} {\bf 30} (2013) 045001, {\tt arXiv:1105.3703 [gr-qc]}.

\bibitem{RovelliQuantumGravity}
C.~Rovelli, {\em {Quantum Gravity}}.
\newblock Cambridge University Press, Cambridge, 2004.

\bibitem{GomesCouplingShapeDynamics}
H.~Gomes and T.~Koslowski, ``{Coupling Shape Dynamics to Matter Gives
  Spacetime},'' {\em General Relativity and Gravitation} {\bf 44} (2012)
  1539--1553, {\tt arXiv:1110.3837 [gr-qc]}.

\bibitem{KieferNonminimallyCoupledScalar}
C.~Kiefer, ``{Non-minimally coupled scalar fields and the initial value problem
  in quantum gravity},'' {\em Physics Letters B} {\bf 225} (1989) 227--232.

\bibitem{EpsteinNonpositivityOfThe}
H.~Epstein, V.~Glaser, and A.~Jaffe, ``{Nonpositivity of the energy density in
  quantized field theories},'' {\em Il Nuovo Cimento} {\bf 36} (1965)
  1016--1022.

\bibitem{FordClassicalScalarFields}
L.~Ford and T.~Roman, ``{Classical scalar fields and the generalized second
  law},'' {\em Physical Review D} {\bf 64} (2001) 024023.

\bibitem{GilbargEllipticPartialDifferential}
D.~Gilbarg and N.~S. Trudinger, {\em {Elliptic Partial Differential Equations
  of Second Order}}.
\newblock Springer, 2nd~ed., 1983.

\bibitem{HenneauxAsymptoticBevaiorAnd}
M.~Henneaux, C.~Mart\'{\i}nez, R.~Troncoso, and J.~Zanelli, ``{Asymptotic
  behavior and Hamiltonian analysis of anti-de Sitter gravity coupled to scalar
  fields},'' {\em Annals of Physics} {\bf 322} (2007) 824--848, {\tt
  arXiv:hep-th/0603185}.

\bibitem{ThiemannKinematicalHilbertSpaces}
T.~Thiemann, ``{Kinematical Hilbert spaces for Fermionic and Higgs quantum
  field theories},'' {\em Classical and Quantum Gravity} {\bf 15} (1998)
  1487--1512, {\tt arXiv:gr-qc/9705021}.

\bibitem{ThiemannQSD5}
T.~Thiemann, ``{Quantum spin dynamics (QSD) V: Quantum Gravity as the Natural
  Regulator of Matter Quantum Field Theories},'' {\em Classical and Quantum
  Gravity} {\bf 15} (1998) 1281--1314, {\tt arXiv:gr-qc/9705019}.

\bibitem{RendallConstantMeanCurvature}
A.~D. Rendall, ``{Constant Mean Curvature Foliations in Cosmological
  Spacetimes},'' {\em Helvetica Physica Acta} {\bf 69} (1996) 490--500.

\bibitem{HojmanGeometrodynamicsRegained}
S.~Hojman, K.~Kuchar, and C.~Teitelboim, ``{Geometrodynamics regained},'' {\em
  Annals of Physics} {\bf 96} (1976) 88--135.

\bibitem{PeldanActionsForGravity}
P.~Peldan, ``{Actions for gravity, with generalizations: A Review},'' {\em
  Classical and Quantum Gravity} {\bf 11} (1994) 1087--1132, {\tt
  arXiv:gr-qc/9305011}.

\end{thebibliography}\endgroup

\end{document}